\documentclass[onecolumn,tighten,iop,numberedappendix]{emulateapj}

\usepackage[english]{babel}

\usepackage{amsmath}
\usepackage{amsfonts}
\usepackage{amssymb}
\usepackage{aas_macros}

\begin{document}

\title{A new multi-dimensional general relativistic neutrino
  hydrodynamics code for core-collapse supernovae \\ I. Method and
  code tests in spherical symmetry}

\author{Bernhard M\"uller\altaffilmark{1},
        Hans-Thomas Janka\altaffilmark{1},
    and Harald Dimmelmeier\altaffilmark{2}}
\email{bjmuellr@mpa-garching.mpg.de}
\email{thj@mpa-garching.mpg.de}
\altaffiltext{1}{Max-Planck-Institut f\"ur Astrophysik, Karl-Schwarzschild-Str. 1, D-85748 Garching, Germany}
\altaffiltext{2}{Physics Department, Aristotle University of Thessaloniki, GR-54124 Thessaloniki, Greece}
\email{harrydee@mpa-garching.mpg.de}
\shorttitle{A NEW GENERAL RELATIVISTIC SUPERNOVA CODE}
\shortauthors{M\"ULLER ET AL.}

\begin{abstract}

We present a new general relativistic code for hydrodynamical
supernova simulations with neutrino transport in spherical and
azimuthal symmetry (1D and 2D, respectively). The code is a
combination of the \textsc{CoCoNuT} hydro module, which is a
Riemann-solver based, high-resolution shock-capturing method, and the
three-flavor, fully energy dependent \textsc{VERTEX} scheme for the
transport of massless neutrinos. \textsc{VERTEX} integrates the
coupled neutrino energy and momentum equations with a variable
Eddington factor closure computed from a model Boltzmann equation and
uses the ``ray-by-ray plus'' approximation in 2D, assuming the
neutrino distribution to be axially symmetric around the radial
direction at every point in space, and thus the neutrino flux to be
radial. Our spacetime treatment employs the Arnowitt-Deser-Misner
(ADM) 3+1 formalism with the conformal flatness condition (CFC) for
the spatial three-metric. This approach is exact for the 1D case and
has previously been shown to yield very accurate results for spherical
and rotational stellar core collapse. We introduce new formulations of
the energy equation to improve total energy conservation in
relativistic and Newtonian hydro simulations with grid-based Eulerian
finite-volume codes. Moreover, a modified version of the
\textsc{VERTEX} scheme is developed that simultaneously conserves
energy and lepton number in the neutrino transport with better
accuracy and higher numerical stability in the high-energy tail of the
spectrum. To verify our code, we conduct a series of tests in
spherical symmetry, including a detailed comparison with published
results of the collapse, shock formation, shock breakout, and
accretion phases.  Long-time simulations of proto-neutron star cooling
until several seconds after core bounce both demonstrate the
robustness of the new \textsc{CoCoNuT-VERTEX} code and show the
approximate treatment of relativistic effects by means of an effective
relativistic gravitational potential as in \textsc{PROMETHEUS-VERTEX}
to be remarkably accurate in spherical symmetry.
\end{abstract}


\section{Introduction}
\label{sec:intro}
The collapse and explosion of massive stars has been an active topic
in astrophysical research for decades, and has provided a major
impetus for advances in computational relativistic astrophysics, in
particular ever since the first numerical simulations of core collapse
were conducted in the 1960s ({Colgate} \& {White} 1966). The general outline of
this rich astrophysical phenomenon is fairly well established by now:
After the collapse and bounce of the central iron core to a compact
neutron star ($M/R \approx 0.1 \ldots 0.2$ in relativistic units) a
shock wave is formed, which stalls a few $\mathrm{ms}$ after
bounce. Afterwards, accretion onto the newly-formed neutron star
continues typically for a few hundreds of $\mathrm{ms}$ until the
shock is re-energized either by neutrino heating (the ``delayed
neutrino-driven mechanism'', {Wilson} 1985), probably aided by
multi-dimensional hydrodynamical instabilities like convection
{Epstein} 1979; {Herant}, {Benz}, \& {Colgate} 1992; {Herant} {et~al.} 1994; {Burrows}, {Hayes}, \&  {Fryxell} 1995; {Janka} \& {M\"uller} 1995, 1996; {Buras} {et~al.} 2006b, 2006a; {Bruenn} {et~al.} 2006; {Burrows} {et~al.} 2006a and the standing
accretion-shock instability (SASI; {Blondin}, {Mezzacappa}, \&  {DeMarino} 2003), or by some
alternative mechanism, such as MHD effects
(e.g.\ {Burrows} {et~al.}, 2007 and references therein), or energy
deposition by acoustic waves ({Burrows} {et~al.} 2006b). The ongoing quest
for the explosion mechanism, with various research groups pursuing a
number of different avenues, is a testimony to the complexity of the
problem, which involves not only multi-dimensional
(magneto-)hydrodynamics, but also kinetic theory (i.e.\ neutrino
transport), nuclear and particle physics, and general relativistic
(GR) gravity. Even for the few hundreds of $\mathrm{ms}$ from the
collapse to the onset of the explosion, there is no unified approach
which fully accounts for all these intricacies. Instead, a number of
quite disparate approaches has been used in the past to investigate
different aspects of core collapse supernovae, all of which are
arguably deficient in the sense that they sacrifice some essential
physical ingredients in favour of others.

Thus, there have been repeated attempts to solve the problem of the
explosion mechanism by means of an accurate treatment of neutrino
transport in spherical symmetry, combined with reasonably accurate
microphysics (i.e.\ the nuclear equation of state, nuclear burning, and
neutrino-matter interactions). These endeavours culminated in the
development of solvers for the neutrino Boltzmann equation
({Mezzacappa} \& {Messer} 1999; {Burrows} {et~al.} 2000a; {Liebend{\"o}rfer} {et~al.} 2001b, 2004; {Rampp} \& {Janka} 2002; {Yamada}, {Janka}, \& {Suzuki} 1999);
but the exclusion of multi-dimensional hydrodynamical effects is
obviously a serious shortcoming. On the other hand, the restriction to
spherical symmetry allows for a full GR treatment with relatively
little additional effort: Relativistic supernova simulations with grey
or multi-group flux-limited diffusion have been conducted since the
1980s in 1D ({Baron} {et~al.} 1989; {Bruenn}, {De Nisco}, \&  {Mezzacappa} 2001), and relativistic Boltzmann
transport became feasible after another few years
({Yamada} {et~al.} 1999; {Liebend{\"o}rfer} {et~al.} 2001b, 2004). As expected, GR
effects were shown to be anything but negligible: Comparisons with the
Newtonian case (see, e.g.\, {Bruenn} {et~al.} 2001) revealed a sizable
influence of GR on the mass of the inner core at shock formation, the
initial shock strength, the long-term evolution of the shock position,
the compactness of the proto-neutron star, and the neutrino
luminosities and energies.

Another complementary approach has focused primarily on the interplay
between neutrino heating/cooling and multi-dimensional
{(magneto-)}hydrodynamic effects, which has turned out to be no less
important for the understanding of the explosion mechanism than the
proper treatment of neutrino transport and general relativity:
Convective instabilities were found to operate both inside the
proto-neutron star ({Epstein} 1979) and the ``hot-bubble region''
behind the stalled shock ({Bethe} 1990). Hot-bubble convection in
particular was shown to be helpful for increasing the neutrino heating
efficiency already in the 1990s
({Herant} {et~al.} 1992, 1994; {Burrows} {et~al.} 1995; {Janka} \& {M\"uller} 1995, 1996), and another
recently identified hydrodynamic instability of the shock, known as
the ``standing accretion shock instability'' (SASI;
{Blondin} {et~al.} 2003), was found to have a similar effect. Nowadays,
the most advanced computational tools used in this approach combine
multi-dimensional Newtonian hydrodynamics with an up-to-date treatment
of the microphysics with multi-group neutrino transport methods of
similar sophistication as in the 1D case
({Livne} {et~al.} 2004; {Buras} {et~al.} 2006b; {Bruenn} {et~al.} 2006; {Liebend{\"o}rfer}, {Whitehouse}, \&  {Fischer} 2009). But some weak
spots remain to be criticized: First, multi-dimensional neutrino
transport has not been treated in its full complexity yet; the
approaches now in use either assume the neutrino flux vector to point
in the radial direction (the ray-by-ray approach;
{Buras} {et~al.} 2006b; {Bruenn} {et~al.} 2006), neglect energy redistribution in
phase space ({Livne} {et~al.} 2004; {Liebend{\"o}rfer} {et~al.} 2009), or involve even more
radical approximations. Second, all serious attempts to model the
development of multi-dimensional instabilities during the post-bounce
phase, while treating the neutrino transport with reasonable accuracy,
have been carried out in the framework of Newtonian hydrodynamics,
which is rather unsatisfactory considering the importance of GR
effects established in 1D studies. So far, attempts to include
relativistic corrections have been limited to the use of modified
gravitational potentials that can be heuristically derived from the
Tolman-Oppenheimer-Volkov equation of stellar structure
({Rampp} \& {Janka} 2002; {Marek} {et~al.} 2006). Although this method can be justified on
the basis of comparisons with results from relativistic 1D neutrino
transport codes and multi-dimensional relativistic hydrodynamics codes
without neutrino transport
({Liebend{\"o}rfer} {et~al.} 2005; {Marek} {et~al.} 2006; {M{\"u}ller}, {Dimmelmeier}, \&  {M{\"u}ller} 2008), it comes along with
several drawbacks: As the equations of Newtonian hydrodynamics are
left unaltered, it may lead to incorrect predictions for some
dynamical properties of the neutron star such as the oscillation
eigenfrequencies ({M{\"u}ller} {et~al.} 2008), which may also result in a
distortion of the gravitational wave spectrum. Moreover, the
``effective potential approach'' ignores the effects of relativistic
kinematics and still allows superluminal velocities -- which have
actually occurred in Newtonian simulations of the explosions of
O-Ne-Mg cores. Finally, it should be noted that, different from the
purely Newtonian case, a conservation law for the total
(i.e.\ kinetic, internal, and potential) energy cannot be formulated
when a modified gravitational potential is used. As a consequence, it
is difficult to exclude the possibility of non-physical energy loss or
generation, in particular as cumulative effect over long
  evolution periods.

On the other hand, the last few years have seen the development of
genuinely relativistic hydrodynamics codes for multi-dimensional
simulations of core collapse, often based on a rather crude treatment
of the microphysics, and without any up-to-date scheme for the
neutrino transport. Typically, simple analytic equations of state
(EoS), such as the hybrid EoS of {Janka}, {Zwerger}, \&  {M{\"o}nchmeyer} (1993) were employed until
recent years; and the polytropic models for the progenitor had only a
limited resemblance to iron cores from stellar evolution
calculations. These limitations have naturally led the numerical
relativity community to focus on the one hand on the only phases
of the supernova for which such severe approximations may be deemed
sufficient, i.e.\ the collapse and bounce of the core, and on the
  other hand on those issues such as black hole formation
({Sekiguchi} \& {Shibata} 2005), gravitational wave emission from rotational core
collapse ({Dimmelmeier}, {Font}, \&  {M{\"u}ller} 2002a, 2002b; {Shibata} \& {Sekiguchi} 2004), and the
development of non-axisymmetric instabilities ({Shibata} \& {Sekiguchi} 2005) that
do not lend themselves readily to a Newtonian analysis, at least not
with the same accuracy.  Obviously, such an approach, which neglects
physics that is unquestionably of paramount importance for the stellar
core collapse problem, is beset with large uncertainties. However, the
first steps towards a more sophisticated treatment of the microphysics
in core collapse supernovae were taken by {Dimmelmeier} {et~al.} (2007) and
{Ott} {et~al.} (2007a), who studied the collapse of rotating iron cores with
a nuclear equation of state and a simple parameterized scheme for the
deleptonization during collapse ({Liebend{\"o}rfer} 2005). The fact
that their findings are rather different from earlier studies with
more radical approximations -- they exclusively find gravitational
wave signals of the so-called type I ({Zwerger} \& {M{\"u}ller} 1997) -- underlines
the need for a reliable treatment of the microphysics also in the
field of relativistic stellar core collapse.

Clearly, the most satisfactory approach to modeling supernovae would
be to sacrifice as little of the essential ingredients (neutrino
transport, equation of state and other microphysics, general
relativity, multi-dimensional hydrodynamics) as possible in the
numerical models. In this paper, we therefore describe a relativistic
generalization of the Newtonian ray-by-ray-plus method of
{Buras} {et~al.} (2006b) for the neutrino transport, which we haved combined
with the so-called conformal flatness condition (CFC) or
Isenberg-Wilson-Mathews approximation ({Isenberg} 1978; {Wilson}, {Mathews}, \&  {Marronetti} 1996)
for the gravitational field equations. The method presented here has
been implemented by building on the existing radiation transport code
\textsc{VERTEX} ({Rampp} \& {Janka} 2002; {Buras} {et~al.} 2006b) and the relativistic
hydrodynamics code \textsc{CoCoNuT}
({Dimmelmeier} {et~al.} 2002a, 2005), and has already been applied
to 2D simulations of the collapse and post-bounce phase, which will be
the subject of future papers. At this stage, however, we shall confine
ourselves to presenting thorough tests of our new
\textsc{VERTEX-CoCoNuT} code in spherical symmetry.

This paper is organized as follows: In Sec.~\ref{sec:grrhd}, we
discuss the governing equations of hydrodynamics and neutrino
transport (``neutrino hydrodynamics'') for the ray-by-ray-plus
approximation, and comment on the their implementation in
\textsc{CoCoNuT} and \textsc{VERTEX}. Tests of the new code in
spherical symmetry are presented in Sec.~\ref{sec:1d_tests}, most
notably a detailed comparison of 1D simulations of the $15 \ M_\odot$
progenitor model of {Woosley} \& {Weaver} (1995) with \textsc{VERTEX-CoCoNuT},
the relativistic Boltzmann code \textsc{AGILE} of
{Liebend{\"o}rfer} {et~al.} (2004), and the Newtonian \textsc{VERTEX} code of
{Rampp} \& {Janka} (2002), in which GR effects are taken into account
approximately by using a modified gravitational potential. In this
context, the issue of energy and lepton number conservation also
receives particular attention. In order to demonstrate the ability of
\textsc{VERTEX-CoCoNuT} to evolve supernova models stably well beyond
$1 \ \mathrm{s}$ after core bounce, we also present results for a 1D
explosion model of an O-Ne-Mg core
({Nomoto} 1984, 1987; {Kitaura}, {Janka}, \&  {Hillebrandt} 2006).

Further technical details of our numerical implementation, which may
also be of relevance for similar radiation hydrodynamics codes, are
given in the Appendices: In App.~\ref{app:enecons} we address the
problem of total energy conservation for a self-gravitating fluid in
Newtonian and relativistic hydrodynamics, and in
App.~\ref{app:cons_dopp} we outline an efficient finite-difference
scheme for the neutrino moment equations that guarantees the
simultaneous conservation of energy and lepton number.

Note that geometrized units are used throughout Sec.~\ref{sec:grrhd},
i.e.\ both the speed of light and the gravitational constant are set to
unity: $G=c=1$. Greek indices run from $0$ to $3$, Latin indices from
$1$ to $3$.

\section{General Relativistic Neutrino Radiation Hydrodynamics}

\label{sec:grrhd}

\subsection{Hydrodynamics}
\label{sec:gr_hydro}
In general relativity, the hydrodynamic evolution of a perfect fluid
is governed by two conservation equations for the baryonic rest mass
current $J^\mu$ and the stress-energy tensor $T^{\mu\nu}$,
\begin{equation}
  \label{eq:ch_grrhd_hydro_eq}
  \nabla_\mu J^{\mu}=0, \quad \nabla_\nu T^{\mu\nu}=0,
\end{equation}
where $\nabla_\nu$ denotes the covariant derivative. For a perfect
fluid, $J^\mu$ and $T^{\mu\nu}$ can be expressed in terms of the
baryon rest-mass density $\rho$ in the local fluid frame, the
four-velocity $u^\mu$, the pressure $P$, the specific enthalpy
$h=1+\epsilon+P/\rho$ (where $\epsilon$ is the specific internal
energy), and the four-metric $g_{\mu\nu}$,
\begin{equation}
  J^\mu = \rho u^\mu, \quad T^{\mu\nu} = \rho h u^\mu u^\nu + P g^{\mu\nu}.
\end{equation}
For the metric, we adopt the Arnowitt-Deser-Misner (ADM) $3 + 1$
formalism ({Lichnerovicz} 1944) of general relativity to foliate the
spacetime into spacelike hypersurfaces. In this approach, the
four-dimensional line element $\mathrm{d} s^2=g_{\mu\nu}\ \mathrm{d} x^\mu \ \mathrm{d}
x^\nu$ is written as follows,
\begin{equation}
  \mathrm{d} s^2=-\alpha^2 \mathrm{d} t^2 +\gamma_{ij} \left(\mathrm{d} x^i+\beta^i \mathrm{d} t\right) \left(\mathrm{d} x^j+\beta^j \mathrm{d} t\right),
\end{equation}
where $\alpha$ is the lapse function, $\beta^i$ is the shift vector,
and $\gamma_{ij}$ is the induced three-metric on each
hypersurface. Using this decomposition of the four-metric, the
equations of hydrodynamics can be formulated in the frame of an
Eulerian observer\footnote{ Since the terminology in the existing
  literature on relativistic hydrodynamics and radiative transfer is
  not uniform, we add some remarks on this issue here.  The ``frame''
  of an observer (or a class of observers) is an orthonormal tetrad
  field whose time-like base vector coincides with the four-velocity
  of the observer. We shall use the terms ``Eulerian frame'' (as in
  {Banyuls} {et~al.} 1997) or ``lab frame'' (as in {Mihalas} \& {Weibel Mihalas} 1984)
  interchangeably to denote the frame of an Eulerian observer as
  described in the main body of the text. It is important to note that
  such an Eulerian observer is \emph{not} a fixed-coordinate observer
  unless the shift vector vanishes.  }  moving orthogonally to the
spacelike hypersurfaces, i.e.\ with a four-velocity
$n^\mu=(\alpha^{-1},\alpha^{-1}\beta^i)$. As in {Banyuls} {et~al.} (1997), we
introduce the following conserved variables\footnote{ $D$, $S_i$ and
  $\tau$ can be obtained form $J^\mu$ and $T^{\mu\nu}$ using $n^\mu$
  and the projection operator $\bot_{\nu}^{\mu}$ onto the
  three-hypersurfaces: $D=n_\mu J^\mu$, $S^i=\bot^i_\mu n_\nu
  T^{\mu\nu}$, and $\tau=n_\mu n_\nu T^{\mu\nu}-D $.},
\begin{equation}
  D=\rho W, \quad S^i=\rho h W^2 v^i, \quad  \tau=\rho h W^2-P-D,
\end{equation}
where $v^i=u^i/(\alpha u^0)+\beta^i/\alpha$ is the three-velocity in
the Eulerian frame, and $W=1/\sqrt{1-v_iv^i}$ is the corresponding
Lorentz factor. The equations of GR hydrodynamics in flux-conservative
form then read,
\begin{eqnarray}
  \label{eq:hydro_1}
  \frac{\partial \sqrt{\gamma} \rho W}{\partial t}+\frac{\partial \sqrt{-g} \rho W \hat{v}^i}{\partial x^i} &=&0 , \\
  \label{eq:hydro_2}
  \frac{\partial \sqrt{\gamma} \rho h W^2 v_j}{\partial t}+
  \frac{\partial \sqrt{-g}\left(\rho h W^2 v_j\hat{v}^i+\delta^i_j P\right)}{\partial x^i}&=&
  \frac{1}{2}\sqrt{-g} T^{\mu\nu}\frac{\partial g_{\mu\nu}}{\partial x^j}
  +\left(\frac{\partial \sqrt{\gamma} S_j}{\partial t}\right)_\mathrm{C},
  \\
  \label{eq:hydro_3}
  \frac{\partial \sqrt{\gamma} \tau}{\partial t}+
  \frac{\partial \sqrt{-g} \left(\tau \hat{v}^i + P v^i\right)}{\partial x^i}&=&
  \alpha \sqrt{-g} \left(T^{\mu 0}\frac{\partial \ln \alpha}{\partial x^\mu} -T^{\mu\nu}\Gamma^0_{\mu\nu}\right) 
  +\left(\frac{\partial \sqrt{\gamma} \tau}{\partial t}\right)_\mathrm{C}
.
\nonumber
\\
\end{eqnarray}
Here, $\hat{v}^i$ is defined as $\hat{v}^i=v^i-\beta^i / \alpha$, and
$g$ and $\gamma$ are the determinants of the four-metric $g_{\mu\nu}$
and the three-metric $\gamma_{ij}$, respectively. Different from the
purely hydrodynamic case discussed by {Banyuls} {et~al.} (1997), the
right-hand sides (RHSs) of Eqs.~(\ref{eq:hydro_2}) and
(\ref{eq:hydro_3}) contain source terms for the exchange of momentum
and energy with neutrinos (which can be related to angular moments of
the collision integral and are therefore denoted by a subscript ``C'')
in addition to the gravitational and geometric source terms. These
source terms will be specified in Sec.~\ref{sec:sources_cfc}.

Eqs.~(\ref{eq:hydro_1}) to (\ref{eq:hydro_3}) must be supplemented
by an additional equation expressing the conservation of electron
number in the absence of weak interactions (which will be discussed
separately),
\begin{equation}
  \label{eq:hydro_4}
  \frac{\partial \sqrt{\gamma} \rho W Y_e}{\partial t}+\frac{\partial \sqrt{-g} \rho W Y_e \hat{v}^i}{\partial x^i}=
  \left(\frac{\partial \sqrt{\gamma} \rho W Y_e}{\partial t}\right)_\mathrm{C},
\end{equation}
where $Y_e$ is the electron fraction (number of electrons minus number
of positrons per baryon). The change of the electron number due to
neutrino emission or absorption is accounted for by the source term
on the RHS, in a similar fashion as energy and momentum exchange with
neutrinos in Eqs.~(\ref{eq:hydro_2}) and (\ref{eq:hydro_3}).

Since we are dealing with a multi-component fluid whose composition is
not uniquely determined by the thermodynamic state $(\rho,T,Y_e)$
unless nuclear statistical equilibrium (NSE) applies, additional
conservation equations for the mass fractions $X_k$ of protons,
neutrons, $\alpha$-particles and heavier nuclei are also needed,
\begin{equation}
  \label{eq:hydro_5}
  \frac{\partial \sqrt{\gamma} \rho W X_k}{\partial t}+\frac{\partial \sqrt{-g} \rho W X_k \hat{v}^i}{\partial x^i}=0.
\end{equation}
Finally, the pressure and the composition (in the NSE regime) in
Eqs. (\ref{eq:hydro_2},\ref{eq:hydro_3},\ref{eq:hydro_5}) have to be provided by an
equation of state (EoS).

We rely on the \textsc{CoCoNuT} code
({Dimmelmeier} {et~al.} 2002a, 2005) to solve the equations of
relativistic hydrodynamics in spherical polar coordinates by means of
a high-resolution shock-capturing (HRSC) scheme, employing piecewise
parabolic (PPM) reconstruction ({Colella} \& {Woodward} 1984), an approximate
Riemann solver, and second-order Runge-Kutta timestepping. Our method
of choice for the Riemann solver is a new hybrid HLLC/HLLE scheme in
\textsc{CoCoNuT} along the lines of {Quirk} (1994), combining the
high resolution of the HLLC solver ({Toro}, {Spruce}, \& {Speares} 1994; {Mignone} \& {Bodo} 2005) with
the robustness of the HLLE solver ({Harten}, {Lax}, \& {van Leer} 1983; {Einfeldt} 1988)
against odd-even decoupling near grid-aligned shocks
({Quirk} 1994; {Liou} 2000; {Kifonidis} {et~al.} 2003; {Sutherland}, {Bisset}, \&  {Bicknell} 2003). In order to treat
the advection equations for the nuclear mass fractions, we have added
a simplified version of the consistent multi-fluid advection (CMA)
method by {Plewa} \& {M{\"u}ller} (1999). Moreover, the computational performance of
the code has been improved significantly by enhancing parallelism in
the axisymmetric (2D) mode.

Most notably, however, our most recent simulations also use a
reformulated energy equation instead of Eq.~(\ref{eq:hydro_3}) to
improve total energy conservation. The new scheme, which is described
in detail in App.~\ref{app:enecons}, is designed to alleviate a
well-known problem with the gravitational source term in the energy
equation: While there is a global conservation law for the ADM energy
(or the total energy in the Newtonian case), the gravitational source
term in the standard form of the energy equation~(\ref{eq:hydro_3})
used in {Dimmelmeier} {et~al.} (2002a, 2005, 2008) does
not have the form of a flux divergence, and therefore a standard
finite-volume discretization will lead to a numerical violation of
global energy conservation.  Nonetheless, as we demonstrate in
App.~\ref{app:enecons}, the gravitational source term can partly be
absorbed in a flux divergence term to reduce this conservation error
considerably. In the case of a stationary spacetime, or in the
Newtonian limit, it is even possible to achieve total energy
conservation up to machine precision.

\subsection{Metric Equations -- The Conformal Flatness Condition}
The \textsc{CoCoNuT} code solves Einstein's field equation using the
conformal flatness condition (CFC), which was introduced by
{Isenberg} (1978) and first used in a pseudo-dynamical context by
{Wilson} {et~al.} (1996).  In the CFC approximation, the spatial three-metric
is assumed to be conformally flat\footnote{ We note in passing that this assumption can always be
  fulfilled by an appropriate gauge choice for a spherically symmetric
  spacetime, and no approximations to the ADM equations are actually
  made by using CFC in that case. See Sec.~\ref{sec:approx_quality}
  for more details.}, i.e.\ it is obtained from the flat
three-metric\footnote{Thus, in spherical polar coordinates we have
  $\hat{\gamma}_{ij}=\mathrm{diag}\left( 1,r^2,r^2 \sin^2\theta
  \right)$.}  $\hat{\gamma}_{ij}$ by a conformal transformation
$\gamma_{ij}=\phi^4 \hat{\gamma}_{ij}$, where $\phi$ is the conformal
factor, thus reducing the number of independent metric quantities to
five.  As a consequence, the four constraint equations for the metric
in the ADM formalism, combined with a slicing condition, determine the
metric completely. In our case, we use maximal slicing, i.~e. we
require the trace of the extrinsic curvature $K^{ij}$ to vanish:
$K=K_i^i=0$. We then obtain a set of five non-linear elliptic
equations for $\alpha$, $\phi$, and $\beta^i$,
\begin{eqnarray}
  \label{eq:cfc_1}
  \hat{\Delta}\phi &=& -2\pi \phi^5 \left(E+\frac{K_{ij}K^{ij}}{16 \pi}\right), \\
  \label{eq:cfc_2}
  \hat{\Delta}(\alpha \phi) &=& -2\pi \alpha \phi^5 \left(E + 2 S + \frac{7 K_{ij}K^{ij}}{16 \pi}\right), \\
  \label{eq:cfc_3}
  \hat{\Delta}\beta^i &=& 16 \pi \alpha \phi^4 S^i + 2 \phi^{10} K^{ij} \hat{\nabla}_j\left(\frac{\alpha}{\phi^6}\right)-
  \frac{1}{3}\hat{\nabla}^i \hat{\nabla}_j \beta^j,
\end{eqnarray}
where $\hat{\Delta}$ and $\hat{\nabla}$ are the Laplace and covariant
derivative operators for a flat three-space.  The total matter-energy
density $E=\tau+D$ (for $\tau$ and $D$ see Sec.~(\ref{sec:gr_hydro})),
the three-momentum density $S_i$ and the trace $S=\gamma_{ij} S^{ij}$
of the spatial components $S_{ij}$ of the stress-energy tensor (as
measured by an Eulerian observer) appear in the non-linear source
terms on the RHS, and comprise contributions both from
matter and (neutrino) radiation. For the matter component we obtain
\begin{eqnarray}
 E &=& \rho h W^2 -P,\\
 S &=& \rho h W^2 v^2 + 3P,\\
 S^i &=& \rho h W^2 v^i,
\end{eqnarray}
while for neutrinos,
\begin{eqnarray}
  E&=&S= 4 \pi W^2\left(J+2 v_r H +v_r^2 K\right),\\
  S_1&=&4 \pi \phi^{-2} W^2\left[H\left(1+v_r^2\right)+ v_r \left(J+K\right)\right],\\
  S_2&=&S_3=0,
\end{eqnarray}
in our approximation. Here, $J$, $H$, and $K$ denote the zeroth, first
and second moments of the neutrino radiation intensity in the comoving
orthonormal frame (see Sec.~\ref{sec:neutrino_transport}), and
$v_r=\phi^2 v^1$ is the radial velocity component in the orthonormal
basis of an Eulerian observer.

We solve the CFC equations (\ref{eq:cfc_1}), (\ref{eq:cfc_2}), and
(\ref{eq:cfc_3}) in the reformulated version of {Cordero-Carri{\'o}n} {et~al.} (2009)
using a fixed point iteration scheme as described in
{Dimmelmeier} {et~al.} (2005); i.e.\ during each iteration step, the
non-linear source terms are updated, and the Laplace operators on the
left-hand side (LHS) of the equations are inverted by means of a multipole
expansion (cf. {M\"uller} \& {Steinmetz} 1995). While this method suffers from a
somewhat low convergence rate compared to the spectral nonlinear
Poisson solver of {Dimmelmeier} {et~al.} (2005), the underlying algorithm is
more easily amenable to parallelization, which is a critical issue for
multi-dimensional neutrino transport simulations.

\subsection{Neutrino Transport in CFC Spacetime}
\label{sec:neutrino_transport}
\subsubsection{Overview}
Up to now, all self-consistent numerical simulations of neutrino
transport in core collapse supernovae are based on the assumptions i)
that a semi-classical treatment of kinetic theory is applicable, and
ii) that neutrino oscillations need not be taken into account (see
{Cardall}, 2008; {Yamada}, 2000; {Strack} \& {Burrows}, 2005 for a discussion of the more
general case). In our approach to the transport problem, we retain
those assumptions, although they may be invalid under certain specific
circumstances (cf. Sec.~\ref{sec:approx_quality}).  Under this
proviso, the evolution of the invariant neutrino distribution function
$f$ (the number of neutrinos per phase-space volume $\mathrm{d}^3 x\,
\mathrm{d}^3 p$) is governed by the relativistic Boltzmann equation
for massless or ultra-relativistic particles, which reads (see
e.g.\ {Lindquist}, 1966; {Ehlers}, 1971; {Stewart}, 1971)
\begin{equation}
  \label{eq:rel_boltzmann_1}
  p^\mu \frac{\partial f}{\partial x^\mu}+\frac{\mathrm{d} p^\mu}{\mathrm{d} \lambda} \frac{\partial f}{\partial p^\mu}=\mathfrak{C}
\end{equation}
in any coordinate basis. Here, $f$ and the collision term
$\mathfrak{C}$ depend on the spacetime coordinates $x^\mu$ and the
four-momentum vector $p^\mu$ (obeying the mass-shell constraint $p_\mu
p^\mu=0$) as measured in the associated holonomic base $\partial_\mu$;
$\lambda$ is the affine parameter.

In general, Eq.~(\ref{eq:rel_boltzmann_1}) describes a six-dimensional
time-evolution problem, which can be scaled down considerably by using
the ray-by-ray-plus approximation ({Buras} {et~al.} 2006b) in the following
manner: First, the equations of neutrino transport are formulated for
the spherically symmetric case (Sec.~\ref{sec:boltzmann_cfc} to
\ref{sec:sources_cfc}), thus reducing the dimensionality of the
individual transport problems to be solved from six to three. By
working with a finite number (i.e.\ two) of angular moments of the
Boltzmann equation, yet another dimension can be eliminated.  The
additional \emph{space} dimensions are then taken into account by
solving these equations independently on different radial ``rays''
(corresponding to the angular zones of the polar grid) with direction
unit vector $\mathbf{n}$, assuming rotational symmetry around
$\mathbf{n}$ for the neutrino distribution function.  In addition,
certain terms from the full transport equations in 2D (or 3D in
general) are taken into account to avoid unphysical behaviour in the
optically thick regime (see {Buras} {et~al.}, 2006b and
Sec.~\ref{sec:ray_by_ray_cfc}). We emphasize that the ray-by-ray-plus
approximation does \emph{not mean} that the transport of neutrinos is
considered to be only in the radial direction. Instead it is assumed
that the neutrino distribution function is axially symmetric around
the radial direction, and that the neutrino pressure tensor is
diagonal.

\subsubsection{Boltzmann Equation in Spherical Symmetry}
\label{sec:boltzmann_cfc}
The precise form of Eq.~(\ref{eq:rel_boltzmann_1}) in a spherically symmetric
spacetime has been discussed in various places in the literature
({Lindquist} 1966; {Castor} 1972; {Schinder} 1988; {Mezzacappa} \& {Matzner} 1989); it depends on the
adopted gauge and slicing conditions, and on the tetrad basis used in momentum
space.  In our case, the spacetime metric is of the CFC type with
$\beta^1=\beta^r$ and $\beta^2=\beta^3=0$,
\begin{equation}
  \mathrm{d} s^2=-\alpha(t,r)^2 \mathrm{d} t^2+\phi(t,r)^4 \left[ \left(\mathrm{d} r+\beta^r(t,r)\  \mathrm{d} t\right)^2+ r^2 \mathrm{d} \theta^2 + r^2 \sin^2\theta \,\mathrm{d} \varphi^2\right].
\end{equation}
Furthermore, we carry out a transformation to the orthonormal frame comoving
with the fluid, in which the collision integral $\mathfrak{C}$ can be
expressed most conveniently; the expressions for the collision integral
provided in App.~A of {Rampp} \& {Janka} (2002) thus remain valid. With the
velocity field chosen as
\begin{equation}
  \left(v^1,v^2,v^3\right)=\left(\phi^{-2} v_r(t,r),0,0\right),
\end{equation}
a possible choice for the basis four-vectors of such a frame in terms
of the coordinate derivatives $\partial_t$, $\partial_r$,
$\partial_\theta$ and $\partial_\varphi$ (see
{Hawking} \& {Ellis}, 1973; {Misner}, {Thorne}, \& {Wheeler}, 1973; {Straumann}, 2004 for this notation) is given by
\begin{eqnarray}
  \mathbf{e_0}&=& \alpha^{-1} W\, \partial_t + W\left(\phi^{-2} v_r-\alpha^{-1} \beta^r \right) \partial_r, \\
  \mathbf{e_1}&=& \alpha^{-1}v_r W\, \partial_t + W\left(\phi^{-2}-\alpha^{-1}\beta^r v_r \right) \partial_r, \\
  \mathbf{e_2}&=& \phi^{-2} r^{-1} \, \partial_\theta, \\
  \mathbf{e_3}&=& \phi^{-2} r^{-1} \sin^{-1} \theta \,\partial_\varphi.
\end{eqnarray}
In spherical symmetry, the distribution function $f$ depends only on $t$, $r$,
the comoving frame energy $\varepsilon=\mathbf{p}\cdot \mathbf{e}_0$ (where
$\mathbf{p}$ is the neutrino four-momentum) and the angle cosine
$\mu=\mathbf{p} \cdot \mathbf{e}_1/\mathbf{p} \cdot \mathbf{e_0}$, and obeys
the following PDE:
\begin{eqnarray}
\lefteqn{
 W
\left[
\frac{\xi}{\alpha} \left(\frac{\partial f}{\partial t}-\beta^r \frac{\partial f}{\partial r}\right)
+
\frac{\nu}{\phi^2} \frac{\partial f}{\partial r}
\right]
-\frac{\varepsilon W^3}{r \alpha \phi^3}
\frac{\partial f}{\partial \varepsilon}
\left\{
\beta^r \phi^3 \left(-\psi-r\mu\frac{\partial v_r}{\partial r}\right)+
v_r^2 \phi
\left[
\beta^r \phi \left(2r \frac{\partial \phi}{\partial r}-\psi \phi\right)
+
\right.
\right.
}
\nonumber
\\
&&\left.
r \left(-\mu \frac{\partial \alpha}{\partial r}+\mu^2 \phi^2 \frac{\partial \beta^r}{\partial r}-\frac{\partial \phi^2}{\partial t}\right)
\right]
+
v_r^3
\left[
r \mu \phi \left(-\mu \frac{\partial \alpha}{\partial r}+\frac{\partial \beta^r \phi^2}{\partial r}- \frac{\partial \phi^2}{\partial t}\right)-
\psi \frac{\alpha}{\phi} \frac{\partial r \phi^2}{\partial r}
\right]
+
\nonumber
\\
&&
\phi
\left[
r \mu \left(\mu \alpha \frac{\partial v_r}{\partial r}+\frac{\partial \alpha}{\partial r} +
\phi^2 \left(-\mu \frac{\partial \beta^r}{\partial r}+\frac{\partial v_r}{\partial t}\right) \right)
+r \frac{\partial \phi^2}{\partial t}
-r \beta^r \frac{\partial \phi^2}{\partial r}
\right]
+
v_r \alpha
\left[
\phi \left(\psi +r \mu \frac{\partial v_r}{\partial r}\right)+
\right.
\nonumber
\\
&&
\left.
\left.
2r \psi \frac{\partial \phi}{\partial r}
+\phi^2 \left(\mu \frac{\partial v_r}{\partial t}-\frac{\partial \beta^r}{\partial r}\right)
+\frac{\partial \phi^2}{\partial t}
\right]
\right\}+
\frac{W^3 \left(1-\mu^2\right)}{r \alpha \phi^3}
\frac{\partial f}{\partial \mu}
\left\{
\alpha
\left[
\phi \left(\frac{\xi}{W^2}-r \nu \frac{\partial v_r}{\partial r}\right)
+2r \frac{\xi}{W^2}\frac{\partial \phi}{\partial r}
\right]
+
\right.
\nonumber
\\
\label{eq:boltzmann_cfc}
&&
\phi
\left[
\beta \phi^2 \left(r \xi \frac{\partial v_r}{\partial r}-\frac{\nu}{W^2}\right)
-
\frac{r}{W^2}
\left( \xi \frac{\partial \alpha}{\partial r}-\nu \phi^2 \frac{\partial \beta^r}{\partial r}\right)
-
\left.
r \xi \phi^2 \frac{\partial v_r}{\partial t}
\right]
\right\}= \mathfrak{C}\left[f\right],
\end{eqnarray}
where $\nu=\mu+v_r$, $\xi=1+\mu v_r$ (bearing in mind that $c=1$) and
$\psi=1-\mu^2$. It should be noted that our result is analytically
equivalent to the one obtained by {Mezzacappa} \& {Matzner} (1989) for the same
gauge and slicing conditions. The transfer equation for the radiation
intensity $\mathcal{I}=h^{-3} c^{-2}\varepsilon^3 f$ can be obtained
by exploiting the relation $\varepsilon^3 \partial f / \partial \varepsilon=\partial
\left(f \varepsilon^3\right) / \partial \varepsilon -3 \varepsilon^2 f$,
and is omitted here for the sake of brevity.

\subsubsection{Exact Moment Equations for CFC Metric in Spherical Symmetry}
\label{sec:momeq_cfc}
Multiplying Eq.~(\ref{eq:boltzmann_cfc}) by $h^{-3} c^{-2}
\varepsilon^3$ and taking the zeroth and first angular moments yields
the moment equations in a spherically symmetric CFC spacetime,
resulting in
\begin{eqnarray}
\label{eq:momeq_cfc_energy_j}
\nonumber
\lefteqn{\frac{\partial W \left(\hat{J}+v_r \hat{H}\right)}{\partial t}
+\frac{\partial}{\partial r}
\left[
\left(W \frac{\alpha}{\phi^2} -\beta^r v_r\right) \hat{H}+
\left(W v_r \frac{\alpha}{\phi^2} -\beta^r\right) \hat{J}
\right]
-}
\\
&&
\frac{\partial}{\partial \varepsilon}
\left\{
W \varepsilon \hat{J}
\left[
\frac{1}{r}\left(\beta^r-\frac{\alpha v_r}{\phi^2}\right)
+2 \left(\beta^r-\frac{\alpha v_r}{\phi^2}\right) \frac{\partial \ln \phi}{\partial r}
-2 \frac{\partial \ln \phi}{\partial t}
\right]
+
\right.
\nonumber
\\
&&
W \varepsilon \hat{H}
\left[
v_r \left(\frac{\partial \beta^r \phi^2}{\partial r}-2 \frac{\partial \ln \phi}{\partial t}\right)
-\frac{\alpha}{\phi^2}\frac{\partial \ln \alpha W}{\partial r}
+\alpha W^2 \left(\beta^r\frac{\partial v_r}{\partial r}-\frac{\partial v_r}{\partial t}\right)
\right]
-
\nonumber
\\
&&
\left.
\varepsilon \hat{K}
\left[
\frac{\beta^r W}{r}-\frac{\partial \beta^r W}{\partial r}
+W v_r r \frac{\partial}{\partial r}\left(\frac{\alpha}{r \phi^2}\right)
+W^3 \left(\frac{\alpha}{\phi^2}\frac{\partial v_r}{\partial r} + v_r\frac{\partial v_r}{\partial t}\right)
\right]
\right\}-
\nonumber
\\
&&
W \hat{J}
\left[
\frac{1}{r}\left(\beta^r-\frac{\alpha v_r}{\phi^2}\right)
+2 \left(\beta^r-\frac{\alpha v_r}{\phi^2}\right) \frac{\partial \ln \phi}{\partial r}
-2 \frac{\partial \ln \phi}{\partial t}
\right]-
\nonumber
\\
&&
W\hat{H}
\left[
v_r \left(\frac{\partial \beta^r \phi^2}{\partial r}-2 \frac{\partial \ln \phi}{\partial t}\right)
-\frac{\alpha}{\phi^2}\frac{\partial \ln \alpha W}{\partial r}
+\alpha W^2 \left(\beta^r\frac{\partial v_r}{\partial r}-\frac{\partial v_r}{\partial t}\right)
\right]
+
\nonumber
\\
&&
\hat{K}
\left[
\frac{\beta^r W}{r}-\frac{\partial \beta^r W}{\partial r}
+W v_r r \frac{\partial}{\partial r}\left(\frac{\alpha}{r \phi^2}\right)
+W^3 \left(\frac{\alpha}{\phi^2}\frac{\partial v_r}{\partial r} + v_r\frac{\partial v_r}{\partial t}\right)
\right]
=
\alpha \hat{C}^{(0)}
,
\end{eqnarray}
for the energy equation and
\begin{eqnarray}
\label{eq:momeq_cfc_energy_h}
\nonumber
\lefteqn{\frac{\partial W \left(\hat{H}+v_r \hat{K}\right)}{\partial t}
+\frac{\partial}{\partial r}
\left[
\left(W \frac{\alpha}{\phi^2} -\beta^r v_r\right) \hat{K}+
\left(W v_r \frac{\alpha}{\phi^2} -\beta^r\right) \hat{H}
\right]
-}
\\
&&
\frac{\partial}{\partial \varepsilon}
\left\{
W \varepsilon \hat{H}
\left[
\frac{1}{r}\left(\beta^r-\frac{\alpha v_r}{\phi^2}\right)
+2 \left(\beta^r-\frac{\alpha v_r}{\phi^2}\right) \frac{\partial \ln \phi}{\partial r}
-2 \frac{\partial \ln \phi}{\partial t}
\right]
+
\right.
\nonumber
\\
&&
W \varepsilon \hat{K}
\left[
v_r \left(\frac{\partial \beta^r \phi^2}{\partial r}-2 \frac{\partial \ln \phi}{\partial t}\right)
-\frac{\alpha}{\phi^2}\frac{\partial \ln \alpha W}{\partial r}
+\alpha W^2 \left(\beta^r\frac{\partial v_r}{\partial r}-\frac{\partial v_r}{\partial t}\right)
\right]
-
\nonumber
\\
&&
\left.
\varepsilon \hat{L}
\left[
\frac{\beta^r W}{r}-\frac{\partial \beta^r W}{\partial r}
+W v_r r \frac{\partial}{\partial r}\left(\frac{\alpha}{r \phi^2}\right)
+W^3 \left(\frac{\alpha}{\phi^2}\frac{\partial v_r}{\partial r} + v_r\frac{\partial v_r}{\partial t}\right)
\right]
\right\}
+
\nonumber
\\
&&
\left(\hat{J}-\hat{K}\right)
\left[
v_r \left(\frac{\beta^r}{r}-\frac{\partial \beta^r}{\partial r}\right)
+\frac{\partial}{\partial r}\left(\frac{W \alpha}{\phi^2}\right)-\frac{W \alpha}{r \phi^2}
+W^3\left(\frac{\partial v_r}{\partial t}- \beta^r \frac{\partial v_r}{\partial r}\right)
\right]
+
\nonumber
\\
&&
\left(\hat{H}-\hat{L}\right)
\left[
\frac{W^3 \alpha}{\phi^2}\frac{\partial v_r}{\partial r}
+\frac{\beta W}{r}-\frac{\partial \beta W}{\partial r}
-W v_r r \frac{\partial}{\partial r}\left(\frac{\alpha}{r \phi^2}\right)
+\frac{\partial W}{\partial t}
\right]-
\nonumber
\\
&&
W \hat{H}
\left[
\frac{1}{r}\left(\beta^r-\frac{\alpha v_r}{\phi^2}\right)
+2 \left(\beta^r-\frac{\alpha v_r}{\phi^2}\right) \frac{\partial \ln \phi}{\partial r}
-2 \frac{\partial \ln \phi}{\partial t}
\right]
-
\nonumber
\\
&&
W  \hat{K}
\left[
v_r \left(\frac{\partial \beta^r \phi^2}{\partial r}-2 \frac{\partial \ln \phi}{\partial t}\right)
-\frac{\alpha}{\phi^2}\frac{\partial \ln \alpha W}{\partial r}
+\alpha W^2 \left(\beta^r\frac{\partial v_r}{\partial r}-\frac{\partial v_r}{\partial t}\right)
\right]
+
\nonumber
\\
&&
\hat{L}
\left[
\frac{\beta^r W}{r}-\frac{\partial \beta^r W}{\partial r}
+W v_r r \frac{\partial}{\partial r}\left(\frac{\alpha}{r \phi^2}\right)
+W^3 \left(\frac{\alpha}{\phi^2}\frac{\partial v_r}{\partial r} + v_r\frac{\partial v_r}{\partial t}\right)
\right]
=
\alpha \hat{C}^{(1)}
.
\end{eqnarray}
for the momentum equation. Here, the angular moments $J$, $H$, $K$ and
$L$ of the specific intensity are given by
\begin{equation}
 \left\{J,H,K,L\right\}=\frac{1}{2} \int\limits_{-1}^{1} \mathrm{d} \mu\, \mu^{\{0,1,2,3\}} \mathcal{I},
\end{equation}
while $C^{(0)}$ and $C^{(1)}$ are the corresponding zeroth and first
moment of the collision integral.  Note that
Eqs.~(\ref{eq:momeq_cfc_energy_j}) and (\ref{eq:momeq_cfc_energy_h})
are written in terms of the densitized moments $\hat{J}$, $\hat{H}$,
$\hat{K}$ and $\hat{L}$ (where $\hat{X}=\sqrt{\gamma} X$ for any
  quantity $X$) which contain the square root of the determinant of
the three-metric $\sqrt{\gamma}= \phi^6 r^2 \sin \theta$ as an additional factor, in order to better reflect
the underlying law of energy conservation. The virtue of such a
formulation is more apparent in the equations governing the evolution
of the neutrino number density and flux,
\begin{eqnarray}
\label{eq:momeq_cfc_number_j}
\nonumber
\lefteqn{\frac{\partial W \left(\hat{\mathcal{J}}+v_r \hat{\mathcal{H}}\right)}{\partial t}
+\frac{\partial}{\partial r}
\left[
\left(W \frac{\alpha}{\phi^2} -\beta^r v_r\right) \hat{\mathcal{H}}+
\left(W v_r \frac{\alpha}{\phi^2} -\beta^r\right) \hat{\mathcal{J}}+
\right]
-}
\\
&&
\frac{\partial}{\partial \varepsilon}
\left\{
W \varepsilon \hat{\mathcal{J}}
\left[
\frac{1}{r}\left(\beta^r-\frac{\alpha v_r}{\phi^2}\right)
+2 \left(\beta^r-\frac{\alpha v_r}{\phi^2}\right) \frac{\partial \ln \phi}{\partial r}
-2 \frac{\partial \ln \phi}{\partial t}
\right]
+
\right.
\nonumber
\\
&&
W \varepsilon \hat{\mathcal{H}}
\left[
v_r \left(\frac{\partial \beta^r \phi^2}{\partial r}-2 \frac{\partial \ln \phi}{\partial t}\right)
-\frac{\alpha}{\phi^2}\frac{\partial \ln \alpha W}{\partial r}
+\alpha W^2 \left(\beta^r\frac{\partial v_r}{\partial r}-\frac{\partial v_r}{\partial t}\right)
\right]
-
\nonumber
\\
&&
\left.
\varepsilon \hat{\mathcal{K}}
\left[
\frac{\beta^r W}{r}-\frac{\partial \beta^r W}{\partial r}
+W v_r r \frac{\partial}{\partial r}\left(\frac{\alpha}{r \phi^2}\right)
+W^3 \left(\frac{\alpha}{\phi^2}\frac{\partial v_r}{\partial r} + v_r\frac{\partial v_r}{\partial t}\right)
\right]
\right\}
=
\alpha \hat{\mathcal{C}}^{(0)}
,
\end{eqnarray}
\begin{eqnarray}
\label{eq:momeq_cfc_number_h}
\nonumber
\lefteqn{\frac{\partial W \left(\hat{\mathcal{H}}+v_r \hat{\mathcal{K}}\right)}{\partial t}
+\frac{\partial}{\partial r}
\left[
\left(W \frac{\alpha}{\phi^2} -\beta^r v_r\right) \hat{\mathcal{K}}+
\left(W v_r \frac{\alpha}{\phi^2} -\beta^r\right) \hat{\mathcal{H}}+
\right]
-}
\\
&&
\frac{\partial}{\partial \varepsilon}
\left\{
W \varepsilon \hat{\mathcal{H}}
\left[
\frac{1}{r}\left(\beta^r-\frac{\alpha v_r}{\phi^2}\right)
+2 \left(\beta^r-\frac{\alpha v_r}{\phi^2}\right) \frac{\partial \ln \phi}{\partial r}
-2 \frac{\partial \ln \phi}{\partial t}
\right]
+
\right.
\nonumber
\\
&&
W \varepsilon \hat{\mathcal{K}}
\left[
v_r \left(\frac{\partial \beta^r \phi^2}{\partial r}-2 \frac{\partial \ln \phi}{\partial t}\right)
-\frac{\alpha}{\phi^2}\frac{\partial \ln \alpha W}{\partial r}
+\alpha W^2 \left(\beta^r\frac{\partial v_r}{\partial r}-\frac{\partial v_r}{\partial t}\right)
\right]
-
\nonumber
\\
&&
\left.
\varepsilon \hat{\mathcal{L}}
\left[
\frac{\beta^r W}{r}-\frac{\partial \beta^r W}{\partial r}
+W v_r r \frac{\partial}{\partial r}\left(\frac{\alpha}{r \phi^2}\right)
+W^3 \left(\frac{\alpha}{\phi^2}\frac{\partial v_r}{\partial r} + v_r\frac{\partial v_r}{\partial t}\right)
\right]
\right\}
+
\nonumber
\\
&&
\left(\hat{\mathcal{J}}-\hat{\mathcal{K}}\right)
\left[
v_r \left(\frac{\beta^r}{r}-\frac{\partial \beta^r}{\partial r}\right)
+\frac{\partial}{\partial r}\left(\frac{W \alpha}{\phi^2}\right)-\frac{W \alpha}{r \phi^2}
+W^3\left(\frac{\partial v_r}{\partial t}- \beta^r \frac{\partial v_r}{\partial r}\right)
\right]
+
\nonumber
\\
&&
\left(\hat{\mathcal{H}}-\hat{\mathcal{L}}\right)
\left[
\frac{W^3 \alpha}{\phi^2}\frac{\partial v_r}{\partial r}
+\frac{\beta W}{r}-\frac{\partial \beta W}{\partial r}
-W v_r r \frac{\partial}{\partial r}\left(\frac{\alpha}{r \phi^2}\right)
+\frac{\partial W}{\partial t}
\right]
=
\alpha \hat{\mathcal{C}}^{(1)}
,
\end{eqnarray}
where
$\{\mathcal{J},\mathcal{H},\mathcal{K},\mathcal{L},\mathcal{C}\}=\{
\varepsilon^{-1}J,\varepsilon^{-1}H,\varepsilon^{-1}K,\varepsilon^{-1}L,\varepsilon^{-1}
C \}$.  Apart from the collision integral, no source terms appear in
the neutrino number equation (\ref{eq:momeq_cfc_number_j}).  Since the
evolved quantity $W \left(\mathcal{J} + v_r \mathcal{H} \right)$ is
proportional to the spectral neutrino number density $N_\mathrm{eul}$
in the Eulerian (coordinate/lab) frame,
\begin{equation}
N_\mathrm{eul} = \frac{4\pi}{c} W \left(\mathcal{J} + v_r \mathcal{H} \right),
\end{equation}
a finite-volume discretization of Eq.~(\ref{eq:momeq_cfc_number_j})
exactly conserves the total neutrino number measured in the laboratory
(coordinate) frame,
\begin{equation}
  \int \mathrm{d}^3 x \, \int \mathrm{d} \varepsilon \,
 \sqrt{\gamma} N_\mathrm{eul}=\frac{4\pi}{c}
  \int \mathrm{d}^3 x \,\mathrm{d} \varepsilon\, 
 \sqrt{\gamma} W \left(\mathcal{J}+v_r \mathcal{H}\right),
\end{equation}
in the absence of neutrino-matter interactions. As a consequence,
numerical conservation of lepton number can be achieved with an
appropriate conservative treatment of the flux terms in the
neutrino moment equations and the source term for the electron
fraction.

With a few minor exceptions, the numerical solution of the neutrino
moment equations in our code proceeds exactly as described by
{Rampp} \& {Janka} (2002). Equations~(\ref{eq:momeq_cfc_energy_j}) and
(\ref{eq:momeq_cfc_energy_h}) are written in a conservative
finite-difference form using second-order space discretization on a
staggered mesh and backward differencing in time; only the neutrino
advection terms
\begin{equation}
  \frac{\partial}{\partial r}
  \left[
    \left(W \frac{\alpha}{\phi^2} v_r -\beta^r\right) \hat{J}
    \right]
  ,
  \quad
  \frac{\partial}{\partial r}
  \left[
    \left(W \frac{\alpha}{\phi^2} v_r -\beta^r\right) \hat{H}
    \right]
\end{equation}
are discretized with upwind differences in space and central
differences in time. The higher moments $K$ and $L$ are computed from
the variable Eddington factors $f_K$ and $f_L$ obtained from the
solution of a model Boltzmann equation
(cf. Sec.~\ref{sec:tangent_ray_cfc}) as $K=f_K J$ and $L=f_L J$ where
necessary.

Our numerical scheme for the moment equations differs from that of
{Rampp} \& {Janka} (2002) in one important respect. {Rampp} \& {Janka} (2002) also solve
Eqs.~(\ref{eq:momeq_cfc_number_j}) and (\ref{eq:momeq_cfc_number_h})
for the neutrino number density and flux in addition to
Eqs.~(\ref{eq:momeq_cfc_energy_j}) and (\ref{eq:momeq_cfc_energy_h})
in order to guarantee both lepton number conservation and energy
conservation. However, this procedure has two important drawbacks:
Solving another pair of equations is computationally expensive, and,
more importantly, the numerical solution can become inconsistent in
the sense that $(J,H)=(\varepsilon \mathcal{J},\varepsilon
\mathcal{H})$ no longer holds. In the high-energy tail of the spectrum
$J_i/mathcal{J}_i$ and $H_i/\mathcal{H}_i$ in a given zone in energy
space can even assume values outside the zone boundaries. In order to
circumvent this problem, we have developed a special second-order
finite-difference representation for the terms governing advection in
energy space accounting for gravitational redshift and Doppler shift.
This allows us to guarantee \emph{both} energy and lepton number
conservation \emph{without} solving the equations for neutrino number
density and flux. Since our improved treatment of the redshift and
Doppler terms is rather involved, we give a complete description of
the algorithm in App.~\ref{app:cons_dopp}.

Finally, the moment equations are supplemented by
two additional equations for the operator-split update of the specific
internal energy and the electron fraction of the matter due to
neutrino-matter interactions,
\begin{eqnarray}
  \label{eq:source_ene}
  \left( \frac{\partial \epsilon}{\partial t}\right)_\nu
  &=&
  -\frac{4\pi \alpha}{\rho W}
  \int\limits_0^\infty \mathrm{d} \varepsilon\,
  \sum_{i \in \left\{\nu_e,\bar{\nu}_e,\nu_{\mu/\tau} \right \} } C_{i}^{(0)},
  \\
  \label{eq:source_y_e}
  \left( \frac{\partial Y_e}{\partial t} \right)_\nu
  &=&
  -\frac{4\pi \alpha m_\mathrm{B}}{\rho W}
  \int\limits_0^\infty \mathrm{d} \varepsilon\,
  \left( \mathcal{C}_{\nu_e}^{(0)}-\mathcal{C}_{\bar{\nu}_e}^{(0)} \right),
\end{eqnarray}
where the factor $\alpha W^{-1}$ takes care of the conversion from
proper time to coordinate time. The resulting non-linear system of
equations for $J$, $H$, $\epsilon$ and $Y_e$ is then solved by
Newton-Raphson iteration. As in {Rampp} \& {Janka} (2002), we treat $\mu$ and
$\tau$ neutrinos and their antiparticles as a single species, and
solve the corresponding moment equations separately from the (fully
coupled) moment equations for the transport of electron neutrinos and
antineutrinos.

\subsubsection{Model Boltzmann Equation}
\label{sec:tangent_ray_cfc}
In the variable Eddington factor technique, the closure relations
required for Eqs.~(\ref{eq:momeq_cfc_energy_j}) and
(\ref{eq:momeq_cfc_energy_h}) are extracted from the formal solution
of a model Boltzmann equation, with the moments of the neutrino
distribution function occurring in the collision integral taken from
the solution of the moment equations in an iterative procedure
({Mihalas} \& {Weibel Mihalas} 1984; {Burrows} {et~al.} 2000b; {Rampp} \& {Janka} 2002). Since the model Boltzmann
equation is only used to compute \emph{normalized} moments of the
radiation intensity, it is sufficient to consider a strongly
simplified equation version of Eq.~(\ref{eq:boltzmann_cfc}). We
therefore neglect terms containing the shift vector or derivatives of
the metric functions $\alpha$ and $\phi$ in
Eq.~(\ref{eq:boltzmann_cfc}), as well as higher-order terms
$\mathcal{O}(v_r^2)$ in the radial velocity. Furthermore, we set $\nu=
\mu+v_r \rightarrow \mu$ and $\xi=1+\mu v_r \rightarrow 1$ in the terms for
advection in energy space. However, the radial advection term is kept
exactly as in the full equation. The model transfer equation for the
radiation intensity thus obtained reads
\begin{eqnarray}
\nonumber
\lefteqn{ 
\frac{\partial \mathcal{I}}{\partial t}
+
\left(\frac{\alpha v_r}{\phi^2} -\beta^r\right)\frac{\partial f}{\partial r}
+
\frac{\alpha \mu}{\phi^2} \frac{\partial f}{\partial r}
+
\mathcal{I}
\left[
\left(3-\mu^2\right) \frac{\alpha v_r}{r \phi^2}+
 \left(1+\mu^2\right) \alpha \phi^2 \frac{\partial v_r}{\partial r} +
2\alpha \mu \frac{\partial v_r}{\partial t}
\right]-
}
\\
\nonumber
&&
\frac{\partial \mathcal{I}}{\partial \varepsilon}
\left\{ \varepsilon
\left[
\left(1-\mu^2\right) \frac{\alpha v_r}{r \phi^2}+
 \mu^2 \frac{\alpha}{\phi^2} \frac{\partial v_r}{\partial r} +
\alpha \mu \frac{\partial v_r}{\partial t}
\right]
\right\}+
\frac{\alpha \left(1-\mu^2\right)}{r \phi^2}
\frac{\partial \mathcal{I}}{\partial \mu}
+
\\
\label{eq:model_boltzmann}
&&
\frac{\partial}{\partial \mu}
\left\{
\left(1-\mu^2\right)
\left[
\mu \left(\frac{\alpha v_r}{r\phi^2}-\frac{\alpha}{\phi^2}\frac{\partial v_r}{\partial r}\right)
-
\mu \alpha \frac{\partial v_r}{\partial t}
\right]
\mathcal{I}
\right\}
=C \left[\mathcal{I}\right].
\end{eqnarray}
Here, $C \left[\mathcal{I}\right]$ denotes the collision term in the
transfer equation, which is related to the collision term
$\mathfrak{C}$ in the Boltzmann equation~(\ref{eq:boltzmann_cfc}) by
$C \left[\mathcal{I}\right]=\varepsilon^3 h^{-3} c^{-2}
\mathfrak{C}[f]$.  Aside from the appearance of prefactors containing
the metric functions $\alpha$ and $\phi$ and of the radial shift
$\beta^r$ in the advection term, Eq.~(\ref{eq:model_boltzmann}) is
identical to the flat-space transfer equation in the
$\mathcal{O}(v/c)$-approximation. Therefore, the same solution method
as in {Rampp} \& {Janka} (2002) can be employed, i.e.\ the radial advection term
and the energy derivatives in Eq.~(\ref{eq:model_boltzmann}) are
treated by a conservative interpolation procedure and a time-explicit
upwind scheme respectively, while the tangent-ray method
({Yorke} 1980; {Mihalas} \& {Weibel Mihalas} 1984) is used for the remaining terms.

\subsubsection{Source Terms}
\label{sec:sources_cfc}
Once the solution of the moment equations is known, the source terms
$Q_{Y_e}$ $Q_E$, and $Q_M$ for electron number, energy, and momentum
in the local fluid frame can be computed from the zeroth and first
angular moments $C^{(0)}$ and $C^{(1)}$ of the collision integral,
\begin{eqnarray}
\label{eq:sourcet_1}
Q_{Y_e}&=&\left(\frac{\mathrm{d} \rho Y_e}{\mathrm{d} \lambda}\right)_\mathrm{C}
=
-4\pi m_\mathrm{B}
\int\limits_0^\infty \mathrm{d} \varepsilon\,
\left( \mathcal{C}_{\nu_e}^{(0)}-\mathcal{C}_{\bar{\nu}_e}^{(0)} \right),
\\
\label{eq:sourcet_2}
Q_E&=&
\left(\frac{\mathrm{d} \rho \varepsilon}{\mathrm{d} \lambda}\right)_\mathrm{C}
=
-4\pi
\int\limits_0^\infty \mathrm{d} \varepsilon\,
\sum_{i \in \left\{\nu_e,\bar{\nu}_e,\nu_{\mu/\tau} \right \} } C_{i}^{(0)},
\\
\label{eq:sourcet_3}
Q_M&=&
-4\pi
\int\limits_0^\infty \mathrm{d} \varepsilon\,
\sum_{i \in \left\{\nu_e,\bar{\nu}_e,\nu_{\mu/\tau} \right \} } C_{\nu}^{(1)}.
\end{eqnarray}
To obtain the correct transformation to the Eulerian frame (in
  which the equations of hydrodynamics are solved), we consider the
  equations of electron number conservation and energy-momentum
  conservation in their covariant form. Source terms due to neutrino
  interactions must enter as scalar and four-vector terms on the
  right-side in these equations:
\begin{equation}
 \nabla_\mu \left(\rho Y_e u^\mu \right)=s, \quad \nabla_\nu
 T^{\mu\nu}=q^\mu.
\end{equation}
Here, $s$ and $q^\mu$ are the scalar electron number density source
term and the covariant energy-momentum vector source term. Since the
transformation of scalars and four-vectors is trivial, we only need to
express $s$ and $q^\mu$ in terms of $Q_{Y_e}$, $Q_E$, and $Q_M$ in the
comoving frame to obtain the correct source terms in any desired
frame:
\begin{equation}
s = Q_{Y_e},
\quad
q^\mu =Q_E \mathbf{e_0} +Q_M \mathbf{e_1}.
\end{equation}
After carrying out a Lorentz transformation\footnote{Since we consider
  a purely radial velocity field at this stage, the Lorentz
  transformation of a four-vector from the comoving frame to the
  Eulerian frame simply reads: $(q^0,q^1,q^2,q^3)\rightarrow W
  (q^0+v_r q^1,q^1+v_r q^0,q^2,q^3)$.} back to the Eulerian frame and
recasting the resulting equations into conservative form, we obtain
the following source terms for the conserved Eulerian quantities that
appear in Eqs.~(\ref{eq:hydro_2}), (\ref{eq:hydro_3}) and
(\ref{eq:hydro_4}):
\begin{eqnarray}
\label{eq:source_terms_radial_1}
\left( \frac{\partial \sqrt{\gamma} S_1}{\partial t}\right)_\mathrm{C}
&=&
\sqrt{\gamma} W \left(v_r Q_E+Q_M\right),
\\
\left( \frac{\partial \sqrt{\gamma} \tau}{\partial t}\right)_\mathrm{C}
&=&
\label{eq:source_terms_radial_2}
\sqrt{\gamma} W \left(Q_E+v_r Q_M\right),
\\
\left(\frac{\partial \sqrt{\gamma}\rho W Y_e}{\partial t}\right)_\mathrm{C}
&=&
\label{eq:source_terms_radial_3}
\sqrt{\gamma} Q_{Y_e}.
\end{eqnarray}
Note that we have multiplied the source terms by the metric
$\sqrt{\gamma}$, because it is the densitized variables $\sqrt{\gamma}
S_i$, $\sqrt{\gamma} \tau$ and $\sqrt{\gamma} D Y_e$ that are actually
evolved in Eqs.~(\ref{eq:hydro_2}), (\ref{eq:hydro_3}) and
(\ref{eq:hydro_4}). This is possible because the metric factor
$\sqrt{\gamma}$ is obviously not changed by neutrino interactions.

\subsubsection{Ray-by-ray-plus Approach for Axisymmetric Problems}
\label{sec:ray_by_ray_cfc}
As noted by {Rampp} \& {Janka} (2002), a treatment of neutrino transport in
spherical symmetry (as described in Secs.~\ref{sec:momeq_cfc} to
\ref{sec:sources_cfc}) can be extended to the multi-dimensional case by
assuming that all the involved physical quantities depend only
parametrically on the coordinates $\theta$ and $\phi$ (using spherical
polar coordinates), while neglecting any lateral derivatives in the
full three-dimensional moment equations.  However, {Buras} {et~al.} (2006b)
pointed out that the lateral advection of neutrinos in the optically
thick regime and the non-radial components of the neutrino pressure
gradient need to be taken into account to avoid unphysical convective
activity in the proto-neutron star. In general relativity the
additional advection terms in the moment equations are,
\begin{eqnarray}
\label{eq:lat_adv_1}
\left(\frac{\partial \sqrt{\gamma} W \left(J+v_r H\right)}{\partial t}\right)_\mathrm{adv}+
\frac{\partial \alpha \sqrt{\gamma} W \hat{v}^2 \left(J+v_r H\right)}{\partial \theta}&=&0, \\ 
\label{eq:lat_adv_2}
\left(\frac{\partial \sqrt{\gamma} W \left(H+v_r K\right)}{\partial t}\right)_\mathrm{adv}+
\frac{\partial \alpha \sqrt{\gamma} W \hat{v}^2 \left(H+v_r K\right)}{\partial \theta}&=&0,
\end{eqnarray}
which are solved in an operator-split step before the radial transport
sweep. We also include terms for the compressional heating of the
neutrinos due to lateral motions, which are added to the moment equations
(\ref{eq:momeq_cfc_energy_j}) and (\ref{eq:momeq_cfc_energy_h}) when
performing the radial transport sweep,
\begin{eqnarray}
\left(\frac{\partial \sqrt{\gamma} \left(J+v_r H\right) }{\partial t} \right)_\mathrm{comp}&=&-
\frac{\partial \alpha \sqrt{\gamma} W \hat{v}^2}{\partial \theta} \frac{J-K}{2} 
-\frac{\partial}{\partial \varepsilon}
\left(\varepsilon \frac{\partial \alpha \sqrt{\gamma} W \hat{v}^2}{\partial \theta} \frac{J-K}{2}\right) ,
\\
\left(\frac{\partial \sqrt{\gamma} \left(H+v_r K\right) }{\partial t} \right)_\mathrm{comp}&=&-
\frac{\partial \alpha \sqrt{\gamma} W \hat{v}^2}{\partial \theta} \frac{K-L}{2} 
-\frac{\partial}{\partial \varepsilon}
\left(\varepsilon \frac{\partial \alpha \sqrt{\gamma} W \hat{v}^2}{\partial \theta} \frac{H-L}{2}\right).
\end{eqnarray}
It should be noted that gravitational redshift associated with lateral
motion is still neglected here; however, this is a tiny effect in
typical supernova environments due to the relatively small asphericity
of the gravitational field.

The lateral component of the neutrino pressure gradient appears as a source term for the
component $S_2$ of the fluid momentum density,
\begin{equation}
\left(\frac{\partial \sqrt{\gamma} S_2}{\partial t}\right)_\mathrm{lat}=-\sqrt{-g}\frac{\partial P_\nu}{\partial \theta},
\end{equation}
where we assume that the neutrino pressure is given by $P_\nu=4/3 \pi
J$ in the optically thick regime. Since the acceleration four-vector
due to the neutrino pressure gradient must be orthogonal to the
four-velocity of the fluid, the source term in the energy equation
(work exerted on the fluid) is given by
\begin{equation}
\left(\frac{\partial \sqrt{\gamma} \tau}{\partial t}\right)_\mathrm{lat}=-\sqrt{-g} v^2 \frac{\partial P}{\partial \theta}.
\end{equation}

Finally, contrary to the Newtonian case, the covariant source vector
$q^\mu$ has to be transformed to the Eulerian frame by a Lorentz boost
which includes the non-radial velocity components. $Q_E$ and $Q_M$
now enter differently into the source term for $S_1$, and additional
source terms for $S_2$ and $S_3$ appear,
\begin{eqnarray}
\label{eq:sourcet_2d_1}
\left( \frac{\partial \sqrt{\gamma} S_1}{\partial t}\right)_\mathrm{C}
&=&
\sqrt{\gamma} \left[ W v_1 Q_E+Q_M\left(1+v_1 v_r\frac{W-1}{v_i v^i}\right)\right],
\\
\label{eq:sourcet_2d_2}
\left( \frac{\partial \sqrt{\gamma} S_2}{\partial t}\right)_\mathrm{C}
&=&
\sqrt{\gamma} \left[ W v_2 Q_E+Q_M v_2 v^r\frac{W-1}{v_i v^i}\right],
\\
\label{eq:sourcet_2d_3}
\left( \frac{\partial \sqrt{\gamma} S_3}{\partial t}\right)_\mathrm{C}
&=&
\sqrt{\gamma} \left[ W v_3 Q_E+Q_M v_3 v_r\frac{W-1}{v_i v^i}\right].
\end{eqnarray}
While the new terms in $Q_M$ are of order $\mathcal{O}(v^2)$ and can
usually be neglected, all the terms in $Q_E$ are of order
$\mathcal{O}(v)$ and need to be included to ensure the correct
evolution of the specific internal energy $\epsilon$. This is
particularly crucial for rapidly rotating configurations, where the
rotational velocity $v_\phi$ can reach a significant fraction of $c$.

\subsection{Quality of our Approximation}
\label{sec:approx_quality}
Our relativistic ray-by-ray-plus method entails several
approximations, which fall into three categories, viz., approximations
in the treatment i) of the microphysics (EoS, neutrino interactions,
etc.), ii) of the neutrino transport, and iii) of the gravitational
field equations. Concerning the microphysics, we rely on the same
up-to-date description as the non-relativistic
\textsc{VERTEX-PROMETHEUS} code ({Rampp} \& {Janka} 2002; {Buras} {et~al.} 2006b), which
includes a number of additional neutrino reactions compared to the
``standard'' set of neutrino opacities used by {Bruenn} (1985) (see
{Burrows} \& {Sawyer} 1998; {Hannestad} \& {Raffelt} 1998; {Buras} {et~al.} 2003; {Itoh} {et~al.} 2004; {Marek} {et~al.} 2005; {Langanke} {et~al.} 2003, 2008
for the extensions to the standard opacities). Neutrino oscillations
are presently neglected despite the fact that the classical MSW effect
and collective neutrino-antineutrino flavor oscillations may possibly
have an important bearing at least on the observable neutrino
properties. However, considering that our understanding of the latter
phenomenon is by no means complete, and that the classical MSW effect
is only relevant far outside the supernova core, it is not sure
whether neutrino oscillations are \emph{dynamically} relevant for the
supernova problem at all. Moreover, no self-consistent dynamical
simulations including these effects have been carried out so far;
recent studies have contented themselves with post-processing
progenitor or simulation data ({Duan} {et~al.} 2008; {Lunardini}, {M{\"u}ller}, \&  {Janka} 2008). Thus,
our treatment of neutrino physics, while certainly not complete, is
state-of-the-art for dynamical supernova simulations, and the same is
true for the different high-density EoSs
({Lattimer} \& {Swesty} 1991; {Shen} {et~al.} 1998; {Hillebrandt} \& {Wolff} 1985) available in our code.

Handling the neutrino transport with the ``ray-by-ray-plus'' scheme of
{Buras} {et~al.} (2006b) introduces approximations on two different levels:
First, the variable Eddington factor technique is not strictly
equivalent to full Boltzmann transport even in spherical symmetry,
since only a simplified ``model'' Boltzmann equation is solved to
provide the closure for the moment equations. This shortcoming is
mitigated by the fact that comparisons with analytic solutions
({Rampp} \& {Janka} 2002) and with Boltzmann solvers based on the $S_N$-method
({Liebend{\"o}rfer} {et~al.} 2005) show that the approximate solution is
entirely adequate in many cases. Second, the ``ray-by-ray-plus''
approach ignores the lateral propagation (but not the lateral
advection) of neutrinos. However, the analysis of {Buras} {et~al.} (2006b)
at least suggests that this approximation does not affect the heating
in the gain layer dramatically, and one may therefore expect that it
captures the overall dynamics during the proto-neutron star accretion
phase and also during the subsequent wind phase quite
adequately. Unfortunately, no detailed comparison of the
``ray-by-ray-plus'' method with true multi-angle transport or
alternative approaches, like multi-group flux-limited diffusion
(MGFLD), which would allow us to better assess the quality of our
method, is available as yet.  So far, the different approaches to
multi-dimensional transport hitherto pursued by supernova modellers
all have their individual strengths and weaknesses: The only true
multi-angle simulations ({Livne} {et~al.} 2004; {Ott} {et~al.} 2008) for example, have
been carried out at the expense of neglecting inelastic scattering
reactions, and MGFLD has been shown to smear out lateral variations of
the radiation field quite considerably ({Ott} {et~al.} 2008), while the
``ray-by-ray-plus'' method probably tends to overestimate such
variations. Still, simulations with ray-by-ray transport show effects
that are qualitatively similar to those seen by {Ott} {et~al.} (2008) with
multi-angle transport. Global asymmetries of the neutrino flux
  are preserved better than with MGFLD, for example. At any rate,
ray-by-ray transport can be classified as a state-of-the-art method in
the field, notwithstanding the fact that it is certainly limited to
scenarios where the anisotropies in the matter distribution (at least
out to the neutrinosphere) do not become exceedingly large. Compared
to other radiative transfer methods used in multi-dimensional
\emph{relativistic} simulations up to now -- like the deleptonization
scheme of {Liebend{\"o}rfer} (2005), which was first used in GR by
{Ott} {et~al.} (2007b), or the grey two-moment formalism (using the
Eddington approximation) by {Farris} {et~al.} (2008) -- it is a much more
ambitious and accurate approach to neutrino transport in core collapse
supernovae.

Given the limitations of the ``ray-by-ray-plus'' method, the use of
the CFC approximation to the gravitational field equations is quite
natural: CFC deteriorates in accuracy for strongly deformed, rapidly
rotating matter configurations, where the ``ray-by-ray-plus''method is
not applicable anyway -- there is little to be gained by a more
accurate treatment of GR in that regime. It is noteworthy, however,
that the CFC approximation has been found to be accurate for a wide
range of (sometimes extreme) astrophysical scenarios, such as rapidly
rotating neutron stars ({Cook}, {Shapiro}, \& {Teukolsky} 1996), rigidly rotating disks
({Kley} \& {Sch{\"a}fer} 1999), and rotational core collapse
({Shibata} \& {Sekiguchi} 2004; {Cerd{\'a}-Dur{\'a}n} {et~al.} 2005; {Ott} {et~al.} 2007a, 2007b; {Dimmelmeier} {et~al.} 2007). In
the core collapse scenario, CFC still appears to be adequate
({Dimmelmeier} {et~al.} 2006) for the extremely rapidly rotating models of
{Shibata} \& {Sekiguchi} (2005); even toroidally deformed neutron stars obviously
pose little problem to CFC. These facts indicate that in typical
supernova simulations, where the neutron star rotates moderately at
most, CFC will not be the weakest link in the chain of approximations,
and that, at least for our purpose, GR effects are treated correctly
in our code. All in all, \textsc{VERTEX-CoCoNuT} can thus be regarded
as a truly relativistic radiation hydrodynamics code for core collapse
supernovae, featuring a sophisticated neutrino transport scheme and an
up-to-date treatment of the microphysics.

\section{Code Tests in Spherical Symmetry}
\label{sec:1d_tests}

\subsection{Simple Test Problems -- Radiating Spheres}
\label{sec:rad_spheres}

\begin{figure}
\plotone{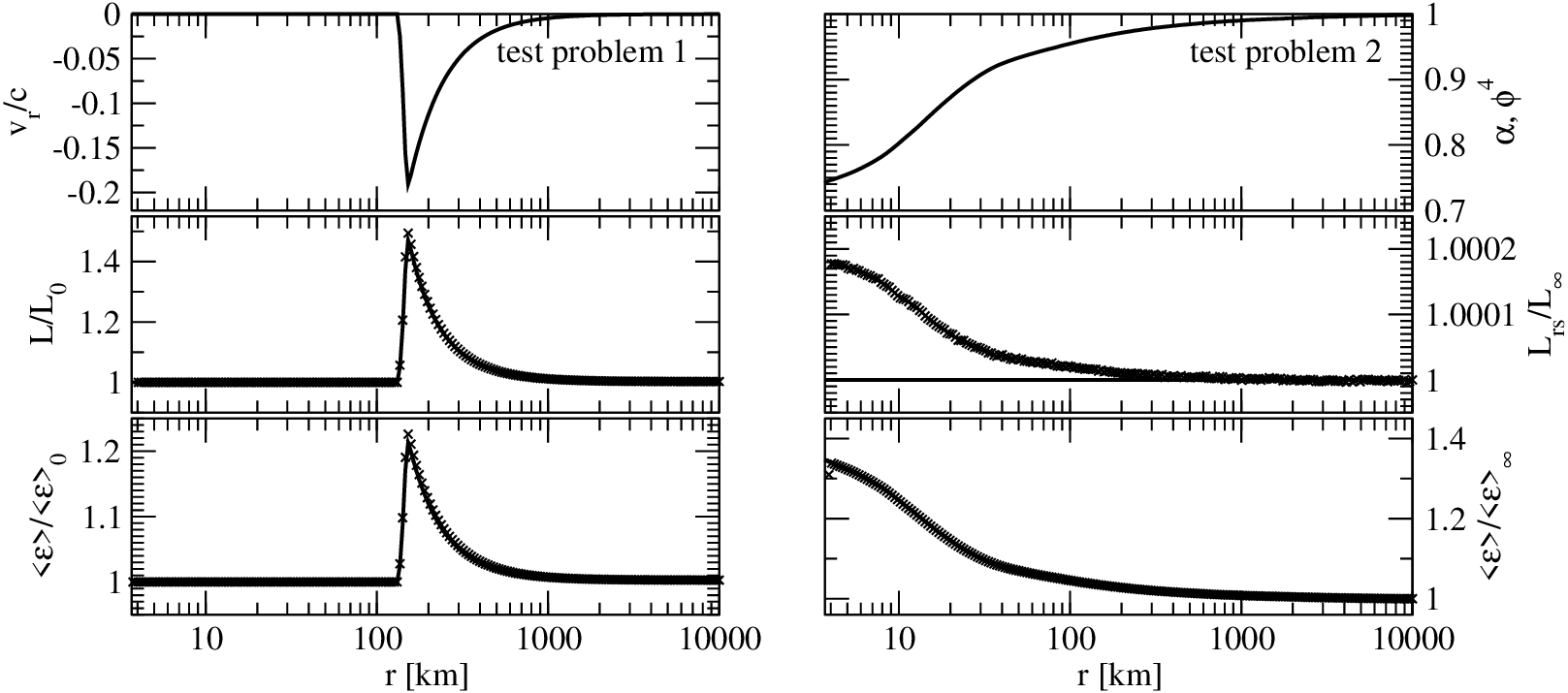}
\caption{Left: Setup and results for test problem 1: The radial
  velocity profile used for the test is shown in the upper panel. The
  middle and bottom panels show the luminosity $L$ and mean energy
  $\left \langle \varepsilon \right \rangle$ in the moving observer
  frame as obtained with \textsc{VERTEX-CoCoNuT} (crosses), as well as
  the analytic solution (solid lines). For convenience, $L$ and $\left \langle
  \varepsilon \right \rangle$ are given in units of their values $L_0$
  and $\left \langle \varepsilon \right \rangle_0$ just outside the
  radiating sphere.  Right: Setup and results for test problem 2: The
  lapse function $\alpha$ and conformal factor $\phi$ are shown in the
  upper panel. The middle and bottom panels show the
  redshift-corrected luminosity $L_\mathrm{rs}$ and the mean energy
  $\left \langle \varepsilon \right \rangle$ in the frame of an
  Eulerian observer as obtained with \textsc{VERTEX-CoCoNuT}
  (crosses), as well as the analytic solution (solid lines).  For
  convenience, $L_\mathrm{rs}$ and $\left \langle \varepsilon \right
  \rangle$ are given in units of their values at infinity $L_\infty$
  and $\left \langle \varepsilon \right \rangle_\infty$.
\label{fig:test_rs_1}
}
\end{figure}

\begin{figure}
\plotone{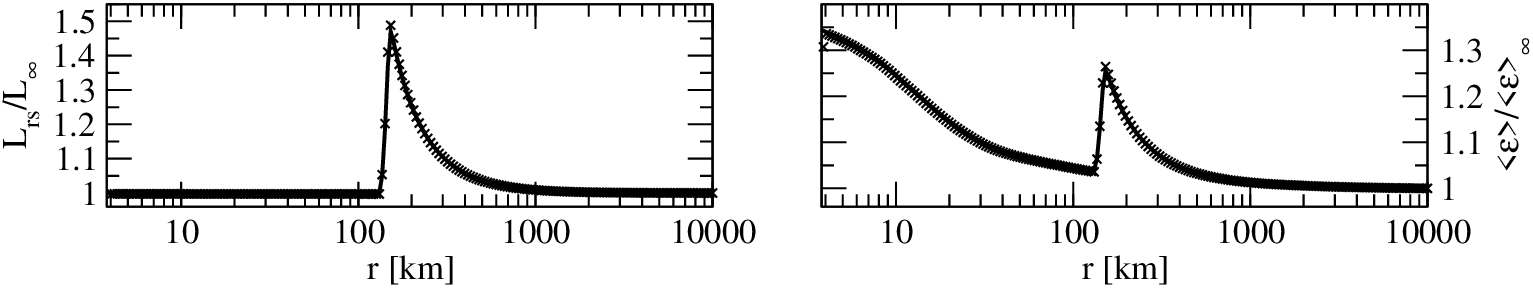}
\caption{Redshift-corrected luminosity $L_\mathrm{rs}$ and mean energy
  $\left \langle \varepsilon \right \rangle$ (both in the moving
  observer frame) for test problem 3. Crosses indicate the results
  obtained with \textsc{VERTEX-CoCoNuT}, the analytic solutions are
  drawn as solid lines.  For convenience, $L_\mathrm{rs}$ and $\left
  \langle \varepsilon \right \rangle$ are given in units of their
  values at infinity $L_\infty$ and $\left \langle \varepsilon \right
  \rangle_\infty$.
\label{fig:test_rs_3}
}
\end{figure}

Stationary solutions for the propagation of radiation from an extended
spherical central source into vacuum furnish one of the simplest kinds
of test problems in radiation hydrodynamics. Although not overly
challenging \emph{per se}, they provide a useful consistency check for
the numerical implementation of the relativistic neutrino moment
equations. The principle underlying these tests may be illustrated by
considering the neutrino energy equation (\ref{eq:momeq_cfc_energy_j})
in flat space, assuming stationarity and specializing to the
laboratory frame (where $v_r\equiv0$),
\begin{equation}
\frac{1}{r^2}\frac{\partial\left(r^2 H_\mathrm{eul}\right)}{\partial r}=C^{(0)}.
\end{equation}
Outside the central source (which may be chosen as a homogeneous
isothermal sphere of radius $R$ with vanishing scattering opacity and
frequency-independent absorption opacity as a very simple case), the
collision integral vanishes, and hence the lab frame luminosity
remains constant,
\begin{equation}
L_\mathrm{eul}=16\pi^2 H_\mathrm{eul} r^2=\mathrm{const.}
\end{equation}
Likewise, since no energy-exchanging reactions occur, and since all the
Doppler terms in the moment equations vanish, the mean energy of the radiation
is independent of the radius $r$ in the vacuum region,
\begin{equation}
\left\langle \varepsilon
\right\rangle_\mathrm{eul}=H_\mathrm{eul}/\mathcal{H}_\mathrm{eul}=\mathrm{const.}
\end{equation}
The luminosity and mean energy as measured by moving observers can be
constructed from the transformation properties of the radiation moments $J$,
$H$, and $K$ (see, e.g.\, {Mihalas} \& {Weibel Mihalas}, 1984). Since $J=H=K$ for $r\gg R$, we have
\begin{equation}
\label{eq:test_rs_1}
L W^2 \left(1+v_r^2+2v_r\right)=
L \frac{1+v_r}{1-v_r}=\mathrm{const},
\quad
\left\langle \varepsilon \right\rangle W \left(1+v_r\right)=\mathrm{const},
\end{equation}
for the luminosity $L=16\pi^2 H/c$ and the mean energy $\left\langle
\varepsilon \right\rangle$ in the frame of a moving observer.

\textbf{Test problem 1:} We have compared the numerical solution
computed with \textsc{VERTEX-CoCoNuT} in the comoving frame to this
analytical solution for the case of a radiating sphere with
$R=4\ \mathrm{km}$ and an observer velocity field with a peak value of
$v=-0.2 c$ at $r=150 \ \mathrm{km}$,
\begin{equation}
v_r=\left\{
\begin{array}{ll}
0, & r<135 \ \mathrm{km} \\
-0.2c\, \frac{r-135 \ \mathrm{km}}{15 \ \mathrm{km}}, & 135 \ \mathrm{km} \leq r < 150 \ \mathrm{km} \\
-0.2c \left(\frac{150\ \mathrm{km}}{r}\right)^2,& 150\ \mathrm{km} \leq r\\
\end{array}
\right.
\end{equation}
This setup mimics the velocity profile typically
encountered in the accretion phase of core collapse supernovae with a
strong shock discontinuity at radii of order $\approx 100
\ \mathrm{km}$.  The left panel of Fig.~\ref{fig:test_rs_1} shows that
the numerical solution is in excellent agreement with
Eq.~(\ref{eq:test_rs_1}), with a maximum deviation of less than
$2\%$. Considering that the finite-difference representation of the
moment equations has not been specifically tuned to reproduce the
analytic result, and that the energy grid is rather coarse ($\Delta
\varepsilon/\varepsilon\approx 0.28$), the accuracy achieved by
\textsc{VERTEX-CoCoNuT} seems more than adequate, and indicates that
the velocity-dependent terms in the moment equations have been
correctly implemented.

\textbf{Test problem 2:} The implementation of the gravitational
redshift terms can be tested in a similar fashion: We now consider the
propagation of radiation from an isothermal sphere in a curved
spacetime, assuming that the metric is diagonal and isotropic, i.e.\
$\mathrm{d} s^2=-\alpha^2\mathrm{d} t^2+\phi^4 \left(\mathrm{d} r^2 +r^2 \mathrm{d}
\theta^2+\sin^2 \theta \mathrm{d} \phi^2\right)$. The frequency-integrated
neutrino energy and number equations (obtained from
Eqs.~(\ref{eq:momeq_cfc_energy_j}) and (\ref{eq:momeq_cfc_energy_h})
after setting $\beta^r \equiv v_r\equiv 0$ and $W \equiv 1$) can then
be solved directly in the vacuum region,
\begin{equation}
\label{eq:test_rs_2}
\alpha \phi^4 \mathcal{H} r^2=\mathrm{const.},
\quad
\alpha^2 \phi^4 H r^2=\mathrm{const.},
\quad
\alpha \left\langle \varepsilon \right\rangle =\mathrm{const.}
\end{equation}
To test our numerical scheme, we have chosen a lapse function
$\alpha(r)$ from a pseudo-relativistic core collapse simulation with
the \textsc{VERTEX-PROMETHEUS} code ({Rampp} \& {Janka} 2002) (with a minimum
value of $\alpha(0) \approx 0.72$), thus mimicking a potential well of
similar strength as in real applications. For convenience, the
conformal factor $\phi$ is set to $\phi=\sqrt{\alpha^{-1}}$ as in the
weak-field limit. Again, the luminosity and mean energy are in very
good agreement with the analytic results for this test case, as shown
in the right panel of Fig.~\ref{fig:test_rs_1}. The redshift-corrected
luminosity\footnote{The uncorrected relativistic luminosity in a CFC
  spacetime, i.e.\ the energy passing through a sphere around the
  origin of radius $r$ per unit coordinate time is just $16\pi \alpha
  \phi^4 H r^2$ (note that in the adopted gauge, the circumferential
  radius is not given by $r$, but by $r \phi^2$). If the spacetime is
  static, gravitational redshift corrections can be taken into account
  analytically by including another factor $\alpha$.}
$L_\mathrm{rs}=\alpha \left(16\pi \alpha \phi^4 H r^2 \right)$, which
should be constant outside the radiating sphere for our specific
choice for $\alpha$ and $\phi$, is in fact constant to within
$0.02\%$.

\textbf{Test problem 3:} Test problem 1 and 2 can be easily combined
to serve as a more stringent check for the implementation of the
moment equations (as more and more non-vanishing terms enter), if we
consider non-Eulerian observers in a curved spacetime. The
redshift-corrected luminosity $L_\mathrm{rs}$ and mean energy as
measured by an observer moving with a velocity $v_r$ then obey the
following simple equations,
\begin{equation}
\label{eq:test_rs_3}
\frac{1+v_r}{1-v_r} L_\mathrm{rs}=
16 \pi \alpha^2 \phi^4 \frac{1+v_r}{1-v_r} H r^2=
\mathrm{const},
\quad
W \alpha \left(1+v_r\right)\left\langle \varepsilon \right\rangle =\mathrm{const.}.
\end{equation}
With the same choice for $v_r$, $\alpha$, and $\phi$ as in test problem 1 and
2, \textsc{VERTEX-CoCoNuT} once more reproduces the analytic solution to very
good accuracy (see Fig.~\ref{fig:test_rs_3}).

\subsection{Core Collapse in Spherical Symmetry}
\label{sec:gr_1d}

\begin{figure}
\plotone{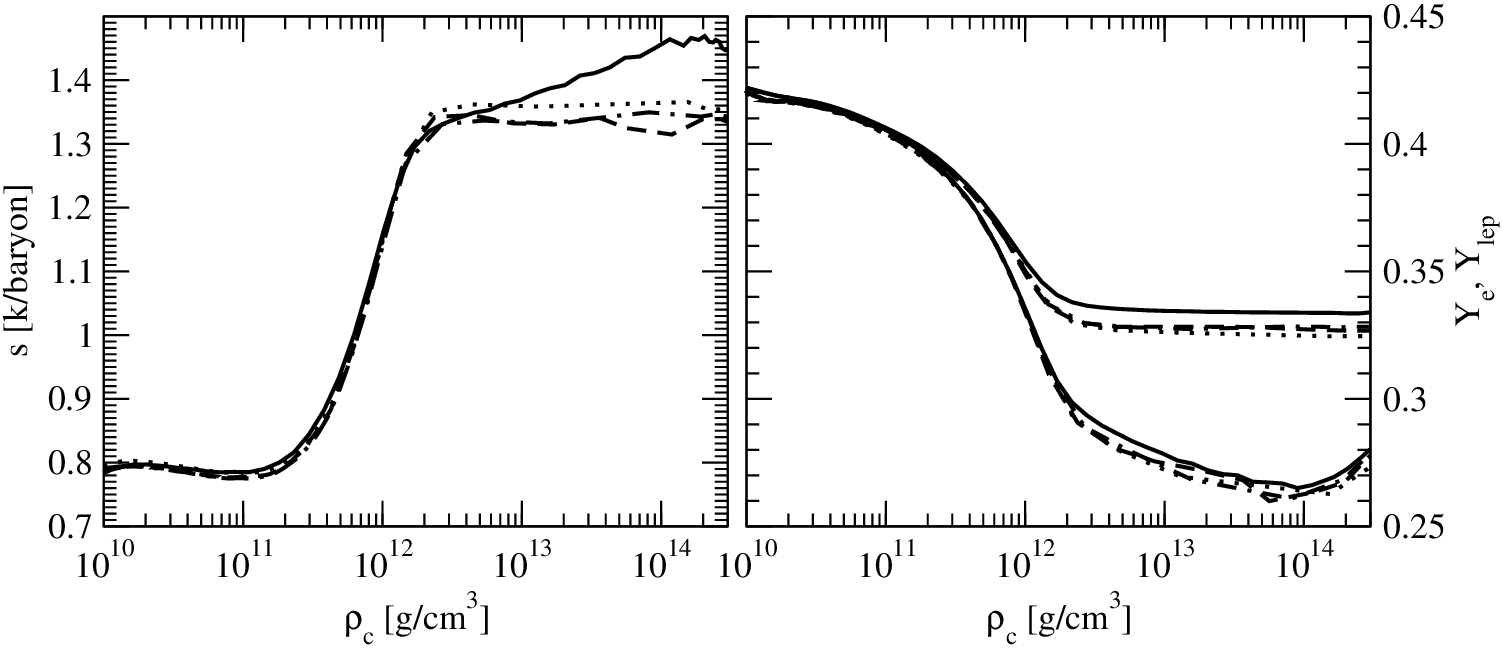} 
\caption{Comparison of the entropy per baryon $s$ (left panel) and the
  electron and lepton fraction $Y_e$ and $Y_\mathrm{lep}$ (right
  panel) versus central density $\rho_\mathrm{c}$ during collapse in
  \textsc{AGILE-BOLTZTRAN} (solid lines), \textsc{VERTEX-CoCoNuT}
  (dashed), and \textsc{VERTEX-PROMETHEUS} (dotted and dash-dotted for
  the effective potentials R and A). A slightly lower central electron
  and lepton fraction than with \textsc{AGILE} is seen for the
  \textsc{VERTEX} codes. After trapping, the central entropy is
  conserved to very good accuracy in \textsc{VERTEX-CoCoNuT}.
\label{fig:comp_collapse}
}
\end{figure}

\begin{figure}
\plotone{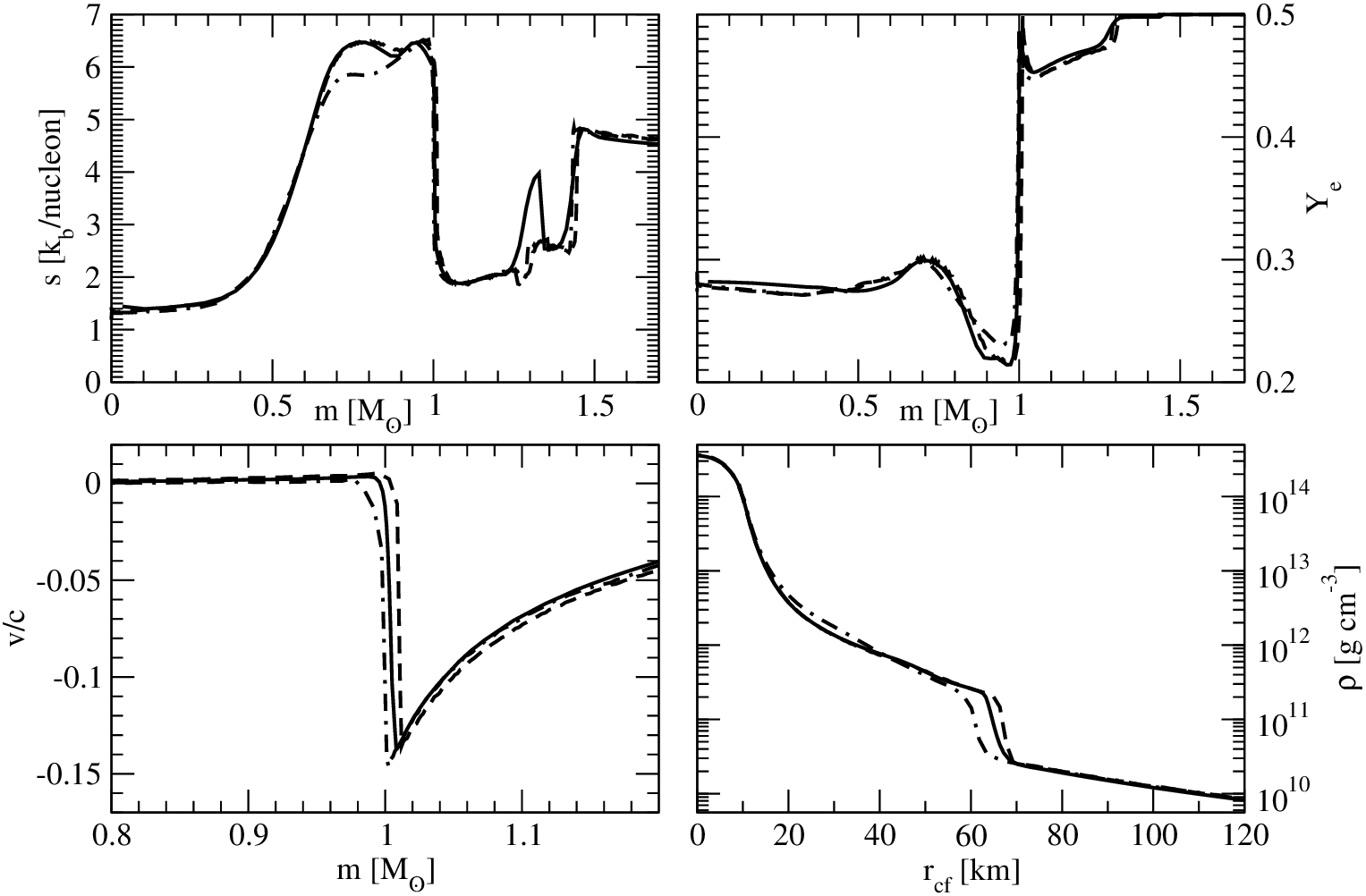} 
\caption{Profiles of the entropy per nucleon $s$ (upper left), the
  electron fraction $Y_e$ (upper right), the radial velocity $v_r$
  (bottom left), and the density $\rho$ (bottom right) for model G15
  at a time of $3 \ \mathrm{ms}$ after bounce, obtained with different
  neutrino transport codes: \textsc{AGILE-BOLTZTRAN} (solid),
  \textsc{VERTEX-CoCoNuT} (dashed), and \textsc{VERTEX-PROMETHEUS}
  (case~A) (dash-dotted). For the sake of clarity, case~R is not
  shown; the profiles are very similar to case~A.
\label{fig:comp_3ms_pb}
}
\end{figure}

\begin{figure}
\plottwo{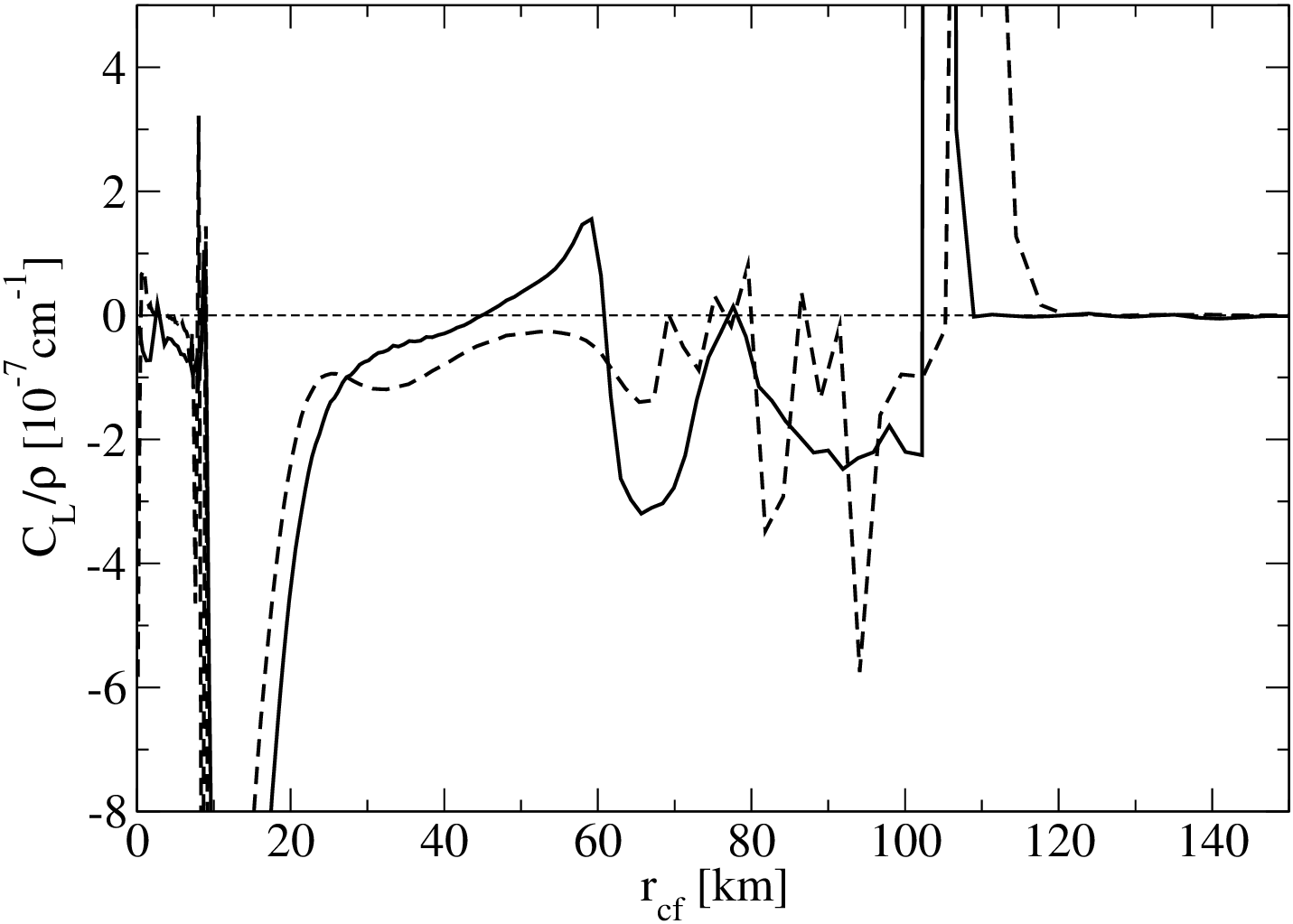}{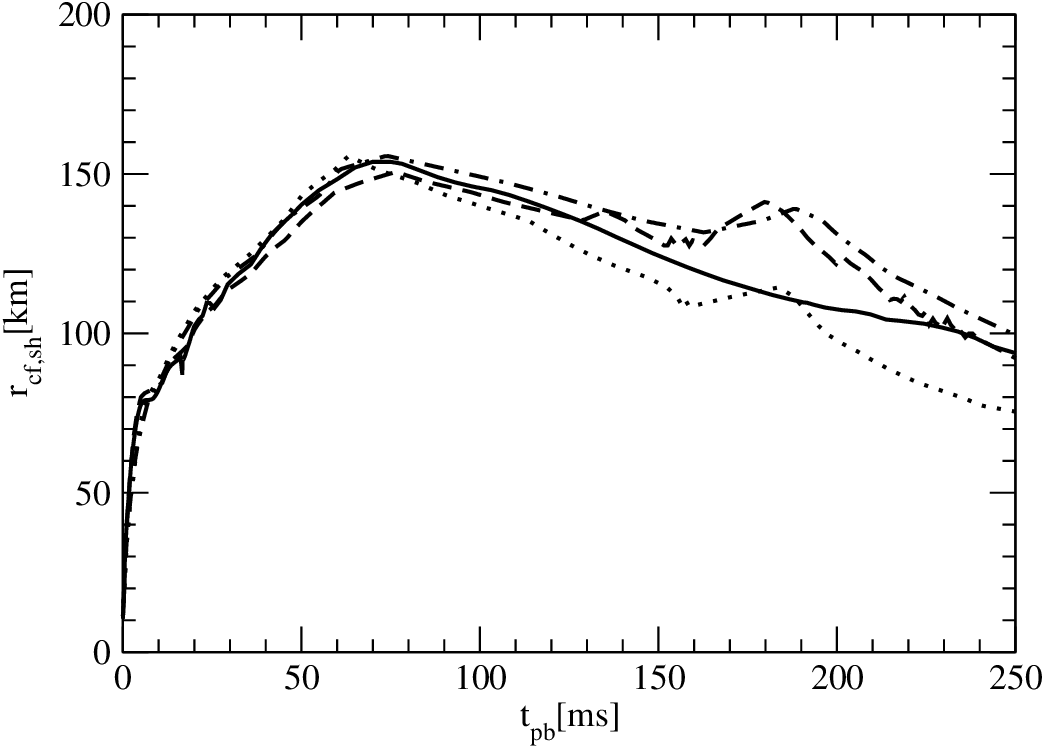}
\caption{Left panel: Ledoux criterion $20 \ \mathrm{ms}$ after bounce for the
  relativistic \textsc{VERTEX-CoCoNuT} run (solid) and for the
  pseudo-relativistic \textsc{VERTEX-PROMETHEUS} (case~A) run
  (dashed). For convenience, $C_\mathrm{L}$ is scaled to the local baryonic
  mass density $\rho $. The region between $r_\mathrm{cf}=45 \ \mathrm{km}$ and
  $r_\mathrm{cf}=60 \ \mathrm{km}$ is convectively unstable in the
  \textsc{CoCoNuT} run, while no extended unstable region is present
  in the \textsc{PROMETHEUS} run.
  Right panel: Time evolution of the shock position for model G15 in
  \textsc{AGILE-BOLTZTRAN} (solid), \textsc{VERTEX-CoCoNuT} (dashed),
  and \textsc{VERTEX-PROMETHEUS} with the effective potentials A and R
  (dotted and dash-dotted, respectively).
\label{fig:ledoux_1d}
\label{fig:shock_1d}
}
\end{figure}

\begin{figure}
\plotone{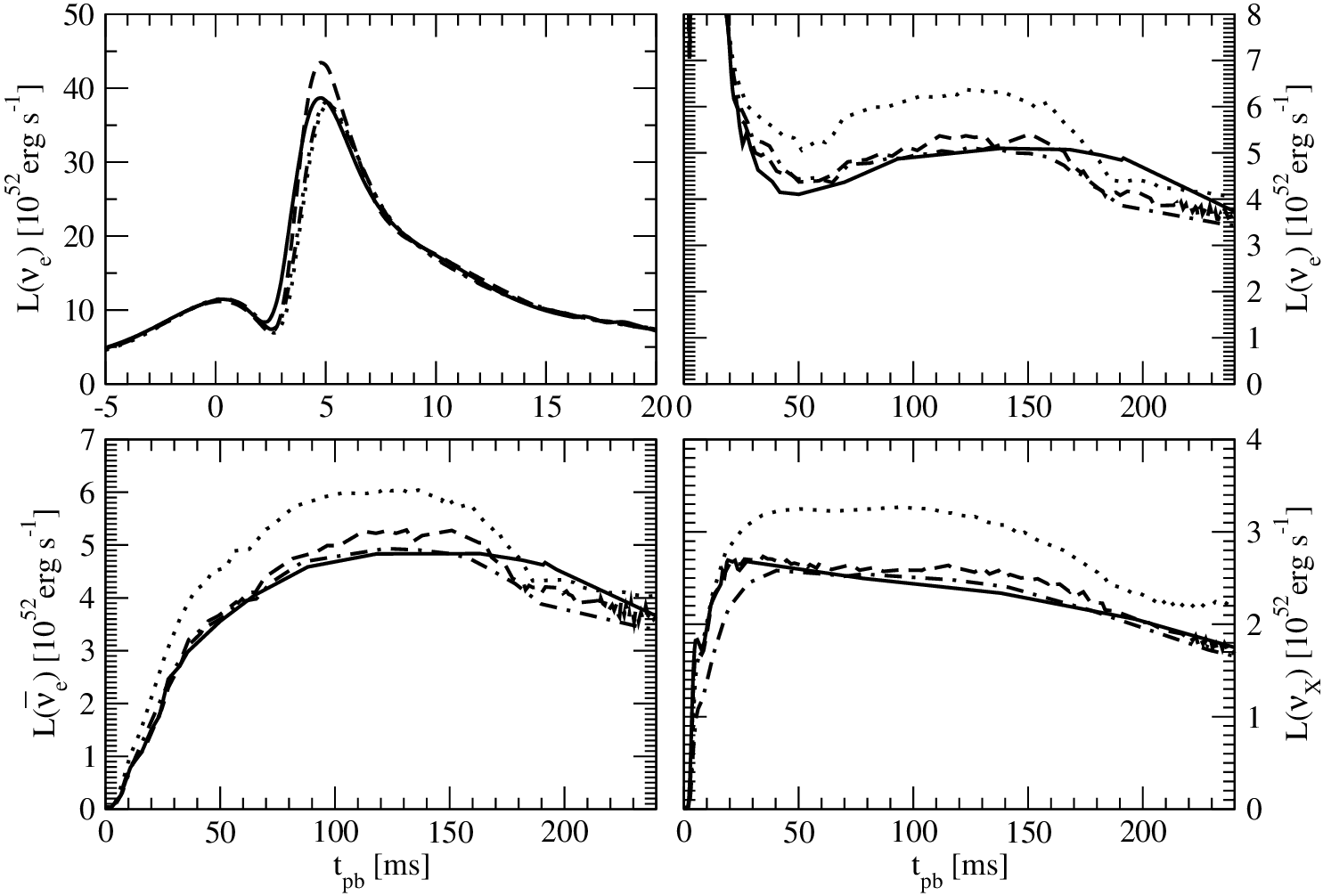} 
\caption{Neutrino luminosities for model G15 obtained with
  \textsc{AGILE-BOLTZTRAN} (solid), \textsc{VERTEX-CoCoNuT} (dashed),
  and VERTEX-PROMETHEUS (dotted and dash-dotted for the effective
  potentials R and A) as measured by an observer at $r=500
  \ \mathrm{km}$ in the comoving frame. Electron neutrino luminosities
  are shown in the upper panels, electron antineutrino and $\mu/\tau$
  neutrino luminosities in the lower left and lower right panel.
\label{fig:lum_1d}
}
\end{figure}

\begin{figure}
\plotone{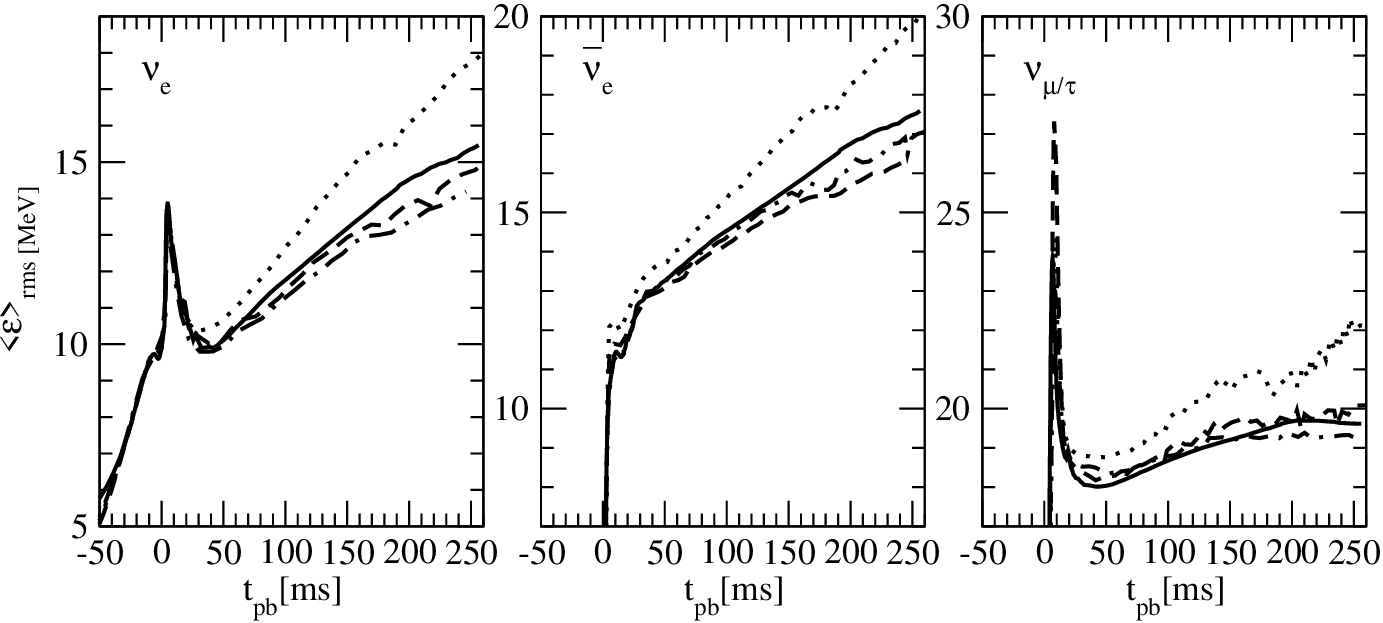}
\caption{Left: Neutrino rms energies for model G15 obtained with
  \textsc{AGILE-BOLTZTRAN} (solid), \textsc{VERTEX-CoCoNuT} (dashed),
  and \textsc{VERTEX-PROMETHEUS} (dotted and dash-dotted for the
  effective potentials R and A), sampled at $r=500 \ \mathrm{km}$ in
  the comoving frame. The three panels show the rms energies for
  electron neutrinos, electron antineutrinos and $\mu/\tau$ neutrinos
  (from left to right).
\label{fig:e_rms_1d}
}
\end{figure}

\begin{figure}
\plotone{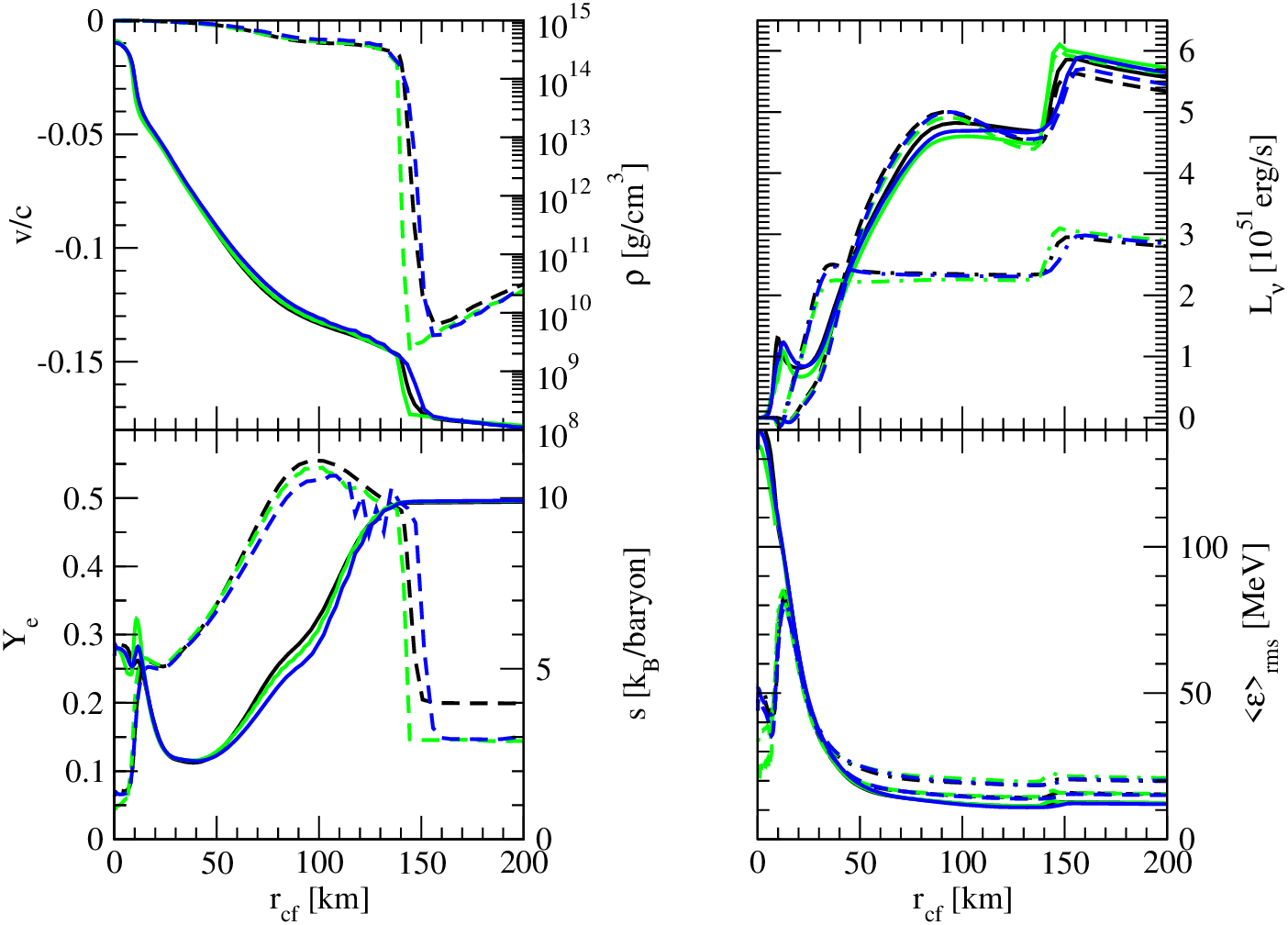} 
\caption{Radial profiles of the velocity $v$ (dashed lines), the
  density $\rho$ (solid lines, upper left), the electron fraction
  $Y_e$ (solid lines), the entropy $s$ per nucleon (dashed lines, bottom
  left), the neutrino luminosities $L_\nu$ (upper right), and the
  neutrino rms energies $\left\langle
  \varepsilon\right\rangle_\mathrm{rms}$ (bottom right) in
  \textsc{AGILE-BOLTZTRAN} (black), \textsc{VERTEX-CoCoNuT} (green),
  and \textsc{VERTEX-PROMETHEUS} with the effective potential A (blue)
  at a post-bounce time of $100 \ \mathrm{ms}$. Neutrino luminosities
  and energies are given in the comoving frame; solid lines are
  used for electron neutrinos, dashed lines for electron
  antineutrinos, and dash-dotted lines for $\mu$/$\tau$ neutrinos.
\label{fig:profile_1d_100ms}
}
\end{figure}

\begin{figure}
\plotone{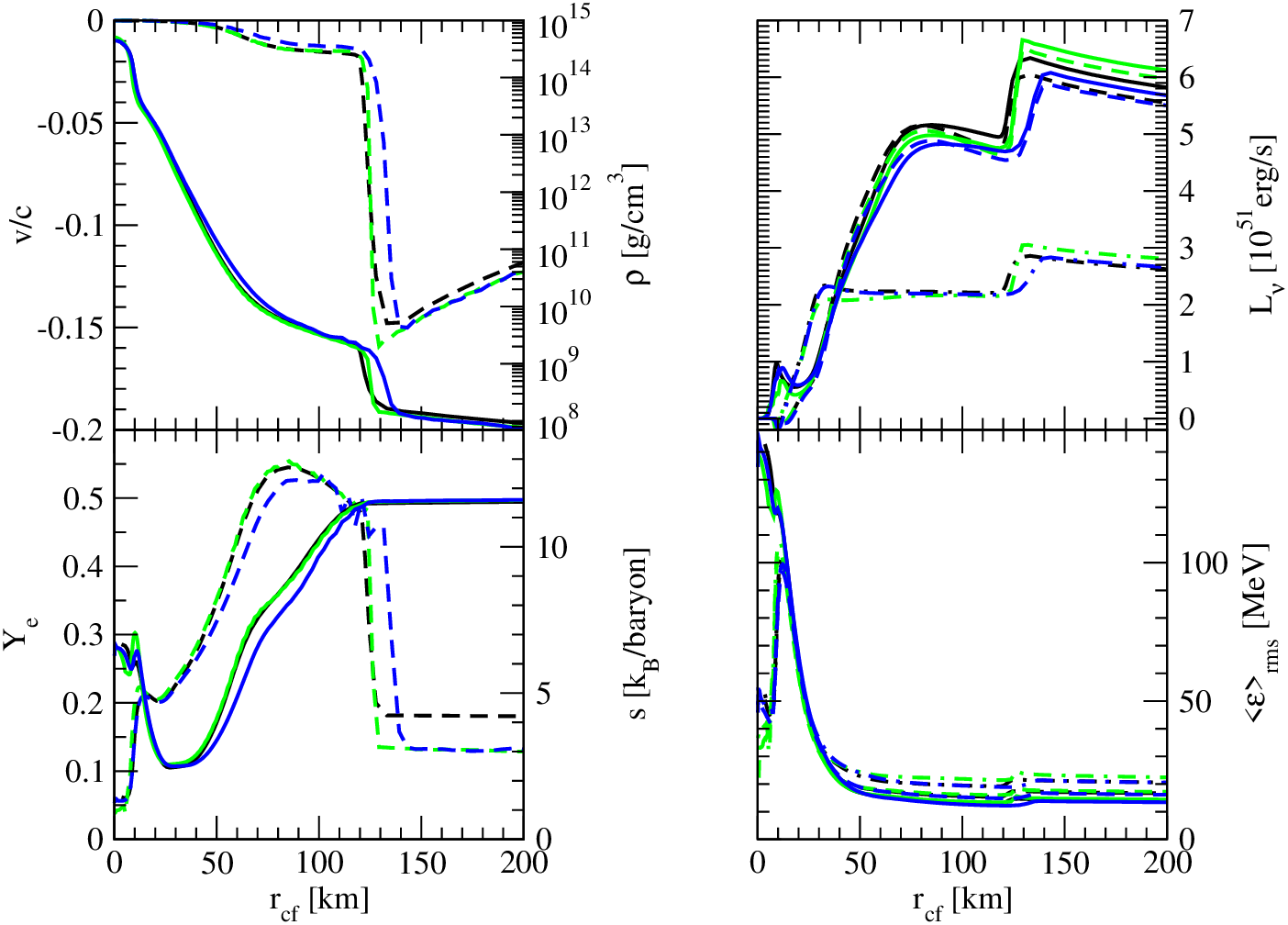} 
\caption{Radial profiles of the velocity $v$ (dashed lines), the
  density $\rho$ (solid lines, upper left), the electron fraction
  $Y_e$ (solid lines), the entropy $s$ per nucleon (dashed lines, bottom
  left), the neutrino luminosities $L_\nu$ (upper right),
  and the neutrino rms energies $\left\langle
  \varepsilon\right\rangle_\mathrm{rms}$ (bottom right) in
  \textsc{AGILE-BOLTZTRAN} (black), \textsc{VERTEX-CoCoNuT} (green),
  and \textsc{VERTEX-PROMETHEUS} with the effective potential A (blue)
  at a post-bounce time of $150 \ \mathrm{ms}$. Neutrino luminosities
  and energies are given in the comoving frame; solid lines are
  used for electron neutrinos, dashed lines for electron
  antineutrinos, and dash-dotted lines for $\mu$/$\tau$ neutrinos.
\label{fig:profile_1d_150ms}
}
\end{figure}

\subsubsection{Aims and Test Setup}
While the accuracy of \textsc{VERTEX-CoCoNuT} for the simple test
problems presented in the last section is reassuring, only a
comparison with existing one-dimensional general relativistic neutrino
transport codes can serve as a true shakedown test for future
applications in ``uncharted territory'' (i.e.\ multi-dimensional GR
neutrino transport). Fortunately, such codes are available
({Yamada} {et~al.} 1999; {Liebend{\"o}rfer} {et~al.} 2001b; {Liebend{\"o}rfer},  {Mezzacappa}, \& {Thielemann} 2001a; {Liebend{\"o}rfer} {et~al.} 2004).
A detailed comparison of the relativistic \textsc{AGILE-BOLTZTRAN} and
the Newtonian \textsc{VERTEX-PROMETHEUS} code (\textsc{AGILE} and
\textsc{PROMETHEUS} for short) has already been carried out by
{Liebend{\"o}rfer} {et~al.} (2005) and {Marek} {et~al.} (2006). Since
{Liebend{\"o}rfer} {et~al.} (2005) provide a useful framework for assessing the
quality of a newly developed code, we have repeated their run G15 with
\textsc{CoCoNuT}, and have compared our results to those obtained with
\textsc{AGILE}. As the neutrino transport module in
\textsc{VERTEX-CoCoNuT} is largely identical to that in
\textsc{VERTEX-PROMETHEUS} (apart from the underlying moment
equations), there is an additional payoff from this testing strategy:
in a number of cases, the origin of the differences between the three
codes can be pinpointed more accurately with a third code available.
In addition to the \textsc{PROMETHEUS} run discussed in
{Liebend{\"o}rfer} {et~al.} (2005), which was carried out using a potential
derived from the TOV equation (``case R''), we also considered a
\textsc{PROMETHEUS} run with a different gravitational potential
(``case~A'') proposed by {Marek} {et~al.} (2006). In total, our analysis
therefore encompasses four different simulations\footnote{The data
  from the \textsc{AGILE} and \textsc{PROMETHEUS} (case R) runs have
  been taken from the online material provided in the electronic
  version of {Liebend{\"o}rfer} {et~al.}, 2005.} of model G15 with
\textsc{AGILE}, \textsc{CoCoNuT}, and \textsc{PROMETHEUS} (with two
different choices for the gravitational potential).

Following {Liebend{\"o}rfer} {et~al.} (2005) as closely as possible, we launch
our simulation from the $15 M_\odot$ progenitor s15s7b2 of
{Woosley} \& {Weaver} (1995), and use the same set of neutrino interactions
rates\footnote{ The details of the implementation differ, however: In
  particular, both \textsc{VERTEX} codes simplify the physics in the
  $\mu$/$\tau$-neutrino sector by not treating neutrinos and
  antineutrinos separately.} and the same high-density EoS of
{Lattimer} \& {Swesty} (1991). Our treatment of the EoS differs slightly from
both \textsc{AGILE} and \textsc{PROMETHEUS} for this test run: While
using the EoS of {Lattimer} \& {Swesty} (1991) above a threshold density of
$6\times 10^7 \ \mathrm{g} \ \mathrm{cm}^{-3}$, nuclear burning is
switched off below $3\times 10^7 \ \mathrm{g} \ \mathrm{cm}^{-3}$ (as
in \textsc{AGILE}), and the composition of the progenitor is advected
with the fluid (different from \textsc{AGILE}, which considers only
one nucleus, $^{28}\mathrm{Si}$, in that regime).  Between those
transition densities, the composition is taken from a 17-species NSE
table\footnote{ This procedure has been adopted to ensure that
  material in grid cells passing from the high-density to the
  low-density regime has a physically reasonable composition.}.  An
Eulerian grid is used both for the hydrodynamics and the neutrino
transport (400 and 234 zones, respectively), and the energy resolution
is the same as in \textsc{PROMETHEUS} (17 zones).

Although the microphysical input is almost identical in all three
codes, it is crucial to bear in mind that they differ considerably
from each other in their approach to neutrino radiation hydrodynamics
before proceeding with a comparison of their results: \textsc{AGILE}
solves the relativistic Boltzmann equations directly by means of a
discrete-angle ($S_N$) method, and relies on an artificial viscosity
scheme with an adaptive (moving) grid for the
hydrodynamics. \textsc{VERTEX-PROMETHEUS} employs a variable Eddington
factor technique for the neutrino transport and the Piecewise
Parabolic Method (PPM) with an exact Riemann solver for evolving the
equations of hydrodynamics; it uses a number of approximations to
incorporate general relativistic effects, while remaining essentially
a Newtonian code.  The key element of its GR treatment is a
modification of the Newtonian gravitational potential that leads to
the same hydrostatic stellar structure equation (TOV equation) as in
general relativity (\textsc{VERTEX-PROMETHEUS} ``case~R'') or to a
sightly modified TOV equation that improves the agreement with the
relativistic case\footnote{ For a heuristic motivation of the
  modifications in Case~A, see {Marek} {et~al.} (2006).}  (``case~A'').
\textsc{VERTEX-CoCoNuT}, while based on the same general approach to
neutrino transport and hydrodynamics as \textsc{VERTEX-PROMETHEUS},
accounts for general relativistic effects exactly in spherical
symmetry (apart from the computation of the Eddington
factors). However, the approximate Riemann solver used in
\textsc{CoCoNuT} (HLLE for this run) is somewhat less accurate than
the exact solver in \textsc{PROMETHEUS}. Moreover, the two
relativistic codes \textsc{CoCoNuT} and \textsc{AGILE} differ in their
choice of gauge and slicing conditions: Ambiguities arising from this
fact must be handled carefully.

In most instances, the different gauges employed by
  \textsc{CoCoNuT} and \textsc{AGILE} do no pose any problem because
  the conversion of the gauge-dependent quantities is usually
  straightforward. Thus, the areal (circumferential) radius
  $r_\mathrm{cf}$, which is used by {Liebend{\"o}rfer} {et~al.} (2005), can
  easily be computed from the isotropic radial coordinate $r$, which
  actually appears as $r$ in the CFC metric:
\begin{equation}
r_\mathrm{cf} =\phi^2 r.
\end{equation}
We also follow {Liebend{\"o}rfer} {et~al.} (2005) in choosing the rate of change
of the areal radius per proper time as radial velocity $v$.  However,
the identification of time slices for the comparison of radial
profiles of density, luminosity, etc. is non-trivial. In principle, a
complete reconstruction of the entire spacetime would be necessary to
allow for a rigorous comparison of the code output.  Since the data
provided in {Liebend{\"o}rfer} {et~al.} (2005) are insufficient for such a
reconstruction, we are forced to identify time slices by the elapsed
coordinate time $t$ (i.e.\ proper time for a non-moving observer at
infinity). This is actually less critical than it might appear if we
synchronize the different codes at bounce and use the post-bounce time
$t_\mathrm{pb}$ as time coordinate, because the system soon settles
down to a quasi-stationary state for which the slicing conditions in
\textsc{AGILE} and \textsc{CoCoNuT} are identical to very good
approximation.

Fortunately, we can even form an estimate of the asynchronicity that
develops between the \textsc{CoCoNuT} and \textsc{AGILE} runs, by
considering the difference of the elapsed proper time
$t_\mathrm{proper}$ at the center of the proto-neutron star after a
given interval $\delta t$ of coordinate time. Since the velocity and
shift vector vanish at the center, $t_\mathrm{proper}$ is given in
terms of the central lapse function $\alpha_\mathrm{c}$ by
\begin{equation}
t_\mathrm{proper}=\int \alpha_\mathrm{c} \mathrm{d} t .
\end{equation}
The elapsed proper time between the \textsc{CoCoNuT} and
\textsc{AGILE} runs therefore differs by
\begin{equation}
\Delta t_\mathrm{proper}=\int
\left(\alpha_\mathrm{c}^\mathrm{CoCoNuT}-\alpha_\mathrm{c}^\mathrm{AGILE}\right) \mathrm{d} t.
\end{equation}
Taking the difference of the central lapse function at bounce
(obtained by extrapolating to $r_\mathrm{cf}=0$) as an upper limit, we
find that the two runs become asynchronous by at most another $2
\ \mathrm{ms}$ every $100 \ \mathrm{ms}$ at the center of the
proto-neutron star.  Since the
$|\alpha_\mathrm{c}^\mathrm{CoCoNuT}-\alpha_\mathrm{c}^\mathrm{AGILE}|$
again becomes smaller after bounce, one may conclude that the two runs
are never more than a few $\mathrm{ms}$ out of sync until the end of
our simulation $250 \ \mathrm{ms}$ after bounce.  Backed by these
findings, we can undertake a detailed comparison of the simulations in
\textsc{CoCoNuT} and \textsc{AGILE} (and, of course, also in
\textsc{PROMETHEUS}) during the collapse, bounce, and post-bounce
phases.

\subsubsection{Collapse, Bounce and Neutrino Burst Phase}
Relativistic effects are of minor importance until the late stages of
collapse when the density reaches several $10^{13} \ \mathrm{g}
\ \mathrm{cm}^{-3}$; only then do the infall velocity $v$ and the
compactness parameter $M/R$ of the iron core reach values of the order
of $0.1$ (in relativistic units).  Consequently, the three codes are
in excellent agreement during most of the collapse phase. Slight
differences in the central electron and lepton fraction after trapping
(i.e.\ at a density of a few $10^{12} \ \mathrm{g}
\ \mathrm{cm}^{-3}$) can be discerned in Fig.~\ref{fig:comp_collapse}.
Since the VERTEX-based codes consistently produce somewhat lower
values, it is likely that this difference stems from small differences
in the treatment of the neutrino interaction rates. It should be noted
that the new \textsc{VERTEX-CoCoNuT} code conserves the central lepton
fraction and entropy to very good accuracy, thus passing another
important test.

The good agreement between the different codes persists until shock
formation (defined as the instant when the specific entropy inside the
sonic point first reaches $3 k_\mathrm{B}/\mathrm{nucleon}$), which
occurs at an enclosed mass of $0.53 M_\odot$ in \textsc{AGILE},
\textsc{CoCoNuT} and \textsc{PROMETHEUS} (case~A), and at a slightly
smaller mass of $0.49 M_\odot$ in \textsc{PROMETHEUS}
(case~R). Interestingly, important special relativistic effects can be
seen shortly after bounce (see Fig.~\ref{fig:comp_3ms_pb}): Within $3
\ \mathrm{ms}$ the shock has propagated to the mass shell $m\approx
1.0 M_\odot$.  The velocity profile for the best effective potential
(case~A) of {Marek} {et~al.} (2006) still matches the profiles from
\textsc{AGILE} and \textsc{CoCoNuT} perfectly, but the entropy profile
left behind by the shock differs appreciably. As long as the shock
travels through an optically thick medium (to neutrinos), the
post-shock entropy is higher by up to $0.5 k_\mathrm{B}$ in the
relativistic case (\textsc{AGILE}, \textsc{CoCoNuT}), indicating that
the shock is initially stronger. The post-shock velocity and the
velocity jump across the shock indeed remain larger in
\textsc{CoCoNuT} than in \textsc{PROMETHEUS} out until $m\approx 0.9
M_\odot$ -- velocity profiles from \textsc{AGILE} during the relevant
phase are not available from {Liebend{\"o}rfer} {et~al.} (2005),
unfortunately. Once the shock reaches $m \approx 0.75 M_\odot$, the
combination of the relativistic jump conditions and the transition to
the semi-transparent regime results in the formation of a shallow
trough in the entropy profile, which is centered around $m \approx 0.9
M_\odot$.  On the other hand, the specific entropy increases almost
monotonically with $m$ out to the shock radius in that region in the
\textsc{PROMETHEUS} run, and the post-shock entropy is even a little
higher than in the GR runs. 

Although a detailed analysis of the cause of these differences is
rather complicated, it is at least possible to account for the higher
post-shock entropies inside $m \approx 0.75 M_\odot$ in the
relativistic case. In this region, the material is still optically
thick, and the different jump conditions thus remain the only
distinguishing factor from the Newtonian case. One can therefore
estimate the relativistic correction to the post-shock entropy by
comparing the Newtonian and relativistic form of the jump conditions
and applying a perturbative analysis: The Newtonian Rankine-Hugoniot
conditions and their relativistic analogue ({Taub} 1948; {Thorne} 1973)
can be recast into equations for the difference of the pre-shock and
post-shock pressure ($P_1$ and $P_2$) and specific enthalpy ($h_1$ and
$h_2$). In the relativistic case, $P_2-P_1$ and $h_2-h_1$ are given
by,
\begin{equation}
  \label{eq:taub1}
  P_2-P_1=\frac{\dot{M}^2}{h_1/\rho_1-h_2/\rho_2},
\end{equation}
and
\begin{equation}
  \label{eq:taub2}
  h_2-h_1=\frac{\dot{M}^2}{h_1+h_2}\left[\left(\frac{h_1}{\rho_1}\right)^2-\left(\frac{h_2}{\rho_2}\right)^2\right].
\end{equation}
Here, $\dot M$ is the baryonic mass flux through the discontinuity in
the rest frame of the shock, which is given in terms of the
three-velocities $\tilde{v}_1$ and $\tilde{v}_2$ and the Lorentz
factors $\widetilde{W}_1$ and $\widetilde{W}_2$ on either side by
\begin{equation}
  \label{eq:taub3}
  \dot{M}=\rho_1 \widetilde{W}_1 \tilde{v}_1=\rho_1 \widetilde{W}_2 \tilde{v}_2.
\end{equation}
The three-velocities and Lorentz factors are measured in the rest
frame of the shock (which is denoted by the tilde). The Newtonian
limit of the jump conditions can then be found by setting
$h_1\rightarrow 1,h_2 \rightarrow 1$ in Eqs.~(\ref{eq:taub1}) and
(\ref{eq:taub2}) and $W_1\rightarrow 1,W_2\rightarrow 1$ in
Eq.~(\ref{eq:taub3}). By linearizing
Eqs.~(\ref{eq:taub1}--\ref{eq:taub3}) around $h=1$ and $W=1$ and
around a Newtonian solution for $h$ and $P$, one obtains an estimate
for the relativistic post-shock enthalpy and pressure (as well as the
other thermodynamic variables). For the post- and pre-shock conditions
in the \textsc{PROMETHEUS} run, this analysis predicts post-shock
entropies that are higher by up to $\Delta s= 0.4 k_\mathrm{B}$ in the
relativistic case, mostly during a very brief period where the infall
velocity $\tilde{v}_1$ is as large as $|\tilde{v}_1|=0.34$ in the rest
frame of shock. This prediction is in rough agreement with the actual
simulation data. As the shock moves further out beyond $m=0.75
M_\odot$, $|\tilde{v}_1|$ decreases and the relativistic correction to
the post-shock entropy soon becomes negligible (e.g.\ $\Delta s
\lesssim 0.03 k_\mathrm{B}$ for $|\tilde{v}_1| \lesssim 0.2$),
resulting in a negative entropy gradient in the range $m \approx 0.8
M_\odot \ldots 0.9 M_\odot$.  For $m \gtrsim 0.9 M_\odot$ the
relativistic post-shock entropy then matches the Newtonian one very
closely. Although neutrino losses from the post-shock region also
start to play a role at this stage, this reduction of $\Delta s$ with
decreasing $|\tilde{v}_1|$ thus provides a qualitative explanation for
the observed trough in the entropy profile.

The effects of relativistic shock propagation are also visible in the
density stratification between the proto-neutron star and the shock
(Fig.~\ref{fig:comp_3ms_pb}), which is more shallow in the region
outside $r_\mathrm{cf}\approx 25 \ \mathrm{km}$ ($m\approx 0.7$),
i.e.\ just outside the region of the first entropy peak in the
relativistic models .  Such small differences may seem unimportant in
the case of spherically symmetric problems -- in particular when they
are washed out after a few tens of ms -- but they can be of some
relevance in multi-dimensional simulations, where they can affect the
growth of hydrodynamic instabilities. Our 1D simulations already give
a strong indication that this is indeed the case: Convective
instability is expected if the Schwarzschild-Ledoux criterion
$C_\mathrm{L}$ (cp. {Schwarzschild}, 1958) is
positive. $C_\mathrm{L}$ is given in terms of the density $\rho$, the
pressure $P$, the specific internal energy density $\epsilon$, and the
speed of sound $c_\mathrm{s}$ by
\begin{equation}
C_\mathrm{L}= \frac{\partial \rho \left(1+\epsilon \right)}{\partial r_\mathrm{cf}}-\frac{1}{c_\mathrm{s}^2}\frac{\partial P}{\partial r_\mathrm{cf}},
\end{equation}
in the relativistic case ({Thorne} 1966), or
\begin{equation}
C_\mathrm{L}= \frac{\partial \rho}{\partial r_\mathrm{cf}}-\frac{1}{c_\mathrm{s}^2}\frac{\partial P}{\partial r_\mathrm{cf}},
\end{equation}
in Newtonian hydrodynamics\footnote{Note that the
  Schwarzschild-Ledoux criterion reduces to the simple Schwarzschild
  criterion $C_\mathrm{s}=\mathrm{d} s/\mathrm{d} r$ for a chemically homogeneous fluid. The
  alternative form $C_\mathrm{L}=\mathrm{d} s / \mathrm{d} r_\mathrm{cf} \left(\partial \rho/\partial
  s\right)_{P,Y_\mathrm{lep}}+\mathrm{d} Y_\mathrm{lep} / \mathrm{d} r_\mathrm{cf} \left(\partial
  \rho/\partial Y_\mathrm{lep}\right)_{s,P}$ ({Buras} {et~al.} 2006b) can
  \emph{not} be applied in the relativistic case. However, the
  relativistic Ledoux criterion can be expressed in terms of the
  spatial derivatives of $s$ and $Y_\mathrm{lep}$ in the following
  manner, $C_\mathrm{L}=\mathrm{d} s / \mathrm{d} r_\mathrm{cf} \left(\partial (\rho+ \rho \epsilon) /\partial
  s\right)_{P,Y_\mathrm{lep}}+\mathrm{d} Y_\mathrm{lep} / \mathrm{d} r_\mathrm{cf} \left(\partial
  (\rho+\rho \epsilon) /\partial Y_\mathrm{lep}\right)_{s,P}$.  }.
The left panel of Fig.~\ref{fig:ledoux_1d} reveals a convectively unstable layer at
$r_\mathrm{cf}\approx 50\ \mathrm{km}$ with a thickness of more than $10
\ \mathrm{km}$ at a time of $20 \ \mathrm{ms}$ after bounce in the
\textsc{CoCoNuT} run, which is not present in the \textsc{PROMETHEUS}
run. 

The stronger deleptonization in the region around $m=0.9 M_\odot$ also
leads to a visibly different evolution of the electron neutrino
luminosities during the neutronization burst (first panel of
Fig.~\ref{fig:lum_1d}). Although the peak luminosity
$L_\mathrm{burst}\approx 3.8\times 10^{53} \ \mathrm{erg}
\ \mathrm{s}^{-1}$ in \textsc{PROMETHEUS} (case~A and R) agrees well
with the one in \textsc{AGILE}, the radiated energy between
$t_\mathrm{pb}=1 \ \mathrm{ms}$ and $t_\mathrm{pb}=8 \ \mathrm{ms}$ is
smaller by a around $10\%$. Conversely, the total energy radiated in
$\nu_e$-s in \textsc{CoCoNuT} during the burst is in perfect agreement
with \textsc{AGILE} despite the higher peak value of
$L_\mathrm{burst}\approx 4.3\times 10^{53} \ \mathrm{erg}
\ \mathrm{s}^{-1}$. The different shape of the burst in \textsc{AGILE}
and \textsc{CoCoNuT} is probably due to the choice of the radial grid,
or the different accuracy of the discretization scheme (first order
vs. second order in space), which both influence the numerical
diffusivity of the transport scheme and thus lead to a different
degree of broadening during the propagation of the neutrino burst from
the decoupling radius to the observer radius at $r_\mathrm{cf}=500
\ \mathrm{km}$ (cp. {Liebend{\"o}rfer} {et~al.}, 2004 for an analysis of the
broadening of the burst).  The treatment of the model Boltzmann
equation in \textsc{CoCoNuT} may also be an important factor, either
because of the approximations in the tangent ray-scheme (which may
affect the transition from the optically thick to the optically thin
regime) or because of the superior angular resolution of the
tangent-ray scheme at large radii compared to the $S_N$-method used in
\textsc{AGILE}, but these factors probably play a minor role.

\subsubsection{Accretion Phase}
The post-bounce evolution of the luminosities of all neutrino flavors
(see also Fig.~\ref{fig:lum_1d}) is again very similar in
\textsc{CoCoNuT} and \textsc{AGILE}. During the first $150 \ \mathrm{ms}$ after
bounce the luminosities in \textsc{CoCoNuT} tend to be slightly higher
than in \textsc{AGILE}. The agreement is about as good as for \textsc{PROMETHEUS}
(case~A) and significantly better than for case~R, where the strength
of the gravitational potential and hence the accretion luminosity is
considerably overestimated.  While \textsc{PROMETHEUS} (case~A) is always very
close to both \textsc{CoCoNuT} and \textsc{AGILE}, it is interesting to note
that only the two last-named codes show a very abrupt transition from
the fast rise in the $\nu_\mu$/$\nu_\tau$ luminosity until $t_{pb}=20
\ \mathrm{ms}$ to the subsequent plateau phase. The small differences
between \textsc{CoCoNuT} and \textsc{AGILE} are well within the range expected
for different codes (cp. the Newtonian run N13 in
{Liebend{\"o}rfer} {et~al.}, 2005) or even for one code at different
resolutions (cp. {Marek}, 2007). As in \textsc{PROMETHEUS}, there is a
noticeable drop in the luminosities of all flavours ($\nu_e$ and
$\bar{\nu}_e$ in particular) between $t_\mathrm{pb}=150 \ \mathrm{ms}$
and $t_\mathrm{pb}=200 \ \mathrm{ms}$ -- i.e.\ at a time, when the
shock reaches the oxygen-rich silicon shell -- which is far less
pronounced in \textsc{AGILE} due to the different treatment of the nuclear
composition in the low-density regime and the possible superiority of
the Riemann solver methods in \textsc{PROMETHEUS} and \textsc{CoCoNuT} in
following the composition discontinuity at the Si-SiO interface.

There is also good agreement between the three codes concerning the
spectral properties of the emitted neutrinos
(see Fig.~\ref{fig:e_rms_1d}). The root mean square (rms) energies of
electron neutrino and antineutrinos as measured by an observer at
$r_\mathrm{cf}=500 \ \mathrm{km}$ (in the comoving frame) in \textsc{CoCoNuT}
differ from those obtained with \textsc{AGILE} by $1 \ \mathrm{MeV}$ at
most. For electron neutrinos and antineutrinos the agreement is
better than in \textsc{PROMETHEUS} (case~R and A), where the rms energies are
either over- or underestimated. On the other hand, the $\mu$ and
$\tau$ neutrinos are a little more energetic in \textsc{CoCoNuT} than
in \textsc{AGILE} during the first $15 \ \mathrm{ms}$ after bounce, whereas
\textsc{PROMETHEUS} (case~A) is very close to \textsc{AGILE}. Given the fact that the
luminosity of $\mu$ and $\tau$ neutrinos during the first $40
\ \mathrm{ms}$ after bounce rises much more slowly than in \textsc{AGILE}, it
is likely that the excellent matching of $\nu_\mu$/$\nu_\tau$ rms
energies between case~A and \textsc{AGILE} is due to a cancellation of the
error introduced by the approximate treatment of general relativity
(leading to lower rms energies) and differences in the implementation
of $\nu_\mu$/$\nu_\tau$ interaction rates or the transport schemes
(leading to higher rms energies). Conceivably, the higher peak value
in \textsc{CoCoNuT} may originate from the different neutrino
transport algorithms, finite-difference schemes, and numerical grids
used in \textsc{VERTEX} and \textsc{AGILE}, which we already suggested
as reason for the slightly higher peak luminosity of the neutrino
burst.

Similar luminosities and spectral properties already suggest similar
neutrino loss rates in the cooling zone and similar heating rates in
the gain region, therefore the dynamical evolution of the
proto-neutron star and the accretion shock in \textsc{CoCoNuT} and
\textsc{AGILE} should not be very different either. The right panel of
Fig.~\ref{fig:shock_1d} shows that this is indeed the case: Until
$t_\mathrm{pb}=150 \ \mathrm{ms}$ the accretion shock in
\textsc{CoCoNuT} is very close to the one in \textsc{AGILE}; the
radial deviation corresponds to only one or two grid
zones. Interestingly, there is a brief transient period of shock
stagnation at $t_\mathrm{pb}\approx 7 \ \mathrm{ms}$ (lasting a few
$\mathrm{ms}$) both in \textsc{AGILE} and \textsc{CoCoNuT}, which is
absent or at least much less pronounced in \textsc{PROMETHEUS}
(cases~A and R). It is conceivable that this feature is connected to
the different relativistic and Newtonian form of the jump conditions,
and to the slightly higher energy loss in electron neutrinos during
the burst in the GR simulations. In all simulations, the maximum shock
radius of $r_\mathrm{cf,max} \approx 150 \ \mathrm{km}$ is reached at
$t_\mathrm{pb}=60 \ \mathrm{ms} \ldots 80 \ \mathrm{ms}$. Afterwards
the shock position in \textsc{CoCoNuT} lies somewhere in between the
two \textsc{PROMETHEUS} runs until $t_\mathrm{pb} \approx 150
\ \mathrm{ms}$, and is a little closer to \textsc{AGILE} than either
of them. The \textsc{CoCoNuT} and \textsc{PROMETHEUS} runs then show a
transient phase of shock expansion when the infalling Si-SiO interface
reaches the accretion front; this feature is absent in \textsc{AGILE}
(as explained before) because of the different nuclear composition of
low-density material. At later times ($t_\mathrm{pb}>220
\ \mathrm{ms}$), the shock in \textsc{CoCoNuT} again recedes to the
same radius as the shock in \textsc{AGILE}. The overall shock
trajectory in \textsc{PROMETHEUS} (case~A, but not for case~R) is very similar to the one
in \textsc{CoCoNuT}, which indicates that the
effects of general relativity can still be adequately captured by an
effective gravitational potential in one-dimensional simulations.

The very close overall agreement between \textsc{CoCoNuT},
\textsc{AGILE} and \textsc{PROMETHEUS} (case~A) and the even better
agreement between \textsc{CoCoNuT} and \textsc{AGILE} is also evident
in Figs.~\ref{fig:profile_1d_100ms} and \ref{fig:profile_1d_150ms},
which show radial profiles of selected physical quantities at $100
\ \mathrm{ms}$ and $150 \ \mathrm{ms}$ after bounce.  There are only
minute differences in the velocity profiles produced by the three
codes; the shock front seems to be resolved a little more sharply in
\textsc{CoCoNuT}, but this is probably incidental, since the sharpness
of the jump changes slightly as the shock slowly moves from zone to
zone.  The effects of genuinely relativistic shock propagation in the
early post-bounce phase, which had resulted in a somewhat different
density stratification in the post-shock region in the GR case, have
now been largely washed out.  This can also be seen in the entropy
profiles, which are very similar during this phase, except for the
fact that \textsc{PROMETHEUS} (case~A) seems to give slightly lower
entropies in the gain region. Interestingly, we now find the
  electron fraction behind the shock to be somewhat lower in
\textsc{PROMETHEUS} (case~A), contrary to the situation near shock
breakout. The luminosity and neutrino rms energy profiles show some
visible, but not worrisome differences between the three codes;
e.g.\ the Doppler jump at the shock is generally a little larger in
\textsc{CoCoNuT} than in \textsc{AGILE}, which is to be expected
however, since there are small differences in the pre- and post-shock
velocities. Considering that the three codes use different radial
grids (and in the case of \textsc{AGILE} also a different grid in
energy space), such small discrepancies seem unavoidable and are
probably not to be ascribed to coding errors in any of them.

\subsubsection{Long-time Tests}
\begin{figure}
\plottwo{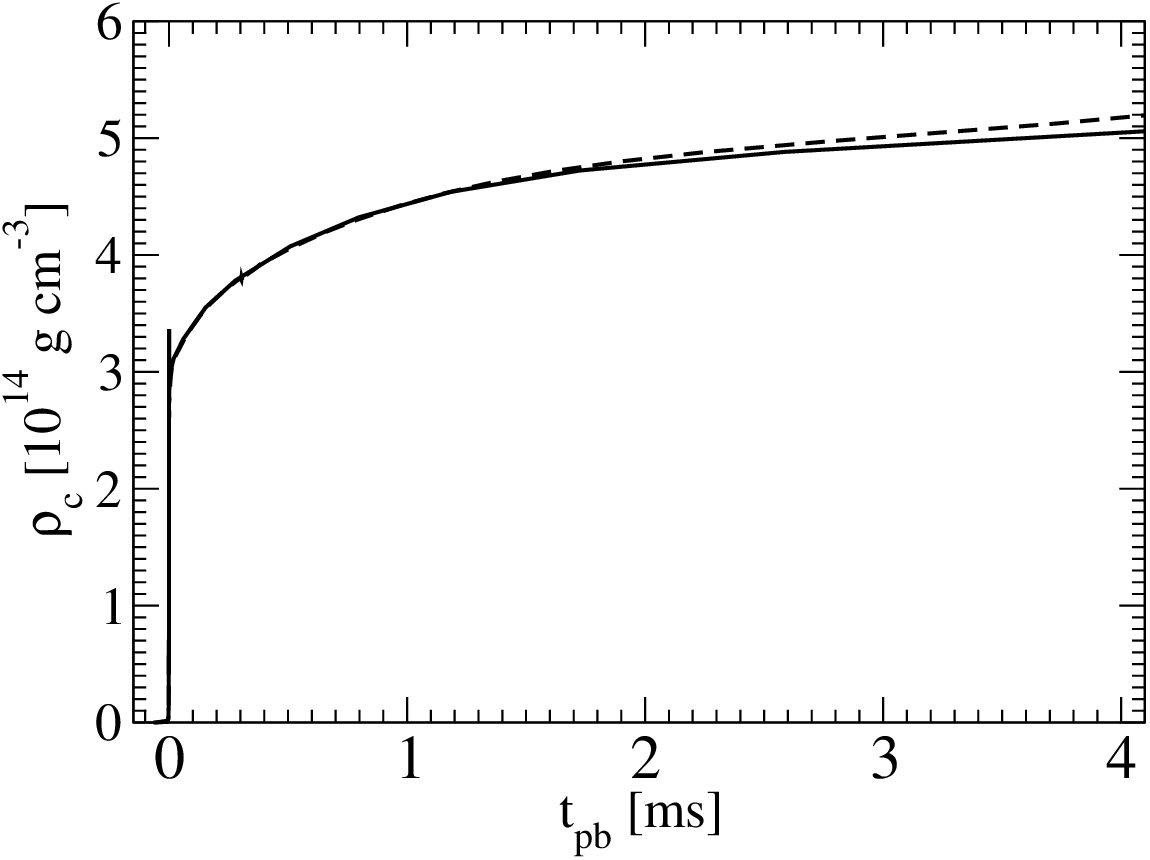}{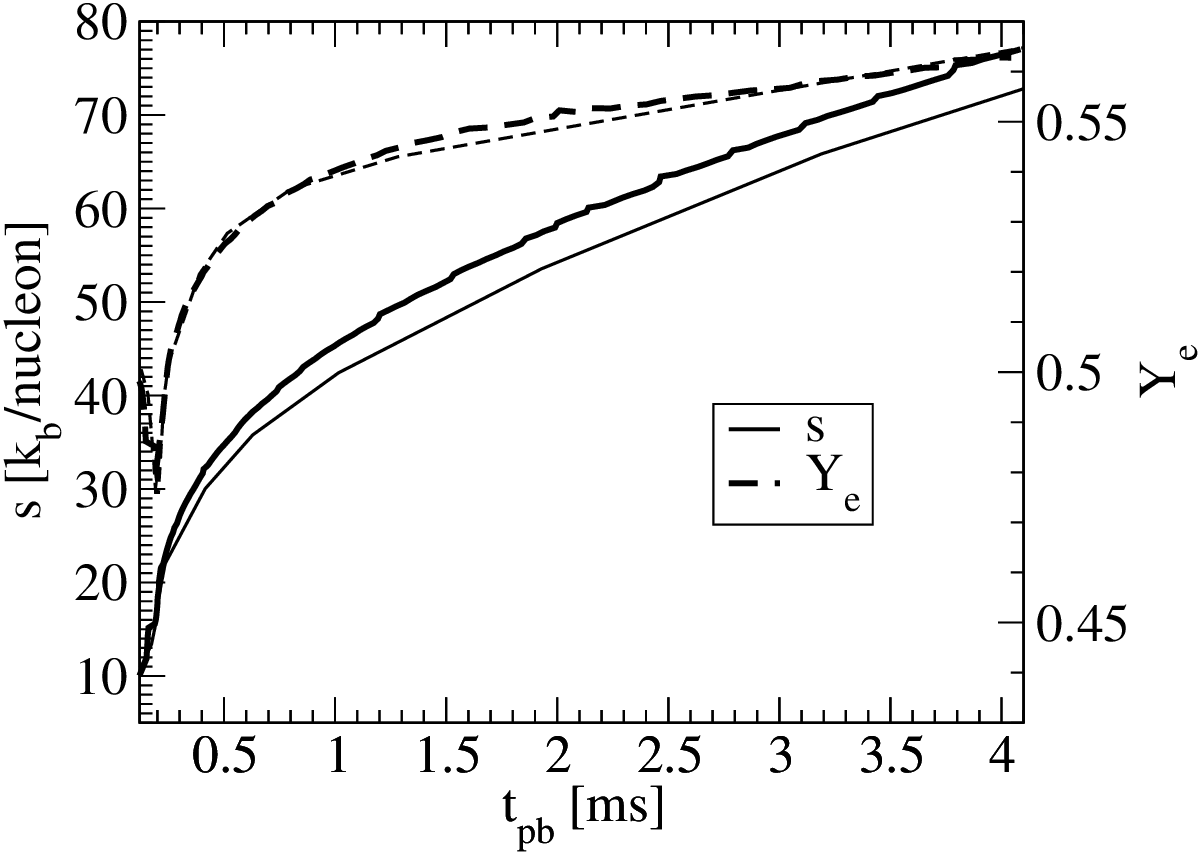}
\caption{Left: Time evolution of the central density $\rho_\mathrm{c}$
  for model G8.8 with \textsc{CoCoNuT} (solid line) and
  \textsc{PROMETHEUS} (dashed line). Right: Specific entropy $s$
  (solid lines) and electron fraction $Y_e$ (dashed lines) in the
  neutrino-driven wind at a fixed radial coordinate of $500
  \ \mathrm{km}$ as a function of time for model G8.8 in the
  \textsc{CoCoNuT} (thick lines) and \textsc{PROMETHEUS} (thin lines)
  runs.
\label{fig:dens_and_wind_onemg}
}
\end{figure}

\begin{figure}
\plotone{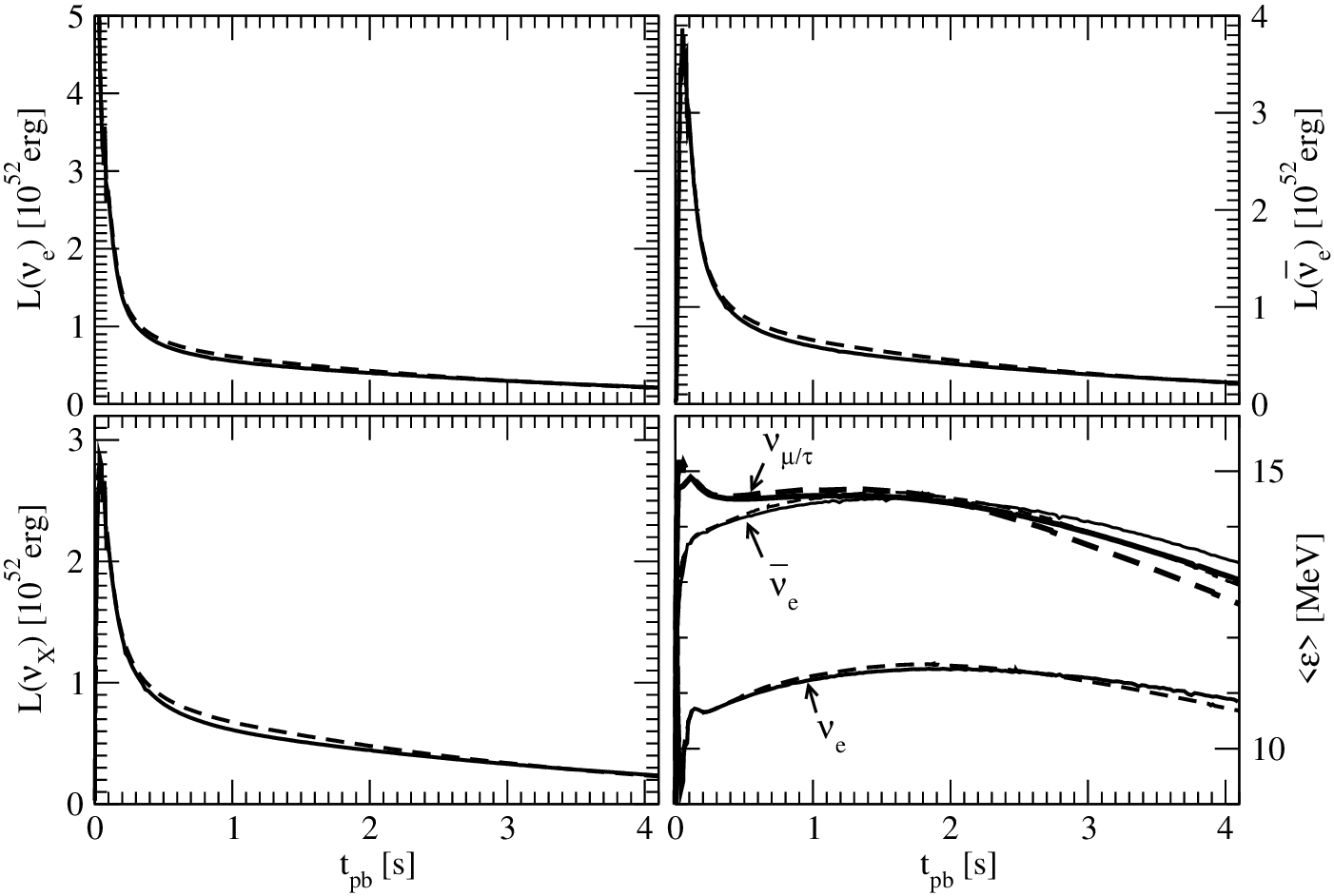}
\caption{Neutrino luminosities and mean energies for model G8.8
  obtained with \textsc{VERTEX-CoCoNuT} (solid lines) and
  \textsc{VERTEX-PROMETHEUS} (case A, dashed lines) as measured at
  $r_\mathrm{cf}=500 \ \mathrm{km}$ in the lab frame. The first three panels show
  the electron neutrino, electron antineutrino and $\mu/\tau$ neutrino
  luminosities; the right lower panel shows the mean energies for
  neutrinos of all flavours ($\nu_e$, $\bar{\nu}_e$,
    $\nu_{\mu/\tau}$) as indicated by arrows. Slightly thicker/thinner
    lines are used for $\nu_{\mu/\tau}$ and $\bar{\nu}_e$,
    respectively, in order to facilitate their visual distinction.
\label{fig:lum_onemg}
}
\end{figure}

Unfortunately a direct verification of our new code beyond $250
\ \mathrm{ms}$ by means of a comparison with \textsc{AGILE-BOLTZTRAN}
is not easily possible, because data for model G15 are not provided by
{Liebend{\"o}rfer} {et~al.} (2005) beyond that point. However, the mere
demonstration that \textsc{VERTEX-CoCoNuT} can stably evolve supernova
models considerably further would be valuable in its own right, and
might also add some insights concerning the reliability of the
effective potential approach used in \textsc{PROMETHEUS} at very late
times. We have therefore also repeated one of the longest-running 1D
simulations carried out with \textsc{VERTEX-PROMETHEUS}, viz., a model
(``G8.8'') of the explosion of the $8.8 M_\odot$ progenitor of
{Nomoto} (1984, 1987) with an O-Ne-Mg core. To this end, we have
included the effect of electron captures on ${}^{20}\mathrm{Ne}$ and
${}^{24}\mathrm{Mg}$ ({Takahara} {et~al.} 1989) and the same approximate
burning treatment as in {Kitaura} {et~al.} (2006) in the \textsc{CoCoNuT}
code. Different from the G15 run and from published long-time
simulations of the same progenitor model
({Fischer} {et~al.} 2009; {H\"udepohl} {et~al.} 2009), we now use a new nuclear EoS kindly
provided by J.~Lattimer (private communication) in the high-density
regime; a more extensive discussion of this new EoS in the context of
supernova simulations will be given in H\"udepohl et al. (2010, in
preparation).

As for the more massive progenitor used in the G15 run,
\textsc{CoCoNuT} and \textsc{PROMETHEUS} give very similar results
throughout collapse and during the short accretion phase of model
G8.8, and we therefore refrain from repeating the same detailed
analysis of these stages as for model G15. Instead, we avail ourselves
of the opportunity to compare a full GR treatment and the effective
potential approach during the post-explosion phase.  Due to the
structure of the progenitor model, the accretion rate drops rapidly as
early as $70\ldots 80 \ \mathrm{ms}$ after bounce when the shock
reaches the edge of the core, and neutrino heating becomes effective
in unbinding the post-shock material, resulting in a weak ($\approx
10^{50} \ \mathrm{erg}$) explosion
(cp. {Kitaura} {et~al.}, 2006; {Janka} {et~al.}, 2008). Interestingly, an important
difference between the \textsc{CoCoNuT} and \textsc{PROMETHEUS} runs
emerges at this point: As the shock propagates through the extremely
steep density gradient at the edge of the core, it accelerates
considerably until it reaches the dilute hydrogen envelope of the
progenitor whose density in the innermost $10^{7} \ \mathrm{km}$ is of
the order of only $10^{-8}\ldots 10^{-7} \ \mathrm{g}
\ \mathrm{cm}^{-3}$.  In the \textsc{PROMETHEUS} run, the maximum
post-shock velocity $v_\mathrm{ps}$ becomes superluminal with a
maximum value of $v_\mathrm{ps}=5.3\ c$, which clearly implies that
the propagation of the shock through the outer layers of the core and
the hydrogen envelope cannot be captured correctly with a Newtonian
code. In the \textsc{CoCoNuT} run all velocities correctly remain
subluminal with a maximum value of $v_\mathrm{ps}=0.9848\ c$
(corresponding to a Lorentz factor of $W=5.76$). However, the shock
becomes relativistic only at densities below $\approx 10
  \ \mathrm{g} \ \mathrm{cm}^{-3}$, and at that point the post-shock
temperatures are already far too low to allow for nuclear burning. The
results of {Janka} {et~al.} (2008) about the nucleosynthesis conditions
behind the shock in O-Ne-Mg supernovae therefore remain valid.

Apart from the different propagation of the shock in the
\textsc{CoCoNuT} and \textsc{PROMETHEUS} runs, the subsequent
evolution of the newly formed remnant in model G8.8 is remarkably
similar in both cases. The contraction of the proto-neutron star due
to the continuing neutrino losses is almost in perfect agreement as
illustrated by the left panel of Fig.~\ref{fig:dens_and_wind_onemg},
which shows the evolution of the central density
$\rho_\mathrm{c}$. Even more than $4 \ \mathrm{s}$ after bounce,
$\rho_\mathrm{c}$ is only $\approx 3 \%$ higher in the
\textsc{PROMETHEUS} run, and the neutron star radii (not shown)
  are almost identical as well, with a maximum deviation of $2 \%$.
Slightly larger differences are observed in the neutrino luminosities
(see Fig.~\ref{fig:lum_onemg}), which are typically a little higher in
the pseudo-Newtonian \textsc{PROMETHEUS} run from $0.5$ to $2
\ \mathrm{s}$ after bounce, but the difference to the relativistic
simulation stays low ($\approx 5 \%$). The mean energies of the
emitted neutrinos also agree to within $3\%$ during the post-explosion
phase, and it is only after more than $2 \ \mathrm{s}$ that the
deviation starts to become larger, reaching a maximum of $0.4
\ \mathrm{MeV}$ for the $\mu/\tau$ neutrinos.  Considering that the
finest energy grid zone has a width of $1.3 \ \mathrm{MeV}$, this
close agreement seems quite remarkable.  However, despite the fact
that the neutrino luminosities and mean energies, as well as the
structure of the proto-neutron star remain extremely similar, there is
a somewhat bigger discrepancy between the two codes concerning
the properties of the neutrino-driven wind (see right panel of
Fig.~\ref{fig:dens_and_wind_onemg}). While the electron fraction $Y_e$
in the wind differs by less than $0.003$ at late times, the
entropy $s$ reached asymptotically in the wind is roughly $7\%$ lower in the
\textsc{PROMETHEUS} run. Since $s$ scales roughly with a small
negative power of the luminosity ($s \propto L^{-1/6}$, see
{Qian} \& {Woosley}, 1996), it seems unlikely that this is a result of the
slightly stronger emission of neutrinos in the \textsc{CoCoNuT} run.
Furthermore, we find the wind entropy to be lower in
\textsc{PROMETHEUS} already at early times when the luminosities are
still almost in perfect agreement with \textsc{CoCoNuT}.  Even if
  the small differences in the mean neutrino energies of the order of
  less than $3 \%$ are also taken into account, the different wind
  entropies still cannot be explained on the basis of the neutrino
  emission alone: Assuming that the dependence of the final entropy
  on the mean energy of electron antineutrinos $\left \langle
  \varepsilon \right \rangle_{\bar{\nu}_e}$ is roughly given by $s
  \propto \left \langle \varepsilon\right \rangle_{\bar{\nu}_e}
  ^{-1/3}$ (see again {Qian} \& {Woosley}, 1996), the combined effect of
  different neutrino energies and luminosities could only account for
  a relative difference of $\lesssim 2 \%$ in $s$ in the most
  optimistic case. This indicates that the Newtonian treatment of
the equations of hydrodynamics in \textsc{PROMETHEUS} is probably the
cause for the observed discrepancy: Although the effective potential
used in \textsc{PROMETHEUS} reproduces the structure of \emph{static}
neutron stars very accurately, it is obviously somewhat less adequate
for \emph{stationary} (but non-static) flows. However, with a
deviation of the specific entropy of the wind of less than $10\%$, the
best effective potential (case~A) of {Marek} {et~al.} (2006) can still be
considered quite reliable even for the wind phase.

These results are very reassuring in two different respects: The
\textsc{VERTEX-CoCoNuT} code has proved a reliable and robust tool
even for long-time supernova simulations well into the proto-neutron
star cooling phase. In addition, the effective potential approach,
though unable to reproduce all the details of the relativistic G8.8
run exactly, has been shown to be very accurate in a regime where it
has hitherto remained untested.

\subsubsection{Energy and Lepton Number Conservation}

\begin{figure}
\plottwo{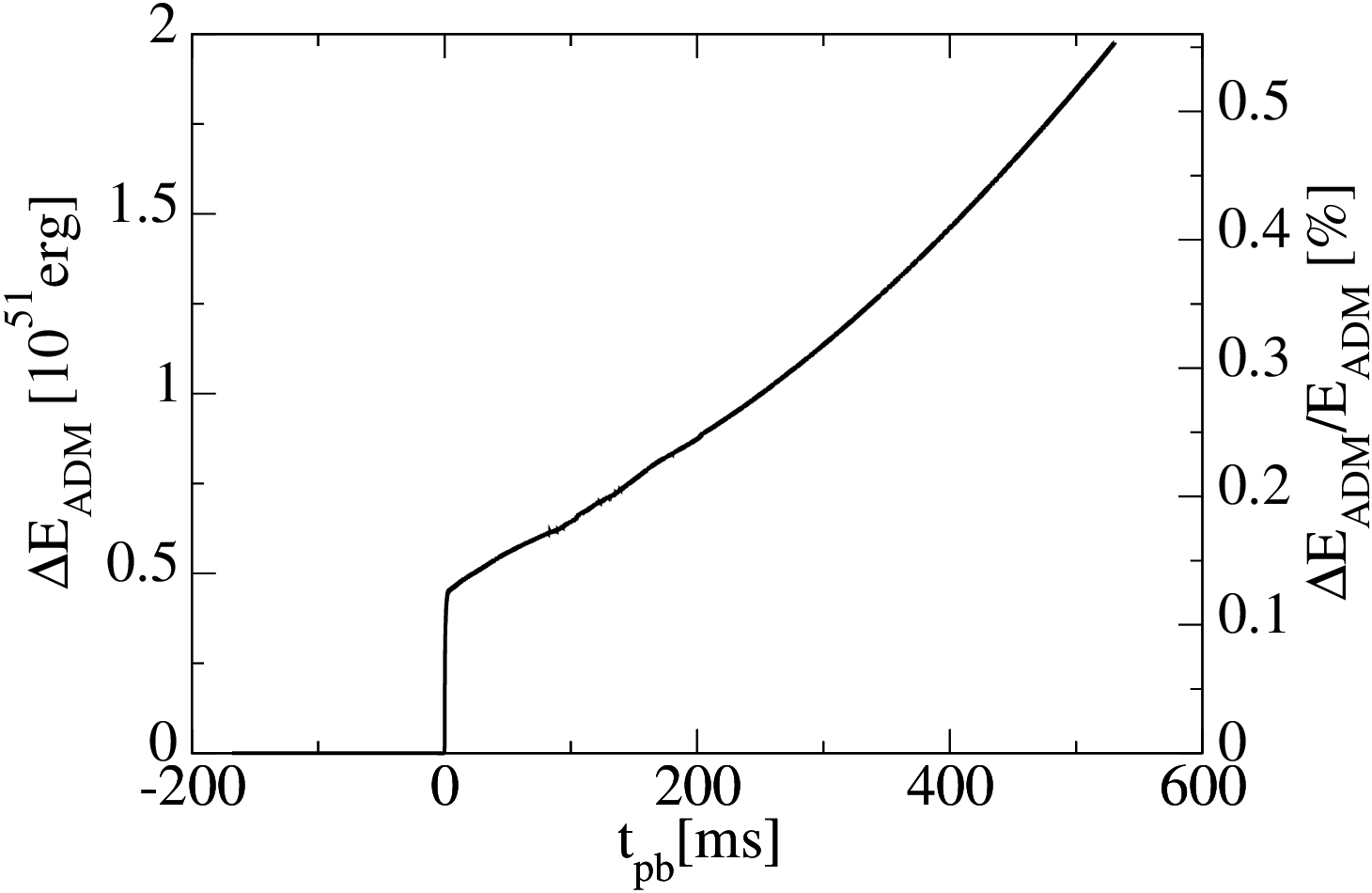}{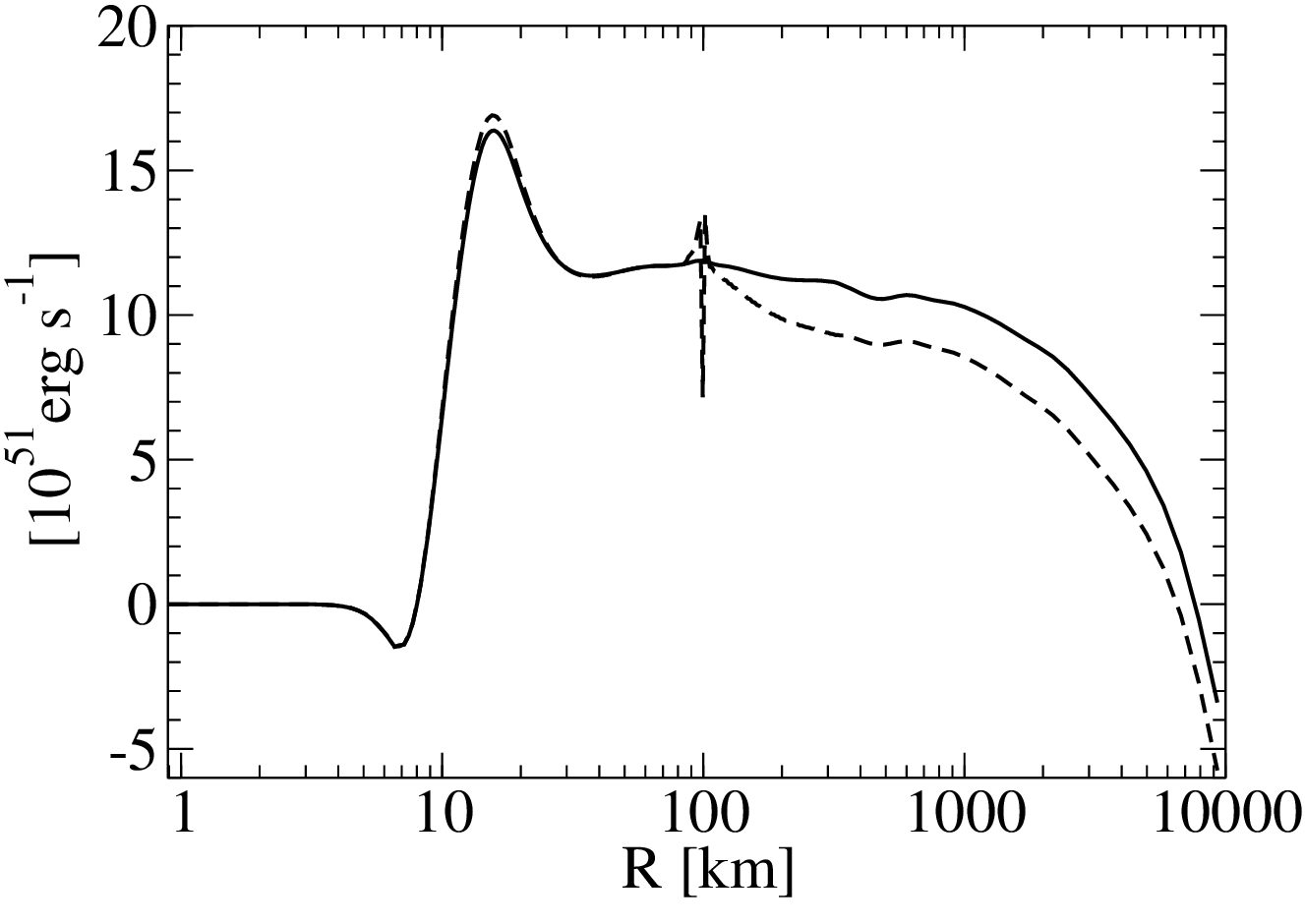} 
\caption{Left panel: Violation of ADM energy conservation $\Delta
  E_\mathrm{ADM}$ as a function of time for model G15. $\Delta
  E_\mathrm{ADM}$ is shown both in absolute values and relative to the
  initial ADM energy. Right panel: Rate of change of the total
  neutrino energy inside a given radius $R$ (solid) compared to the
  contributions from source terms and surface fluxes (dashed) at
  $t_\mathrm{pb}=230 \ \mathrm{ms}$.
\label{fig:enecons_total}
\label{fig:enecons_neutrino}
}
\end{figure}

Our test runs also provide an opportunity to check the ability of
\textsc{VERTEX-CoCoNuT} to conserve total energy and lepton number.
The usefulness of our improved numerical treatment of the relativistic
energy equation presented in App.~\ref{app:enecons} and the
finite-difference scheme for redshift and Doppler terms in the
neutrino moment equation described in App.~\ref{app:cons_dopp} can
thus be tested beyond the level of simplified test problems.  It
should be noted that the very fact that the new code allows us to
quantify the numerical violation of energy conservation unambiguously
is a major advantage compared to the pseudo-relativistic approach of
\textsc{VERTEX-PROMETHEUS}: For the effective potentials of
{Marek} {et~al.} (2006) and {M{\"u}ller} {et~al.} (2008), a conservation for the total
energy cannot even be formulated. On the other hand, in the framework
of the $3 + 1$ formalism there is a conservation law for the ADM
energy ({Landau} \& {Lifschitz} 1997; {Misner} {et~al.} 1973), which is given by
(cp. Eq.~(\ref{eq:cfc_1}))
\begin{equation}
E_\mathrm{ADM} =
\int\limits_0^\infty
\int\limits_0^\pi
\int\limits_0^{2\pi}
\phi^5 \left(E+ \frac{K_{ij} K^{ij}}{16 \pi}\right)
r^2 \sin \theta
\, \mathrm{d} \varphi
\, \mathrm{d} \theta
\, \mathrm{d} r
,
\end{equation}
in the case of a CFC spacetime. The accuracy of total energy
conservation can therefore be checked by monitoring $E_\mathrm{ADM}$
(taking into account surface flux terms where necessary) at least in
spherical symmetry, where the conformal flatness condition does not
involve any approximations to the Einstein equations. In the
multi-dimensional case, the CFC system (\ref{eq:cfc_1} --
\ref{eq:cfc_3}) is not equivalent to the full Einstein equations, and
$E_\mathrm{ADM}$ is therefore not conserved in general. Even though,
$E_\mathrm{ADM}$ can still be computed and remains a valuable
diagnostic quantity for two reasons: First, $E_\mathrm{ADM}$ should
still be conserved to good accuracy as long as the deviations from the
conformal flatness remain sufficiently small, which is probably the
case for the generic supernova problem
(cp. Sec.~\ref{sec:approx_quality}). Second, whether the
non-conservation of $E_\mathrm{ADM}$ is indicative of a numerical
problem with energy conservation or of some failure of the CFC
approximation is actually of minor importance; an overly large deviation
$\Delta E_\mathrm{ADM}$ from the initial value would be a sign of
trouble either way.

However, in the one-dimensional case $E_\mathrm{ADM}$ ought to remain
constant (neglecting surface fluxes), and $\Delta E_\mathrm{ADM}$ can
be directly used to measure the numerical energy conservation error.
The time evolution of $\Delta E_\mathrm{ADM}$ is shown in
Fig.~\ref{fig:enecons_total}: After a very rapid increase by $0.5
\times 10^{51} \ \mathrm{erg}$ around bounce, $\Delta E_\mathrm{ADM}$
gradually grows by another $1.5 \times 10^{51} \ \mathrm{erg}$ during
the next $500 \ \mathrm{ms}$ after bounce\footnote{ Interestingly, the
  evolution of $\Delta E_\mathrm{ADM}$ is somewhat different from the
  relativistic test case in App.~\ref{app:enecons}, where $\Delta
  E_\mathrm{ADM}$ becomes negative at bounce, and afterwards settles
  down at around $-1 \times 10^{50} \ \mathrm{erg}$. Imperfect energy
  conservation in the neutrino sector as described in this section may
  partly account for this. However, the different dynamical evolution
  of the GR model in App.~\ref{app:enecons} (where neutrino energy
  losses are neglected completely) may also be of relevance.  }. In
two-dimensional simulations to $\approx 300 \ \mathrm{ms}$ after
bounce, we also achieve about the same accuracy. The significance of
these numbers is not immediately clear without further analysis. On
the one hand the fact that $\Delta E_\mathrm{ADM}$ is of the order of
the typical explosion energy may seem worrisome, on the other hand
$\Delta E_\mathrm{ADM}$ is small compared to the gravitational binding
energy. Fortunately, the proper scale of reference for $\Delta
E_\mathrm{ADM}$ can be worked out by determining the cause of the
violation of energy conservation more precisely. As shown in
App.~\ref{app:enecons} for the purely hydrodynamic case, $|\Delta
E_\mathrm{ADM}|$ should not exceed a few $10^{50} \ \mathrm{erg}$
during the bounce phase for our improved implementation of the
gravitational source term in the energy equation (\ref{eq:hydro_3}),
and the additional error accumulated after the formation of the
proto-neutron star is even smaller. This suggests that the secular
increase of $\Delta E_\mathrm{ADM}$ during the accretion phase occurs
in the neutrino transport sector. To verify this, we specialize to the
case of a static background spacetime, where the relativistic
evolution equations for the component $T^{00}$ of the stress-energy
tensor $T$ of the matter (or the neutrinos) can be written in a fully
conservative form by means of the following manipulations,
\begin{eqnarray}
\frac{1}{\sqrt{-g}}\frac{\partial }{\partial t}\left(\sqrt{-g} T^{00}\right)+
\frac{1}{\sqrt{-g}}\frac{\partial }{\partial x^j}\left(\sqrt{-g} T^{0j}\right)+
\Gamma^0_{\nu \lambda} T^{\nu \lambda}&=&Q,
\\
\frac{\partial }{\partial t}\left(\sqrt{-g} T^{00}\right)+
\frac{\partial }{\partial x^j}\left(\sqrt{-g} T^{0i}\right)+
\sqrt{-g} g^{00}\frac{\partial}{\partial x^i}\left(g_{00}\right) T^{0i}&=&\sqrt{\gamma}Q,
\\
\frac{\partial }{\partial t}\left(\sqrt{-g} T^{00}\right)+
\frac{\partial }{\partial x^j}\left(\sqrt{-g} T^{0i}\right)+
\sqrt{-g} \frac{\partial \ln \alpha}{\partial x^i} T^{0i}&=&\sqrt{\gamma}Q,
\\
\label{eq:cons_static}
\frac{\partial }{\partial t}\left(\alpha \sqrt{-g} T^{00}\right)+
\frac{\partial }{\partial x^j}\left(\alpha \sqrt{-g} T^{0i}\right)&=&\alpha \sqrt{\gamma}Q.
\end{eqnarray}
Here, $T^{\mu\nu}$ may denote either the stress-energy tensor of the
matter or of the neutrinos, and $Q$ is the source term due to
neutrino-matter interactions, which enters with different signs in
both cases ($Q_\mathrm{matter}=-Q_\mathrm{neutrinos}$). The time-like
components $T_\mathrm{tot}^{0\mu}$ of the total stress-energy tensor
therefore obey the following conservation law,
\begin{equation}
\label{eq:cons_static_total}
\frac{\partial }{\partial t}\left(\alpha \sqrt{-g} T_\mathrm{tot}^{00}\right)+
\frac{\partial }{\partial x^j}\left(\alpha \sqrt{-g} T_\mathrm{tot}^{0i}\right)=0.
\end{equation}

Since our implementation of the gravitational source term in the
energy equation (\ref{eq:hydro_3}) is essentially equivalent to a
conservative discretization of Eq.~(\ref{eq:cons_static}), any
spurious energy loss or generation in the case of a static background
spacetime must originate from our finite-difference representation of
the neutrino moment equations. This can be verified by checking
whether the integral version of Eq.~(\ref{eq:cons_static}) for the
neutrinos is fulfilled numerically. More specifically, we consider the
rate of change of the total neutrino energy inside a given radius $R$,
which ought to be given by
\begin{equation}
\label{eq:cons_static_int}
\frac{\partial }{\partial t}
\int\limits_0^R \alpha \sqrt{-g} T^{00} \, \mathrm{d} r=
-\left[\alpha \sqrt{-g} T^{01}\right]_0^R+
\int\limits_0^R \alpha \sqrt{\gamma}Q \, \mathrm{d} r.
\end{equation}
In other words, the rate of change of the total neutrino energy
$\int_0^R \alpha \sqrt{-g} T^{00} \, \mathrm{d} r$ inside $R$ is given by a
surface flux term $-\left[\alpha \sqrt{-g} T^{01}\right]_0^R$ and a
volume-integrated source term $\int_0^R \alpha \sqrt{\gamma}Q \, \mathrm{d}
r$.  By plotting the LHS and RHS of
Eq.~(\ref{eq:cons_static_int}) as functions of the radius $R$ as in
Fig.~\ref{fig:enecons_neutrino}, one may determine directly where
energy conservation is actually violated: Ideally, the numerically
evaluated LHS and RHS (solid and dashed lines in
Fig.~\ref{fig:enecons_neutrino}) should be identical. Inside the
shock at $R\approx 100 \ \mathrm{km}$ the agreement is indeed very
good, the maximum deviation being $5 \times 10^{50} \ \mathrm{erg}
\ \mathrm{s}^{-1}$. Numerical energy conservation is violated more
strongly at the shock and in the pre-shock region, and the accumulated
error amounts to approximately $2 \times 10^{51} \ \mathrm{erg}
\ \mathrm{s}^{-1}$, which corresponds nicely to the rate of change of
the ADM energy at $t_\mathrm{pb}=230 \ \mathrm{ms}$.

The underlying reason for the numerical violation of neutrino energy
conservation can also be worked out easily: When deriving
Eq.~(\ref{eq:cons_static}) from the frequency-integrated moment
equations in the comoving frame on a static spacetime background, all
the Doppler and gravitational redshift terms are cancelled by
corresponding terms arising from the Lorentz transformation to the lab
frame and Eq.~(\ref{eq:cons_static}). For the finite-differenced
moment equations, these cancellations are no longer exact. Further
analysis shows that the most critical terms in this respect are the
velocity-dependent terms of $\mathcal{O}(v)$, which explains why
energy conservation is violated at the shock and the pre-shock region,
where the absolute velocities are highest. In this region, however,
neutrino heating and cooling are negligible (except for the early
post-bounce phase, where there may be some pre-heating), and the
dynamically relevant error in the heating rate resulting from energy
non-conservation in the neutrino sector is therefore much smaller.
Thus, the dynamical evolution of the accreting proto-neutron star and
its environment should not be significantly affected by the secular
increase of the ADM energy. Furthermore, since
Eq.~(\ref{eq:cons_static_int}) is violated mainly in the optically
thin region, it is quite natural to interpret the numerical increase
of $E_\mathrm{ADM}$ of the order of a few $10^{51} \ \mathrm{erg}
\ \mathrm{s}^{-1}$ as a luminosity error of a few percent and not as
spurious energy input. Considering that uncertainties in the
microphysics and the discretization error probably limit the accuracy
of the neutrino luminosities to a similar level anyway, the
conservation error observed in simulation G15 does not seem very
disturbing. Moreover, the problem with the finite-difference
representation of the Doppler terms in the moment equations becomes
less acute in the post-explosion phase once the shock has left the
grid: In model G8.8, the spurious change of the ADM energy between
$t_\mathrm{pb}=1 \ \mathrm{s}$ and $t_\mathrm{pb}=4 \ \mathrm{s}$ is
only $3.7\times 10^{51} \ \mathrm{erg}$. However, since the velocities
in the neutrino-driven wind can still reach a few percent of the speed
of light, the energy error still does not vanish completely.

Somewhat surprisingly, \emph{lepton number conservation} is also not
fulfilled exactly in models G15 and G8.8; there is a secular drift of
the order of $1 \% \ldots 2 \% $ of the net lepton number luminosity.
At first sight, this seems puzzling, since the treatment of Doppler
and gravitational redshift terms described in App.~\ref{app:cons_dopp}
allows us to solve the neutrino energy
equation~(\ref{eq:momeq_cfc_energy_j}) in such a way that lepton
number is conserved numerically. However, this apparent contradiction
can be easily resolved: In order to simplify the structure of the
Jacobian of the discretized moment equations, we compute the flux
terms $\partial /\partial \varepsilon (\ldots)$ using the flux factor
$f_H$ from the solution of the model Boltzmann equation and the zeroth
moment of the radiation intensity $J$, while the first moment $H$ is
used directly in the corresponding terms outside the derivative
$\partial /\partial \varepsilon$. If $f_H J=H$ held, the cancellation
of these terms described in App.~\ref{app:cons_dopp} would still work,
and lepton number conservation would be guaranteed; but we find $f_H J
\neq H$ for two reasons: First, our Boltzmann equation is not fully
consistent with the moment equations (\ref{eq:momeq_cfc_energy_j}) and
(\ref{eq:momeq_cfc_energy_h}) in the sense that some terms of the
underlying Boltzmann equation (\ref{eq:boltzmann_cfc}) are omitted,
and hence the flux factors computed from the solutions of the model
Boltzmann and moment equations do not agree perfectly, particularly in
the vicinity of the shock. Second, $f_H J$ and $H$ are defined on cell
centers and cell interfaces, respectively, so that the finite
interpolation accuracy will, in general, lead to $f_H J \neq H$. It
should be pointed out, however, that such problems are far less severe
in the optically thick regime, because there $H \approx 0$.  These
arguments suggest that the observed imperfect conservation of lepton
number conservation is again essentially equivalent to a small error
in the neutrino (number) luminosity.

\section{Conclusions}
We have presented a new method for general relativistic neutrino
radiation hydrodynamics in core collapse supernovae that combines
multi-group neutrino transport, an elaborate treatment of the
microphysics, multi-dimensional GR hydrodynamics and the conformal
flatness (CFC) approximation for the gravitational field
equations. Our approach is a generalization of the variable Eddington
factor technique as implemented by {Rampp} \& {Janka} (2002), and relies on the
ray-by-ray-plus approach ({Buras} {et~al.} 2006b) for treating the
multi-dimensional case. As in the Newtonian approximation
({Rampp} \& {Janka} 2002; {Buras} {et~al.} 2006b), the equations of neutrino transport
in spherical symmetry are solved on different angular bins (``rays'')
independently, assuming a radial neutrino flux and a parametric
dependence of the matter and spacetime background on
latitude. However, terms for lateral neutrino advection and
compressional heating of the neutrino gas are also included in the
correct relativistic form, as is the acceleration of the matter due to
lateral neutrino pressure gradients. Although we have considered only
the axisymmetric case in this paper, the method can be generalized to
3D in a straightforward manner.

Our approach involves a number of approximations to the full
six-dimensional relativistic neutrino transport problem on different
levels.  While the variable Eddington factor technique is well tested
in spherical symmetry in the context of core collapse supernovae
({Rampp} \& {Janka} 2002; {Liebend{\"o}rfer} {et~al.} 2004), no direct comparisons of the
ray-by-ray-plus method with multi-angle transport in 2D or 3D are
available as yet. However, there is at least qualitative agreement
with multi-angle transport (see {Ott} {et~al.}, 2008 for results obtained
with such an approach and {Marek} \& {Janka}, 2009 for a discussion of
similarities/dissimilarities to ray-by-ray transport),
even in situations where alternative approaches like multi-dimensional
flux-limited diffusion suffer from excessive lateral smearing of the
radiation field. Although ray-by-ray transport may not be the method
of choice for extremely aspherical configurations (such as accretion
tori around black holes), it is probably adequate for the typical core
collapse scenario where the proto-neutron star rotates only at
moderate speeds and is therefore not overly aspherical. As far as the
GR field equations are concerned, the introduction of CFC is well
justified, since it is an excellent approximation even for rapidly
spinning compact objects for which the ray-by-ray approach would break
down. We also note that the relativistic ray-by-ray method could
easily be generalized to the case where the Einstein equations are
solved in spherical polar coordinates without any approximations.

We have implemented our relativistic ray-by-ray-plus method using the
framework of two existing codes, the time-explicit relativistic
hydrodynamics code \textsc{CoCoNuT} ({Dimmelmeier} {et~al.} 2002a) and the
implicit neutrino transport code \textsc{VERTEX}
({Rampp} \& {Janka} 2002; {Buras} {et~al.} 2006b). Re-utilizing most of the routines of
\textsc{VERTEX} allows us to include the same up-to-date and
well-tested treatment of the neutrino microphysics as in the recent
pseudo-Newtonian simulations of our group
({Buras} {et~al.} 2006b, 2006a; {Kitaura} {et~al.} 2006; {Marek}, {Janka}, \& {M{\"u}ller} 2009; {Marek} \& {Janka} 2009; {H\"udepohl} {et~al.} 2009), which
greatly facilitates comparisons between the Newtonian and relativistic
cases.

A few improvements were made to the \textsc{CoCoNuT} code, such as the
implementation of the approximate Riemann solver HLLC
({Mignone} \& {Bodo} 2005).  More significantly, an improved scheme for the
energy equation was developed, which greatly reduces the numerical
violation of total energy conservation for self-gravitating
systems. The new scheme has also been added to the most recent version
of the \textsc{PROMETHEUS} code, and may be of use for other
grid-based Eulerian finite-volume codes as well. We also devised a
more efficient and consistent treatment for the advection of neutrinos
in energy space in the VERTEX module. The new method eliminates the
need to solve both the neutrino energy \emph{and} number equations to
maintain energy and lepton number conservation.

In this paper, we have limited ourselves to code tests in spherical
symmetry before moving on to the first multi-dimensional applications
of \textsc{VERTEX-CoCoNuT}. While the analytically tractable radiation
transport problems presented in Sec.~\ref{sec:rad_spheres} merely
serve as a consistency check for the implementation of the neutrino
moment equations, we also conducted a full simulation (``G15 run'') of
the collapse, bounce and post-bounce phases for the progenitor model
s15s7b2 of {Woosley} \& {Weaver} (1995), and compared our results to the
relativistic 1D Boltzmann solver \textsc{AGILE-BOLTZTRAN}
({Liebend{\"o}rfer} {et~al.} 2004), building on the detailed comparative study
of {Liebend{\"o}rfer} {et~al.} (2005). Since our code is largely an adapted
version of the Newtonian \textsc{VERTEX-PROMETHEUS} code of
{Rampp} \& {Janka} (2002), this also allows us to better assess the reliability
of the effective potential approach for GR effects
({Rampp} \& {Janka}, 2002; {Marek} {et~al.}, 2006; now also used e.g.\ by
{Messer} {et~al.}, 2007). For this reason, we have also included
\textsc{PROMETHEUS} runs with two different gravitational potentials
(case~R and case~A) in our comparison. Moreover, we performed
long-time proto-neutron star cooling simulations of the post-explosion
remnant of the $8.8 M_\odot$ progenitor model of
{Nomoto} (1984, 1987) with \textsc{VERTEX-CoCoNuT} and
\textsc{PROMETHEUS} (``G8.8 run'').

The results of our tests may be summarized as follows: On
the one hand, \textsc{VERTEX-CoCoNuT} is in very good agreement with
analytic solutions to simple test problems and also with AGILE in the
context of a fully-fledged 1D simulation from core collapse well into
the accretion phase. The agreement with AGILE is slightly better than
for the \textsc{PROMETHEUS} code with the best gravitational potential
currently available. On the other hand, our simulations of the $8.8
M_\odot$ model of {Nomoto} (1984, 1987) show that
\textsc{PROMETHEUS} still produces results extremely similar to
\textsc{CoCoNuT} well beyond the point up to which a direct comparison
to the full GR case was hitherto available in the work of
{Liebend{\"o}rfer} {et~al.} (2005).

The conclusions that can be drawn from these tests are twofold: First,
the \textsc{VERTEX-CoCoNuT} emerges as a reliable neutrino transport
code. Its accuracy has been demonstrated by a thorough comparison of
results with a well established relativistic 1D neutrino transport
code. The G8.8 run, covering more than 4~seconds of the proto-neutron
star cooling phase, also shows it to be sufficiently robust for
long-time simulations. For an explicit hydro code with an implicit
transport solver, the ability to stably evolve a quasi-stationary
proto-neutron star over thousands of dynamical time-scales and to
maintain total energy conservation adequately over such long evolution
times cannot be taken for granted, but our simulations with
\textsc{CoCoNuT} prove that these requirements can be met with an
up-to-date higher-order Eulerian scheme for GR
hydrodynamics. Naturally, the accuracy and robustness of
\textsc{CoCoNuT} in the 1D case will also serve as a point of
reference for future multi-dimensional numerical studies of core
collapse supernovae in general relativity.

The very good agreement of the results of \textsc{CoCoNuT} and
\textsc{PROMETHEUS}, both for the G15 and G8.8 run, is also noteworthy
because it provides further evidence that, at least in 1D, the
evolution of proto-neutron stars can be adequately captured by
approximating GR effects in a pseudo-Newtonian framework by means of
an effective gravitational potential. The usefulness and accuracy of
the best effective potential of {Marek} {et~al.} (2006) has now also been
established for the proto-neutron star cooling phase, and for much
longer evolution times and significantly more compact neutron stars
than in {Liebend{\"o}rfer} {et~al.} (2005) and {Marek} {et~al.} (2006). Despite the
smaller differences to the GR case seen during the early post-bounce
phase, this also gives further credence to 2D simulations using
effective potentials
({Buras} {et~al.} 2006b, 2006a; {Marek} {et~al.} 2009; {Marek} \& {Janka} 2009; {Bruenn} {et~al.} 2006; {Messer} {et~al.} 2007).

However, with the advent of \textsc{VERTEX-CoCoNuT}, the choice
between rather crude approximations for GR effects or a drastic
simplification of the microphysics in multi-dimensional supernova
studies is now obsolete. The role of GR effects in the
multi-dimensional case can now be investigated directly with
up-to-date microphysics and a sophisticated neutrino transport
scheme. There are several prominent topics in supernova physics where
these new capabilities of our code may prove highly valuable, and
which we plan to address in follow-up papers. In particular, we intend
to reinvestigate the dynamics of rotational core collapse and the
development of multi-dimensional hydrodynamic instabilities during the
post-bounce phase. The determination of accurate gravitational wave
signals -- which is connected to these issues -- is obviously another
important goal as well.

\appendix

\section{Energy Conservation in Newtonian and Relativistic Hydrodynamics}
\label{app:enecons}

\subsection{Importance of Energy Conservation in the Supernova Problem}
\label{sec:enecons_intro}
Conservative discretization schemes for the Euler equations of gas
dynamics are one of the cornerstones in modern computational
astrophysics, both because they have appealing mathematical properties
({LeVeque} 1998), and because they conserve important integral
quantities (total rest mass, total energy, etc.) to machine precision.
Unfortunately, the standard formulation of the Euler equations for
self-gravitating fluids is not strictly conservative despite the fact
that there exists an integral conservation law for the total kinetic,
internal, and potential energy for any \emph{localized}\footnote{This
  restriction is crucial, and has tangible implications: It is not
  possible, for example, to formulate a conservation law for the
  energy in a general cosmological context ({Straumann} 2004).} matter
distribution.

In the supernova problem, the non-conservative formulation of
gravitational source terms in the Euler equations raises some critical
questions: Can a secular drift of the total energy on the order of
$10^{51} \ \mathrm{erg}$ over a few $100 \ \mathrm{ms}$ (which is a
typical value for many simulations; see
{Liebend{\"o}rfer} {et~al.}, 2004; {Burrows} {et~al.}, 2006a; {Murphy} \& {Burrows}, 2008) have a qualitative
influence on the propagation of the shock and the development of the
explosion, as the explosion energy is typically of the same order?  Is
there an appreciable effect in the first few seconds of the neutron
star cooling phase? The former question may probably be answered in
the negative, because the accumulated error in the gravitational
energy liberated during the collapse of the iron core and the
contraction of the proto-neutron star makes itself felt primarily as
a small excess or deficit (on the percent level or below) in the
internal energy of the compact central object (which contains most
of the mass on the grid), and not as a large relative
error in the region around the shock. On the other hand, a secular
drift of the total energy can qualitatively affect neutron star
cooling in the post-explosion phase: Once spurious generation of
energy due to numerical inaccuracies outweighs the slowly decreasing
neutrino losses, the proto-neutron star does not cool any longer (such
an effect was actually observed in one of our pseudo-Newtonian
simulations).

In order to alleviate the problems associated with the
non-conservative standard formulation of gravitational source terms in
the energy equation, we have developed an alternative scheme for
treating the energy equation that significantly decreases the secular
drift of the total energy. The new scheme can easily be integrated
into any Eulerian finite-volume method, and is applicable (with only
small variations) both in Newtonian and relativistic hydrodynamics.
In the Newtonian case, the new scheme also has clear advantages when
working with effective gravitational potentials that do not obey the
Poisson equation, despite the fact that a conservation law for the
total energy cannot be formulated in this case: Using a modified
gravitational potential, it turns out that the energy released during
the formation of the neutron star is in good agreement with the
binding energy computed from the Tolman-Oppenheimer-Volkoff equation,
at least for neutron stars of moderate compactness
(cp. {H\"udepohl} {et~al.}, 2009), while systematic energy conservation
errors originating from the discretization of the gravitational source
term may qualitatively change the outcome of numerical simulations.

\subsection{Energy Equation in Newtonian Hydrodynamics}
\subsubsection{Description of the Algorithm}
\label{sec:enecons_algo_newton}
Usually the energy equation for a self-gravitating fluid is written
as
\begin{equation}
\label{eq:enecons_eq_1}
 \frac{\partial e}{\partial t}+\frac{\partial\left(e+P\right)v^i}{\partial x^i}=-\rho v^i \frac{\partial \Phi}{\partial x^i}.
\end{equation}
Here, $e=\rho v^2/2+\rho \epsilon$ is the total energy density, $P$ is
the pressure, $v^i$ is the fluid velocity, $\rho$ is the (baryonic)
mass density, $\epsilon$ is the specific internal energy, and $\Phi$ is
the Newtonian gravitational potential.  Although
Eq.~(\ref{eq:enecons_eq_1}) is not written in the form of a pure
conservation equation, there is a conservation law for the volume
integral of the total (i.e.\ kinetic, internal, and potential) energy
(see, e.g.\, {Shu}, 1992): In order to obtain this integral
conservation law, we absorb the source term $-\rho v^i \partial \Phi/\partial
x^i$ into the flux term on the LHS,
\begin{equation}
\label{eq:enecons_eq_2}
 \frac{\partial e}{\partial t}+\frac{\partial\left(e+P+\rho \Phi \right)v^i}{\partial x^i}=
\Phi \frac{\partial \rho v^i}{\partial x^i},
\end{equation}
and use the continuity equation $\partial \rho /\partial t +\nabla\cdot( \rho
\mathbf{v})=0$ to eliminate the divergence of the mass flux in the new
source term,
\begin{equation}
\label{eq:enecons_eq_3}
 \frac{\partial e}{\partial t}+\frac{\partial\left(e+P+\rho \Phi \right)v^i}{\partial x^i}=
-\Phi \frac{\partial \rho}{\partial t}.
\end{equation}
Integrating Eq.~(\ref{eq:enecons_eq_3}) over the entire spatial domain
then yields
\begin{eqnarray}
\label{eq:enecons_eq_4}
 \int \frac{\partial e}{\partial t}\, \mathrm{d} V
&=&
-\int \Phi \frac{\partial \rho}{\partial t} \, \mathrm{d} V =
\frac{1}{4 \pi G} \int \Phi \frac{\partial \Delta \Phi}{\partial t} \,\mathrm{d} V =
-\frac{1}{4 \pi G} \int \nabla \Phi \frac{\partial \nabla \Phi}{\partial t} \,\mathrm{d} V =
\\
\nonumber
&=&
-\frac{1}{8 \pi G} \frac{\partial}{\partial t} \int \nabla \Phi \cdot \nabla \Phi \,\mathrm{d} V =
\frac{1}{8 \pi G} \frac{\partial}{\partial t} \int \Phi \Delta \Phi \,\mathrm{d} V =
- \frac{1}{2} \frac{\partial}{\partial t} \int \rho \Phi \,\mathrm{d} V.
\end{eqnarray}
Here, we have used Green's first identity, the Poisson equation
$\Delta \Phi=4 \pi G \rho$, and the fact that the surface integral,
\begin{equation}
\int \Phi \nabla \Phi \, \mathrm{d} A,
\end{equation}
vanishes if the domain of integration is extended to
infinity. Eq.~(\ref{eq:enecons_eq_4}) implies that the total energy is
conserved,
\begin{equation}
\int \left(e+\frac{1}{2}\rho\Phi \right) \mathrm{d} V = \mathrm{const.}
\end{equation}

However, the \emph{discretized} representation of the total energy,
\begin{equation}
\label{eq:def_etot}
E_\mathrm{tot}=\sum_{i,j,k} \left(e_{i,j,k}+\frac{1}{2}\rho_{i,j,k} \Phi_{i,j,k} \right) \Delta V_{i,j,k},
\end{equation}
will not be automatically conserved, if the numerical implementation of
the gravitational source term is based on Eq.~(\ref{eq:enecons_eq_1}),
because the discretized source term cannot be manipulated in the same
fashion as in the steps from Eq.~(\ref{eq:enecons_eq_1}) to
(\ref{eq:enecons_eq_3}).  One way to eliminate any secular drift of the
total energy resulting from this, and to avoid the unwanted
consequences described in Sec.~\ref{sec:enecons_intro}, would be to
use a strictly conservative form of the energy equation
({Noonan} 1984),
\begin{equation}
\label{eq:enecons_noonan}
\frac{\partial}{\partial t}
\left[
e+\rho \Phi
+\frac{1}{8\pi G} \left(\frac{\partial \Phi}{\partial x^i}\right)^2
\right]
+\frac{\partial}{\partial x^i}
\left[\left(e+P+\rho \Phi\right)v^i + \frac{1}{4\pi G} \Phi \frac{\partial^2 \Phi}{\partial t \, \partial x^i}\right]
=0.
\end{equation}
Unfortunately, this form of the energy equation has several drawbacks:
The ``gravitational self-energy'' contribution $1/(8\pi G) \left(\partial
\Phi/\partial x^i\right)^2$ to the total energy density can exceed the
kinetic, internal, and binding energy of the matter by a large factor,
resulting in considerable round-off errors in these quantities.
Furthermore, the flux term contains a mixed spatial and temporal
derivative, which may introduce numerical inaccuracies.  It is also
impossible to formulate an analogue of Eq.~(\ref{eq:enecons_noonan}) for
the effective relativistic potentials commonly used in
\textsc{VERTEX-PROMETHEUS} ({Marek} {et~al.} 2006; {M{\"u}ller} {et~al.} 2008) because these do not
obey the Poisson equation.

However, there is a simple and efficient solution to the problem of
numerical energy conservation that avoids these shortcomings and
requires only a minimal modification of the ``standard'' scheme based
on Eq.~(\ref{eq:enecons_eq_1}). The starting point is the
reformulation of the energy equation in Eq.~(\ref{eq:enecons_eq_2}):
Using the continuity equation, the divergence of $\rho v^i$ can be
replaced by the time derivative of the density, which is in turn
partially absorbed in the time derivative of $\rho \Phi$,
\begin{equation}
\label{eq:enecons_new_eq}
 \frac{\partial \left(e+\rho \Phi\right)}{\partial t}+\frac{\partial\left(e+P+\rho \Phi \right)v^i}{\partial x^i}=
\rho \frac{\partial \Phi}{\partial t}.
\end{equation}
The remaining source term now contains the time derivative of the
gravitational potential. Eq.~(\ref{eq:enecons_new_eq}) can be conveniently solved with an
operator-split scheme in three steps.

\textbf{Step 1:} First, the equations of
hydrodynamics are solved with the additional potential energy terms $\rho
\Phi$, where $\Phi$ itself is kept constant,
\begin{equation}
\label{eq:enecons_op_0}
 \frac{\partial \left(e+\rho \Phi\right)}{\partial t}+\frac{\partial\left(e+P+\rho
   \Phi \right)v^i}{\partial x^i}=0 ,\quad \Phi=\mathrm{const.}
\end{equation}
Since $\rho \Phi$ may be considerably larger than the internal energy
density $e$, using $e+\rho \Phi$ as conserved quantity instead of $e$
may lead to large round-off errors in $e$ and produce unwanted
numerical noise (in particular spurious entropy loss or
production). To circumvent this problem, the time derivative $\partial \rho
\Phi /\partial t$ can be re-expressed in terms of the mass flux (as
$\Phi=\mathrm{const.}$), i.e.\ $\partial \rho \Phi /\partial t= \Phi \partial \rho
/\partial t =-\Phi \nabla\cdot\left(\rho \mathbf{v}\right)$,
\begin{equation}
\label{eq:enecons_op_1}
 \frac{\partial e}{\partial t}+\frac{\partial\left(e+P+\rho
   \Phi \right)v^i}{\partial x^i}-\frac{\partial \rho v^i}{\partial x^i} \Phi=0 , \quad \Phi=\mathrm{const.}
\end{equation}
The value $e^{(n+1/3)}$ of the energy density after the first
fractional step can be obtained by integrating
Eq.~(\ref{eq:enecons_op_1}) over the cell volume $V$. It is
instructive to compare the final formula for the update of $e$ with
``standard'' implementations of the gravitational source term. Working
with forward differences in time\footnote{First-order forward
  differences are only adopted for illustration at this point. The
  implementation in higher-order time integration schemes is
  straightforward.}, we now have
\begin{eqnarray}
\label{eq:enecons_op_1_disc}
e^{(n+1/3)}
=e^{(n)} -
\frac{\Delta t}{V}\left[\int_{\partial V} \left(e+P\right) \mathbf{v}\cdot \mathrm{d} \mathbf{A}
+\sum_k \int_{A_k} \left(\Phi_\mathrm{c}-\Phi_k \right) \rho \mathbf{v}\cdot \mathrm{d} \mathbf{A}_k
\right],
\end{eqnarray}
(where $A_k$ denotes the $k$-th cell interface, while
$\Phi_\mathrm{c}$ and $\Phi_k$ are the cell-center and
-interface values of $\Phi$) instead of
\begin{eqnarray}
\label{eq:enecons_alt_disc}
e^{(n+1/3)}
=e^{(n)}-
\frac{\Delta t}{V}
\left[
\int_{\partial V} \left(e+P\right) \mathbf{v}\cdot \boldmath{d} \mathbf{A}
\int\limits_V \nabla \Phi \cdot \rho \mathbf{v} \, \mathrm{d} \mathbf{V}
\right]
\end{eqnarray}
for the standard formulation.  Since $\int
\left(\Phi_k-\Phi_\mathrm{c} \right) \boldmath{d} \mathbf{A}_k/V$ is the
component of $\nabla \Phi$ pointing into the direction of $\boldmath{d}
\mathbf{A}_k$, the update prescription (\ref{eq:enecons_op_1_disc})
can be viewed as an alternative version of
Eq.~(\ref{eq:enecons_alt_disc}), in which the cell-averaged mass flux
$\rho \mathbf{v}$ has been replaced by an average of the mass flux at
cell interfaces. \emph{Thus, gravitational energy is only released or
  stored when mass is actually exchanged between different cells.}
This is the key ingredient for eliminating a secular drift of the
total energy.

When the gravitational field is stationary, no further fractional
steps are required, and the value of $e$ after one full time-step is
just $e^{(n+1)}=e^{(n+1/3)}$. If the gravitational potential varies
with time, two additional steps are necessary once $\Phi$ has been
updated from its old value $\Phi^{(n)}$ to $\Phi^{(n+1)}$ at the end
of the time-step. 

\textbf{Step 2:} In order to ensure that Eq.~(\ref{eq:enecons_op_0})
is fulfilled after the update of the gravitational potential
  $\Phi$, the energy density $e^{(n+2/3)}$ after the second
fractional step must be calculated according to
\begin{equation}
\label{eq:enecons_step_2}
 e^{(n+2/3)}=e^{(n+1/3)}+\rho^{(n+1)}\left(\Phi^{(n)}-\Phi^{(n+1)}\right).
\end{equation}

\textbf{Step 3:} Finally, the source term $\rho \partial \Phi/\partial t$ has to be accounted for
in a third step,
\begin{equation}
 \frac{\partial \left(e+\rho \Phi\right)}{\partial t}=\rho \frac{\partial \Phi}{\partial t}.
\end{equation}
If centred differences in time are used to obtain second-order
accuracy, the energy density at the end of the time-step reads
\begin{equation}
\label{eq:enecons_step_3}
 e^{(n+1)}=e^{(n+2/3)}+\frac{1}{2}\left(\rho^{(n)}+\rho^{(n+1)}\right)\left(\Phi^{(n+1)}-\Phi^{(n)}\right).
\end{equation}
Steps 2 and 3 can easily be combined into a compact formula for the
update of $e$ after each recalculation of the gravitational potential:
\begin{equation}
 e^{(n+1)}=e^{(n+1/3)}+\frac{1}{2}\left(\rho^{(n)}-\rho^{(n+1)}\right)\left(\Phi^{(n+1)}-\Phi^{(n)}\right).
\end{equation}

\subsubsection{Condition for the Exact Conservation of Energy}
\label{sec:exact_conservation}
Under certain conditions, the scheme described in
Sec.~\ref{sec:enecons_algo_newton} conserves the total energy exactly.
To demonstrate this for the special case of a spherical polar grid, we
consider the total energy balance after each sub-step of the algorithm:
Steps 1 and 2 conserve the (discretized) total volume integral of
$e+\rho \Phi$, provided that the flux $(e+P+\rho \Phi) \mathbf{v}$
vanishes at the boundaries of the computational domain,
\begin{equation}
\label{eq:bilanz_1}
\sum_{i,j,k} \left(e_{(i,j,k)}^{(n+2/3)}+\rho_{(i,j,k)}^{(n+1)}
\Phi_{(i,j,k)}^{(n+1)} \right) \Delta V_{i,j,k} = \sum_{i,j,k}
\left(e_{(i,j,k)}^{(n)}+\rho_{(i,j,k)}^{(n)} \Phi_{(i,j,k)}^{(n)}
\right) \Delta V_{i,j,k}.
\end{equation}
If $e$ is then updated according to Eq.~(\ref{eq:enecons_step_3}),
we obtain, by combination with Eq.~(\ref{eq:bilanz_1}),
\begin{eqnarray}
\lefteqn{\sum_{i,j,k} \left(e_{(i,j,k)}^{(n+1)}+\rho_{(i,j,k)}^{(n+1)}
\Phi_{(i,j,k)}^{(n+1)}\right) \Delta V_{i,j,k}
=}
\\
\nonumber
&&
\sum_{i,j,k}
\left[e_{(i,j,k)}^{(n)}
+\frac{1}{2}
\left(
\rho_{(i,j,k)}^{(n)} \Phi_{(i,j,k)}^{(n)}-
\rho_{(i,j,k)}^{(n+1)} \Phi_{(i,j,k)}^{(n)}+
\rho_{(i,j,k)}^{(n+1)} \Phi_{(i,j,k)}^{(n+1)}+
\rho_{(i,j,k)}^{(n)} \Phi_{(i,j,k)}^{(n+1)}
\right)
\right]
\Delta V_{i,j,k}.
\end{eqnarray}
Hence, the total energy after the $(n+1)$-th time-step is according to the definition in Eq.~(\ref{eq:def_etot}),
\begin{equation}
\label{eq:enecons_bilanz_1}
E_\mathrm{tot}^{(n+1)}= E_\mathrm{tot}^{(n)}
+\sum_{i,j,k} \rho_{(i,j,k)}^{(n)} \Phi_{(i,j,k)}^{(n+1)}  \Delta V_{i,j,k} 
-\sum_{i,j,k} \rho_{(i,j,k)}^{(n+1)} \Phi_{(i,j,k)}^{(n)} \Delta V_{i,j,k},
\end{equation}
The sums on the RHS quantify the violation of total energy
conservation during one time-step. If they were replaced by integrals,
we could manipulate them using Green's first identity and the Poisson
equation $\nabla \cdot \nabla \Phi=4 \pi G \rho $ to obtain an exact
cancellation:
\begin{equation}
\label{eq:enecons_bilanz_1b}
\int \rho^{(n)} \Phi^{(n+1)} \, \mathrm{d} V 
-\int \rho^{(n+1)} \Phi^{(n)} \, \mathrm{d} V=
\frac{1}{4 \pi G} \int \left(\nabla\Phi^{(n)}\right) \cdot \left(\nabla\Phi^{(n+1)}\right) \, \mathrm{d} V 
-\frac{1}{4 \pi G} \int \left(\nabla\Phi^{(n+1)}\right) \cdot \left(\nabla\Phi^{(n)}\right) \, \mathrm{d} V=0.
\end{equation}
Using a special finite-difference representation of the Poisson equation,
we can manipulate the sums in Eq.~(\ref{eq:enecons_bilanz_1}) in
very much the same way as by applying Green's first identity.
Let us assume that the source density $4 \pi G \rho$ is
expressed as the divergence of the gravitational acceleration
$\mathbf{g}$ in the following natural manner,
\begin{eqnarray}
4 \pi G \rho \Delta V_{(i,j,k)}
&=&
 g_{(i+1/2,j,k)} \Delta A_{(i+1/2,j,k)}
-g_{(i-1/2,j,k)} \Delta A_{(i-1/2,j,k)},
\nonumber
\\
&+&g_{(i,j+1/2,k)} \Delta A_{(i,j+1/2,k)}
-g_{(i,j-1/2,k)} \Delta A_{(i,j-1/2,k)},
\nonumber
\\
\label{eq:fd_potential_1}
&+&g_{(i,j,k+1/2)} \Delta A_{(i,j,k+1/2)}
-g_{(i,j,k-1/2)} \Delta A_{(i,j,k-1/2)}.
\end{eqnarray}
Here, $\Delta A_{(i+1/2,j,k)}$ denotes the area of the interface
between the cells $(i,j,k)$ and $(i+1,j,k)$, and $g_{(i+1/2,j,k)}$
denotes the perpendicular component of $\mathbf{g}$ thereon; $\Delta
A_{(i,j+1/2,k)}$, etc. are defined analogously. Let us further suppose
that the values of $\mathbf{g}$ on any cell interface are obtained by
multiplying the \emph{difference} of $\Phi$ in the adjacent cells by a
time-independent factor $\zeta_{i,j,k}$, as in the following natural
finite-difference representation,
\begin{eqnarray}
\label{eq:fd_potential_2a}
g_{(i+1/2,j,k)}&=&
\zeta_{i+1/2,j,k}\left(\Phi_{(i,j,k)} -\Phi_{(i+1,j,k)}\right)=:
\frac{1}{r_{i+1}-r_{i}}\left(\Phi_{(i,j,k)} -\Phi_{(i+1,j,k)}\right),
\\
\label{eq:fd_potential_2b}
g_{(i,j+1/2,k)}&=&
\zeta_{i,j+1/2,k}\left(\Phi_{(i,j,k)} -\Phi_{(i,j+1,k)}\right)=:
\frac{1}{r_i\left(\theta_{j+1}-\theta_j\right)}\left(\Phi_{(i,j,k)} -\Phi_{(i,j+1,k)}\right),
\\
\label{eq:fd_potential_2c}
g_{(i,k+1/2)}&=&
\zeta_{i,j,k+1/2}\left(\Phi_{(i,j,k)} -\Phi_{(i,j,k+1)}\right)=:
\frac{1}{r_i \sin\theta_j \left(\varphi_{k+1}-\varphi_k\right)}\left(\Phi_{(i,j,k)} -\Phi_{(i,j,k+1)}\right).
\end{eqnarray}
At the outer boundary, we may set
$g_{(M+1/2,j,k)}=\zeta_{M+1/2,j,k} \Phi_{(m,j,k)}=:\Phi_{(m,j,k)} r_m/r_{m+1/2}^2$ to obtain the correct
monopole moment of the gravitational field.

The first sum term in the energy balance equation (\ref{eq:enecons_bilanz_1})
can now be written as
\begin{eqnarray}
\nonumber
\sum_{i,j,k} 
\rho_{(i,j,k)}^{(n)} \Phi_{(i,j,k)}^{(n+1)} \Delta V_{i,j,k}
=\frac{1}{4 \pi G}
&&
\sum_{i,j,k}
 \left(g_{(i+1/2,j,k)}^{(n)} \Delta A_{(i+1/2,j,k)} -g_{(i-1/2,j,k)}^{(n)} \Delta A_{(i-1/2,j,k)}\right)
 \Phi_{(i,j,k)}^{(n+1)}
\\
\nonumber
&&
+
\sum_{i,j,k}
\left(g_{(i,j+1/2,k)}^{(n)} \Delta A_{(i,j+1/2,k)} -g_{(i,j-1/2,k)}^{(n)} \Delta A_{(i,j-1/2,k)}\right)
 \Phi_{(i,j,k)}^{(n+1)}
\\
&&
+
\sum_{i,j,k}
\left(g_{(i,j,k+1/2)}^{(n)} \Delta A_{(i,j,k+1/2)} -g_{(i,j,k-1/2)}^{(n)} \Delta A_{(i,j,k-1/2)}\right)
 \Phi_{(i,j,k)}^{(n+1)}
.
\label{eq:fdiff_energy}
\end{eqnarray}
The sums on the RHS of Eq.~(\ref{eq:fdiff_energy}) can be rearranged
via summation by parts using the finite-difference representation for
$\mathbf{g}$ in
Eqs.~(\ref{eq:fd_potential_2a}--\ref{eq:fd_potential_2c}) . For the
sum containing the radial components $g_{i+1/2,j,k}$ of the
gravitational acceleration $\mathbf{g}$ on cell interfaces, we have
(bearing in mind that $A_{(1/2,j,k)}=0$, as $i=1/2$ corresponds to the
origin):
\begin{eqnarray}
\nonumber
\lefteqn{
\sum_{i,j,k}
 \left(g_{(i+1/2,j,k)}^{(n)} \Delta A_{(i+1/2,j,k)} -g_{(i-1/2,j,k)}^{(n)} \Delta A_{(i-1/2,j,k)}\right)
 \Phi_{(i,j,k)}^{(n+1)}
}
\\
\nonumber
&=&
\sum_{i=1}^M
\sum_{j=1}^N
\sum_{k=1}^O
g_{(i+1/2,j,k)}^{(n)} \Delta A_{(i+1/2,j,k)} \Phi_{(i,j,k)}^{(n+1)}
-
\sum_{i=1}^M
\sum_{j=1}^N
\sum_{k=1}^O
g_{(i-1/2,j,k)}^{(n)} \Delta A_{(i-1/2,j,k)} \Phi_{(i,j,k)}^{(n+1)}
\\
&=&
\sum_{i=1}^M
\sum_{j=1}^N
\sum_{k=1}^O
g_{(i+1/2,j,k)}^{(n)} \Delta A_{(i+1/2,j,k)} \Phi_{(i,j,k)}^{(n+1)}
 -
\sum_{i=1}^{M-1}
\sum_{j=1}^N
\sum_{k=1}^O
g_{(i+1/2,j,k)}^{(n)} \Delta A_{(i+1/2,j,k)} \Phi_{(i+1,j,k)}^{(n+1)}
\\
\nonumber
&=&
\sum_{i=1}^{M-1}
\sum_{j=1}^N
\sum_{k=1}^O
g_{(i+1/2,j,k)}^{(n)} \Delta A_{(i+1/2,j,k)} \left(\Phi_{(i,j,k)}^{(n+1)}-\Phi_{(i+1,j,k)}^{(n+1)}\right)
+
\sum_{j=1}^N
\sum_{k=1}^O
g_{(M+1/2,j,k)}^{(n)} \Delta A_{(M+1/2,j,k)} \Phi_{(M,j,k)}^{(n+1)}
\\
\nonumber
&=&
\sum_{i=1}^{M-1}
\sum_{j=1}^N
\sum_{k=1}^O
\frac{1}{\zeta_{i+1/2,j,k}}
g_{(i+1/2,j,k)}^{(n)} \Delta A_{(i+1/2,j,k)} g_{(i+1/2,j,k)}^{(n+1)}
+
\frac{1}{\zeta_{M+1/2,j,k}}
\sum_{j=1}^N
\sum_{k=1}^O
g_{(M+1/2,j,k)}^{(n)} \Delta A_{(M+1/2,j,k)} g_{(M+1/2,j,k)}^{(n+1)}.
\end{eqnarray}
The remaining terms on the RHS of Eq.~(\ref{eq:fdiff_energy}) can be
manipulated in a similar fashion\footnote{Note that in spherical polar
  coordinates further ``surface contributions'' vanish since
  $A_{(i,1/2,k)}=A_{(i,N+1/2,k)}=0$, $\Phi_{(i,j,0)}=\Phi_{(i,j,O)}$,
  and $\Phi_{(i,j,O+1)}=\Phi_{(i,j,1)}$.}, and we finally obtain an
expression that is completely symmetric with respect to the time
indices $n$ and $n+1$:
\begin{eqnarray}
\nonumber
\sum_{i,j,k} 
\rho_{(i,j,k)}^{(n)} \Phi_{(i,j,k)}^{(n+1)} \Delta V_{i,j,k}
= \frac{1}{4 \pi G} 
&&
\left(
\frac{1}{\zeta_{i+1/2,j,k}}
\sum_{i=1}^{M-1} \sum_{j=1}^{N} \sum_{k=1}^{O} g_{(i,j+1/2,k)}^{(n)} \Delta A_{(i+1/2,j,k)} g_{(i+1/2,j,k)}^{(n+1)}
\right.
\\
\nonumber
&&
+ 
\frac{1}{\zeta_{i,j+1/2,k}}
\sum_{i=1}^{M} \sum_{j=1}^{N-1} \sum_{k=1}^{O} g_{(i,j+1/2,k)}^{(n)} \Delta A_{(i,j+1/2,k)} g_{(i,j+1/2,k)}^{(n+1)}
\\
\nonumber
&&
+ 
\frac{1}{\zeta_{i,j,k+1/2}}
\sum_{i=1}^{M} \sum_{j=1}^{N} \sum_{k=1}^{O-1} g_{(i,j,k+1/2)}^{(n)} \Delta A_{(i,j,k+1/2)} g_{(i,j,k+1/2)}^{(n+1)}
\\
&&
\left.
+
\frac{1}{\zeta_{M+1/2,j,k}}
\sum_{j=1}^{N} \sum_{k=1}^{O} g_{(M+1/2,j,k)}^{(n)} \Delta A_{(m+1/2,j,k)} g_{(m+1/2,j,k)}^{(n+1)}
\right)
.
\end{eqnarray}
In effect, we have obtained a finite-difference analogue of Green's
first identity with non-vanishing surface contributions. The second
sum in Eq.~(\ref{eq:enecons_bilanz_1}) can be rearranged in precisely the
same way and thus cancels the first sum exactly. Hence, the total
internal, kinetic and gravitational energy is conserved,
\begin{equation}
E_\mathrm{tot}^{(n+1)}= E_\mathrm{tot}^{(n)}.
\end{equation}

\subsection{Energy Equation in General Relativity}
In GR, the situation is somewhat different from the
Newtonian case: At first sight, the fact that no sources appear in the
local conservation law $\nabla_\nu T^{\mu\nu}=0$ for the stress-energy
tensor $T^{\mu\nu}$ might suggest that the energy and momentum
equations can be formulated in a strictly flux-conservative form
(i.e.\ without sources).  However, an integral conservation law can only be
formulated (via Gauss' theorem) for divergence-free vector fields on a
differentiable and orientable manifold ({Wald} 1984; {Straumann} 2004); such
fields can in general be constructed from the energy momentum tensor
only if there exists a Killing vector field\footnote{Special cases
  comprise: i) the flat Minkowski spacetime with ten Killing fields
  (implying the conservation of energy, momentum and angular
  momentum), ii) stationary spacetimes with a time-like Killing field
  (implying energy conservation), and iii) axisymmetric spacetimes
  with a space-like Killing field that creates an $SO(2)$ isometry
  group (implying the conservation of one angular momentum
  component).}.  On the other hand, a conservation law can
be formulated for the sum of $T^{\mu\nu}$ and the Landau-Lifshitz
pseudo-tensor $t^{\mu\nu}$ of the gravitational field
({Landau} \& {Lifschitz} 1997; {Straumann} 2004),
\begin{equation}
\label{eq:landau_lifschitz}
\frac{\partial \left(-g\right) \left(T^{\mu\nu}+t^{\mu\nu}\right)}{\partial x^\nu}=0.
\end{equation}
Unfortunately, the practical use of Eq.~(\ref{eq:landau_lifschitz}) is
limited because $t$ is a complicated (and non-unique) function of the
metric and its derivatives. Furthermore, it is impossible to interpret
$t^{\mu 0}$ as the local energy-momentum vector density of the
gravitational field due to its non-uniqueness and non-tensorial
character; only the integral $E_\mathrm{ADM}=\int \left(T^{\mu
  0}+t^{\mu 0}\right) \, \mathrm{d} V$ (known as ADM energy) over the entire
spatial domain in an asymptotically flat spacetime has a well-defined
physical meaning. In general, Eq.~(\ref{eq:landau_lifschitz}) will only
be useful for special gauge choices and symmetry assumptions, where
the resulting source-free energy equation is not overly complicated
(see e.g.\ {Romero} {et~al.}, 1996; {Liebend{\"o}rfer} {et~al.}, 2001a). The total volume
integral $E_\mathrm{ADM}$ (which can also be expressed in different
ways) is, of course, an important \emph{diagnostic} quantity for
assessing the quality of energy conservation in a numerical code.

Because of these obstacles we construct an improved scheme for the
energy equation in a similar way as in the Newtonian case and do not
attempt to eliminate source terms from the energy equation
altogether. We start from the formulation used by {Dimmelmeier} {et~al.} (2008),
\begin{equation}
\label{eq:ene_cfc_1}
  \frac{\partial \sqrt{\gamma} \tau}{\partial t}+
  \frac{\partial \sqrt{-g} \left(\tau \hat{v}^i + P v^i\right)}{\partial x^i}=
  \sqrt{-g} \left[ T^{00} \left(K_{ij} \beta^i \beta^j - \beta^i \frac{\partial \alpha}{\partial x^i}\right) +
  T^{0i} \left(2 K_{ij} \beta^j -\frac{\partial \alpha}{\partial x^i}\right) +
  T^{ij} K_{ij}\right].
\end{equation}
For convenience, we add the continuity equation, and subsume most of the source terms
on the RHS under the variable $Q$ to arrive at an equation very similar to Eq.~(\ref{eq:enecons_eq_1}),
\begin{equation}
\label{eq:ene_cfc_2}
  \frac{\partial \sqrt{\gamma} \left(\tau+D\right)}{\partial t}+ \frac{\partial
    \sqrt{-g} \left(\tau \hat{v}^i + D \hat{v}^i+ P v^i\right)}{\partial
    x^i}= - \sqrt{-g} \left[\left(\tau + D\right) \hat{v}^i+P
    v^i\right] \frac{\partial \alpha}{\partial x^i} + Q.
\end{equation}
The first source term on the RHS corresponds closely to the Newtonian
source term $\rho v^i \partial \Phi/\partial x^i$. It is the product of the
total energy flux (rest mass flux in Newtonian gravity) and the
flat-space gradient of the lapse function (corresponding to the
Newtonian potential).  This source term can be eliminated by
multiplying Eq.~(\ref{eq:ene_cfc_2}) by $\alpha$,
\begin{equation}
  \alpha \frac{\partial \sqrt{\gamma} \left(\tau+D\right)}{\partial t}+
  \alpha \frac{\partial \sqrt{-g} \left(\tau \hat{v}^i + D \hat{v}^i+ P v^i\right)}{\partial x^i}=
  - \alpha \sqrt{-g} \left[\left(\tau + D\right) \hat{v}^i+P v^i\right] \frac{\partial \alpha}{\partial x^i} + \alpha Q,
\end{equation}
and by pushing the lapse function into the space and time derivatives,
\begin{equation}
  \frac{\partial \sqrt{\gamma} \alpha \left(\tau+D\right)}{\partial t}+
  \frac{\partial \sqrt{-g} \alpha \left(\tau \hat{v}^i + D \hat{v}^i+ P v^i\right)}{\partial x^i}=
   \sqrt{\gamma} \left(\tau+D\right) \frac{\partial \alpha}{\partial t}+ \alpha Q.
\end{equation}
Again, the new source term $\left(\tau+D\right) \partial \alpha/\partial t$
corresponds closely to $\rho \partial \Phi/\partial t$ in the Newtonian case.
Since it is numerically advantageous to separate the baryonic mass
contributions to the total energy (due to the reduction of round-off
errors), we subtract the continuity equation, and finally arrive
at the following alternative energy equation:
\begin{equation}
\label{eq:rel_ene}
  \frac{\partial}{\partial t}\left[ \sqrt{\gamma} \alpha \left(\tau+D\right)
    -\sqrt{\gamma} D \right]+ \frac{\partial}{\partial x^i} \left[ \sqrt{-g}
    \left(\alpha \tau \hat{v}^i + \alpha D \hat{v}^i - D \hat{v}^i +
    \alpha P v^i\right)\right] = \sqrt{\gamma} \left(\tau+D\right)
  \frac{\partial \alpha}{\partial t}+ \alpha Q.
\end{equation}

The numerical solution within a finite-volume scheme can again be
split into three sub-steps, the first of which consists in solving the
energy equation (including only $\alpha Q$ as source term) while keeping
the lapse function $\alpha$ fixed,
\begin{equation}
  \frac{\partial}{\partial t}\left[ \sqrt{\gamma} \alpha \tau+ \sqrt{\gamma} \left(\alpha-1\right) D \right]+
  \frac{\partial}{\partial x^i} \left[ \sqrt{-g} \left(\alpha \tau \hat{v}^i + \alpha D \hat{v}^i - D \hat{v}^i + \alpha P v^i\right)\right] =
   \alpha Q.
\end{equation}
In order to avoid working with $\sqrt{\gamma} \alpha \tau+
\sqrt{\gamma} \left(\alpha-1\right) D $ as conserved quantity, we note
that the change in $\sqrt{\gamma} \left(\alpha-1\right) D$ inside a
cell volume $V$ during the time-step $\Delta t$ can be expressed in
terms of the integral of the flux $D \hat{v}^i$ over the cell boundary
$A=\partial V$. The finite-volume version of the energy equation for this
cell can therefore be written as
\begin{eqnarray}
\label{eq:fv_energy_1}
\int_V \alpha \sqrt{\gamma} \tau^{(n+1/3)} \, \mathrm{d} V&=&
\int_V \alpha \sqrt{\gamma} \tau^{(n)} \, \mathrm{d} V 
+\Delta t \left[
-\int_{\partial V} \sqrt{\gamma} \alpha^2 \left(\tau \hat{v}^i + P v^i\right) \, \mathrm{d} A 
\right.
\\
\nonumber
&&\left.
- \int_{\partial V} \sqrt{\gamma} \alpha \left(\alpha-1\right) D \hat{v}^i \, \mathrm{d} A
+ \left(\alpha-1\right) \int_{\partial V} \alpha \sqrt{\gamma} D \hat{v}^i \, \mathrm{d} A
+ \int_V \alpha Q \, \mathrm{d} V \right] 
.
\end{eqnarray}
The surface integrals in the second line of Eq.~(\ref{eq:fv_energy_1})
can be combined into a single term containing the difference between
the lapse function $\alpha_\mathrm{c}$ at the cell center and the lapse
function $\alpha_k$ at the k-th cell interface,
\begin{eqnarray}
\int_V \sqrt{\gamma} \tau^{(n+1/3)} \, \mathrm{d} V&=&
\int_V \sqrt{\gamma} \tau^{(n)} \, \mathrm{d} V 
+\Delta t \int_V Q \, \mathrm{d} V 
\\
\nonumber
&&
-\frac{\Delta t}{\alpha}
\left[
\int_{\partial V} \sqrt{\gamma} \alpha^2 \left(\tau \hat{v}^i + P v^i\right) \, \mathrm{d} A +
\sum_k \int_{A_k} \sqrt{\gamma} \alpha_k \left(\alpha_\mathrm{c}-\alpha_k\right) \, \mathrm{d} A
\right].
\end{eqnarray}
This formulation requires only a minimal modification of the scheme
for the unmodified energy equation~(\ref{eq:ene_cfc_1}).

In a second step, the lapse function is updated, while the conserved quantity
$\sqrt{\gamma} \left[\alpha \tau +\left(\alpha-1\right) D\right]$ is kept constant,
\begin{equation}
\label{eq:fv_energy_2}
\frac{\partial}{\partial t}\left[\alpha \hat{\tau} + \left(\alpha-1\right)
  \hat{D}\right]=0.
\end{equation}
Note that it is convenient to work with the densitized quantities
$\hat{\tau}=\sqrt{\gamma}\tau$ and $\hat{D}=\sqrt{\gamma}D$ here
because $\sqrt{\gamma}$ may be updated at the same time as $\alpha$.
The discretized version of Eq.~(\ref{eq:fv_energy_2}),
\begin{equation}
\alpha^{(n+1)} \hat{\tau}^{(n+2/3)} + \left(\alpha^{(n+1)}-1\right)\hat{D}^{(n+1)}=
\alpha^{(n)}\hat{\tau}^{(n+1/3)} + \left(\alpha^{(n)}-1\right)\hat{D}^{(n+1)},
\end{equation}
has the solution
\begin{equation}
\sqrt{\gamma} \tau^{(n+2/3)} =
\sqrt{\gamma} \tau^{(n+1/3)} + 
\left(\tau^{(n+1/3)}+D^{(n+1)}\right)
\left(\frac{\alpha^{(n)}}{\alpha^{(n+1)}}-1\right).
\end{equation}
In a third operator-split step, the source term containing
$\sqrt{\gamma} \left(\tau+D\right) \partial \alpha/\partial t$ in
Eq.~(\ref{eq:rel_ene}) is taken care of,
\begin{equation}
\frac{\partial}{\partial t}\left[\sqrt\gamma \alpha \tau + \sqrt{\gamma} \left( \alpha-1 \right) D\right]=
\sqrt{\gamma} \left(\tau+D\right) \frac{\partial \alpha}{\partial t}.
\end{equation}
As in the Newtonian case, the second and third step can be merged, and
if centred differences in time are used, $\tau$ has to be updated according to
\begin{equation}
\hat{\tau}^{(n+1)}=\hat{\tau}^{(n+1/3)}+
\frac{\hat{\tau}^{(n+1/3)}+\hat{\tau}^{(n)}+D^{(n+1)}+D^{(n)}}{2}
\frac{\alpha^{(n)}-\alpha^{(n-1)}}{\alpha^{(n)}}.
\end{equation}

\begin{figure}
\plotone{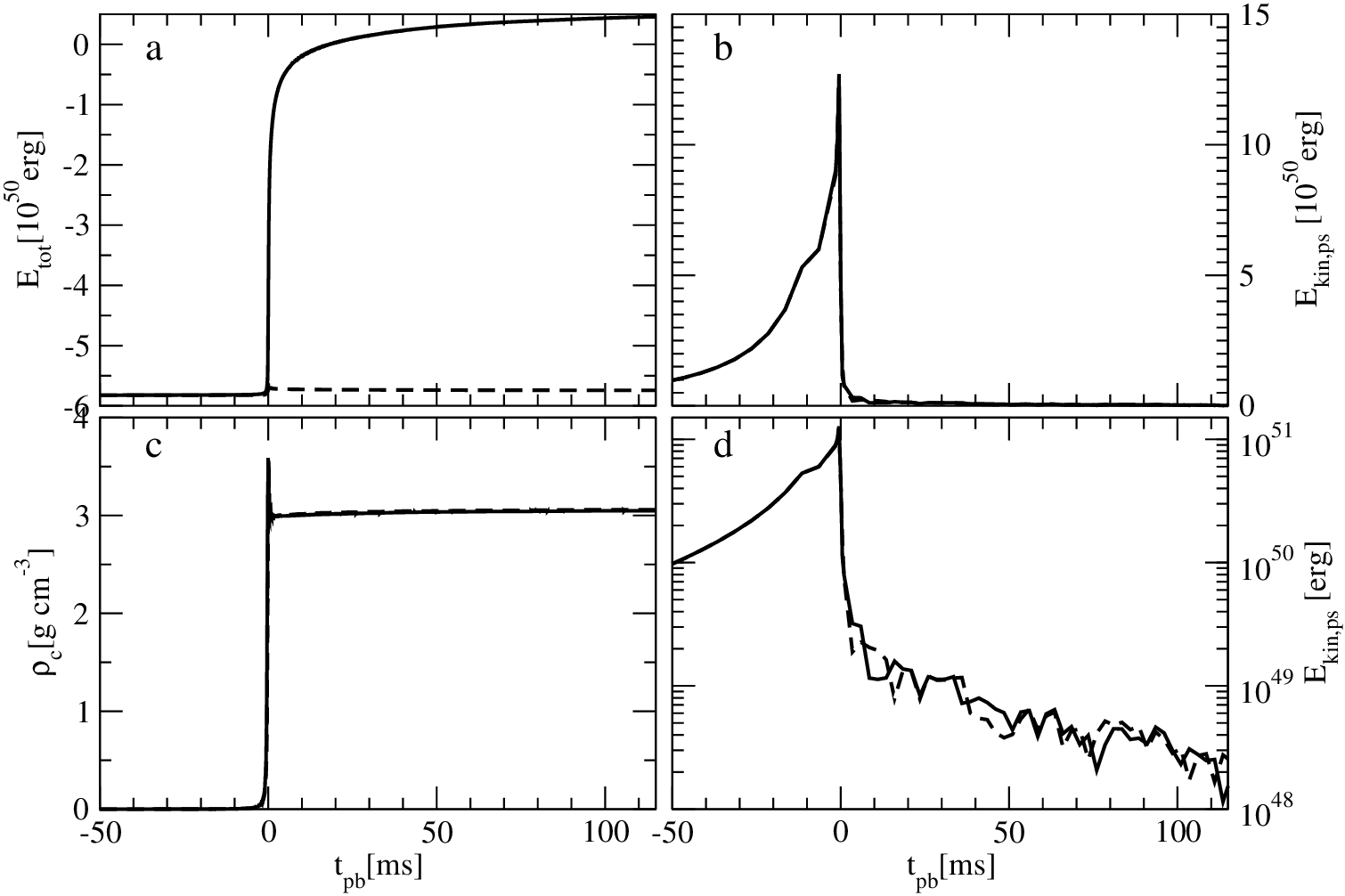} 
\caption{Total energy $E_\mathrm{tot}$, central density
  $\rho_\mathrm{c}$ (panel c), and total kinetic energy
  $E_\mathrm{kin,ps}$ inside the shock radius (panels b and d,
  on a linear and logarithmic scale, respectively) for a
  core collapse runs with \textsc{CoCoNuT} in Newtonian gravity, which
  has been carried out using the deleptonization scheme of
  {Liebend{\"o}rfer} (2005), and the progenitor model s20.0 of
  {Woosley}, {Heger}, \& {Weaver} (2002); the energy lost in the form of neutrinos is
  added to the total energy budget. Results obtained with the
  ``standard'' and improved scheme for the energy equation are drawn
  as solid and dashed lines, respectively.
\label{fig:enecons_newton}
}
\end{figure}

\begin{figure}
\plottwo{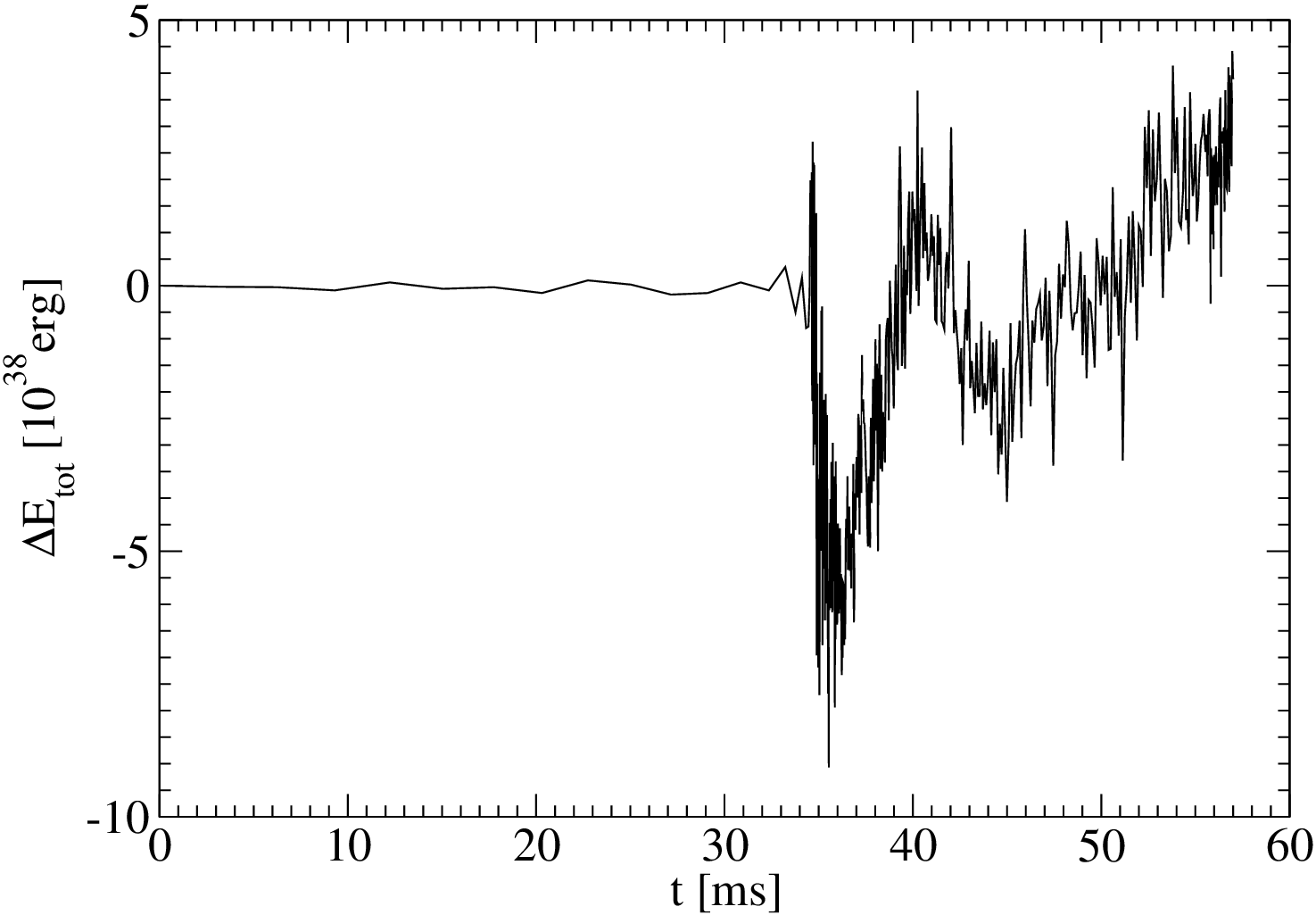}{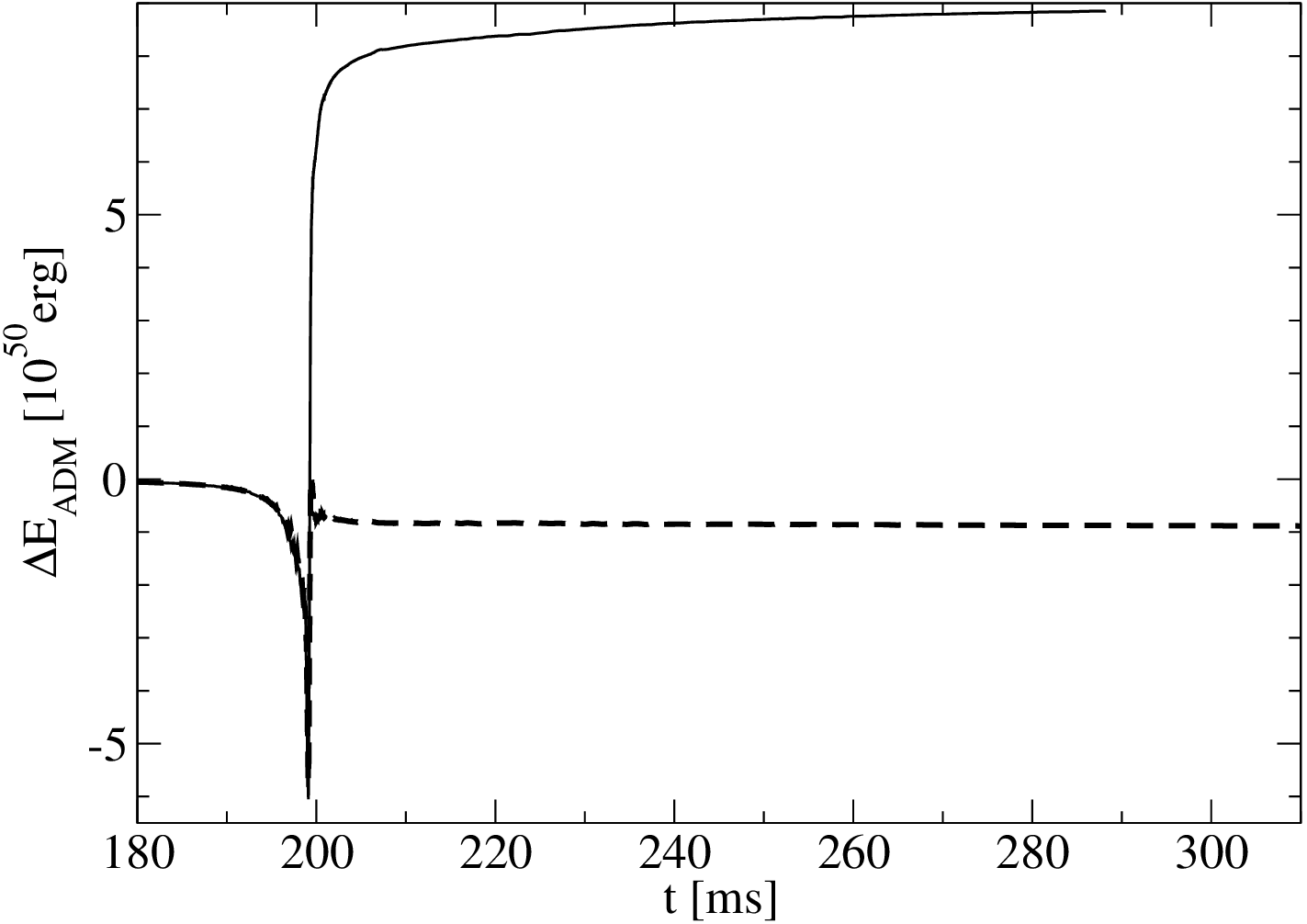} 
\caption{Left panel: Violation of total energy conservation for the
  collapse simulation of a 4/3-polytrope with the TVD test
  code. Initially the total energy of the polytrope is $-1.08 \times
  10^{51} \ \mathrm{erg}$.  Right panel: Violation of ADM energy
  conservation for a general relativistic core collapse run with
  \textsc{CoCoNuT} with the ``standard'' scheme (solid) and the
  improved scheme (dashed) for the relativistic energy equation.  The
  run has been carried out using the parameterized $Y_e(\rho)$
  trajectory of {Liebend{\"o}rfer} (2005), and the progenitor model
  s20.0 of {Woosley} {et~al.} (2002). Neutrino energy losses are neglected
  during the run.
\label{fig:enecons_tvd_and_cfc}
}
\end{figure}

\subsection{Numerical Tests}

To illustrate the virtues of the new scheme for the energy equation,
we have conducted several numerical tests in spherical symmetry with
different codes. Newtonian core collapse simulations have been carried
out with the Newtonian version of \textsc{CoCoNuT} and with a
self-written second-order accurate hydrodynamics code employing TVD
reconstruction and the Kurganov-Tadmor central scheme
({Kurganov} \& {Tadmor} 2000).

For the \textsc{CoCoNuT} run, we employ the deleptonization scheme of
{Liebend{\"o}rfer} (2005), and take neutrino energy losses into
account in the total energy budget, i.e.\ we subtract the
(negative) time- and volume-integrated neutrino energy source term from
the total energy measured at any given time. The left upper panel
of Fig.~\ref{fig:enecons_newton} shows that the new implementation
still produces a residual violation of energy conservation of about $2
\times 10^{49} \ \mathrm{erg}$ around bounce.  However, the total
energy changes by less than $1.7 \times 10^{48} \ \mathrm{erg}$ after
$t_\mathrm{pb}=10 \ \mathrm{ms}$, which is a considerable improvement
compared to the old implementation, where the accumulated error
exceeds the absolute value of the initial binding energy ($\approx 5.8
\times 10^{50} \ \mathrm{erg}$) within a few tens of $\mathrm{ms}$
after bounce. Nonetheless, the reduction of the numerical error is of
little relevance dynamically: The proto-neutron settles down at a
slightly higher central density, which is consistent with the smaller
total energy obtained with the new scheme. The total kinetic energy in
the post-shock region (right panels in Fig.~\ref{fig:enecons_newton})
is essentially identical in both cases as well. We also note that by
evolving $e$ instead of $e+\rho \Phi$, round-off errors in the
internal energy and the specific entropy can indeed be kept small; we
did not find any unphysical entropy production during collapse.

Although a change of the total energy by around $0.3\%$ as in the
\textsc{CoCoNuT} simulation is probably more than satisfactory for
core collapse simulations, this level of accuracy does not exhaust the
full potential of our new scheme, which can in principle conserve the
total energy exactly, as explained in
Sec.~\ref{sec:exact_conservation}. In order to verify this, we have
simulated the collapse of a $4/3$-polytrope with our self-written TVD
test code using the hybrid EoS of {Janka} {et~al.} (1993). Different from
\textsc{CoCoNuT}, the gravitational potential and acceleration are
calculated according to Eqs.~(\ref{eq:fd_potential_1} --
\ref{eq:fd_potential_2c}), hence the condition for exact energy
conservation is satisfied.  Fig.~\ref{fig:enecons_tvd_and_cfc} (left
panel) shows that the total energy changes by less than $10^{39}
\ \mathrm{erg}$ (!), which corresponds to a relative error of less
than $10^{-12}$ in terms of the initial energy
($E_\mathrm{tot,i}=-1.08\times 10^{51} \ \mathrm{erg}$) and less than
$10^{-14}$ in terms of the internal energy at the end of the
simulation. Thus our improved scheme can indeed preserve the total
energy almost to machine precision provided that the gravitational
potential is determined appropriately.

In addition to these tests in Newtonian gravity, we have also
simulated the collapse of model s20.0 in GR using \textsc{CoCoNuT}. We
employ the Liebend{\"o}rfer scheme for deleptonization with one minor
modification: While reducing $Y_e$ according to a
$Y_e(\rho)$-parameterization, we neglect energy losses due to neutrino
emission, because this simplifies the total energy budget
considerably\footnote{Setting the neutrino energy source term to zero
  while retaining a source term for $Y_e$ is of course unphysical,
  but, as far is our test problem is concerned, the prescription
  adopted here is only employed as a means of initiating the collapse.
  A non-zero neutrino energy source term would complicate the energy
  budget considerably: Any update of the internal energy of the fluid
  (or, equivalently, the entropy) would not only change the terms $h$
  and $P$ in the integrand in Eq.~\ref{eq:adm_energy} \emph{locally},
  but would also affect the conformal factor $\phi$ and the extrinsic
  curvature $K_{ij}$ \emph{non-locally}. On the other hand, this
  problem does not arise with full neutrino transport, because the
  neutrino source terms do not change the \emph{total} (matter and
  neutrino) energy-momentum tensor $T_\mathrm{tot}^{\mu\nu}$ in that
  case, and therefore do not affect the gravitational
  field.}. Ideally, we would expect the ADM energy $E_\mathrm{ADM}$,
which can be written as
\begin{equation}
\label{eq:adm_energy}
E_\mathrm{ADM}=-2 \pi \int \left( \phi^5 \rho h W^2-P+
\frac{K_{ij}K^{ij}}{16 \pi} \right)\,\mathrm{d} V
\end{equation}
in a CFC spacetime, to be conserved in such a
situation. Fig.~\ref{fig:enecons_tvd_and_cfc} (right panel) shows that
neither the ``standard'' scheme nor our new scheme manage to conserve
$E_\mathrm{ADM}$ perfectly. In both cases, conservation is violated by
$\left|\Delta E_\mathrm{ADM}\right| \approx 6 \times 10^{50}
\ \mathrm{erg}$ around bounce, which is not surprising considering the
complicated non-linear nature of the field equation and the appearance
of gravitational self-energy terms in the sources and in the ADM
energy. However, $\left|\Delta E_\mathrm{ADM}\right|$ settles down to
a value of less than $10^{50} \ \mathrm{erg}$ within $1 \ \mathrm{ms}$
after bounce in the simulation using our new scheme, and the secular
drift is only $10^{49} \ \mathrm{erg}$ for the next $100
\ \mathrm{ms}$ afterwards. The old scheme, on the other hand, leads to
a violation of around $9 \times 10^{50} \ \mathrm{erg}$ and a small
but clearly visible drift of $E_\mathrm{ADM}$ during the post-bounce
phase.

The relativistic test thus confirms our findings in Newtonian gravity:
Our reformulation considerably improves numerical energy conservation
in simulations of self-gravitating systems. The secular drift during
the quasi-stationary proto-neutron star evolution is not completely
eliminated in all cases, but can be easily reduced by an order of
magnitude, which demonstrates the usefulness of our new scheme
particularly for long-time simulations covering many dynamical
time-scales.

\section{Simultaneous Conservation of Energy and Lepton Number}
\label{app:cons_dopp}

\begin{figure}
\plotone{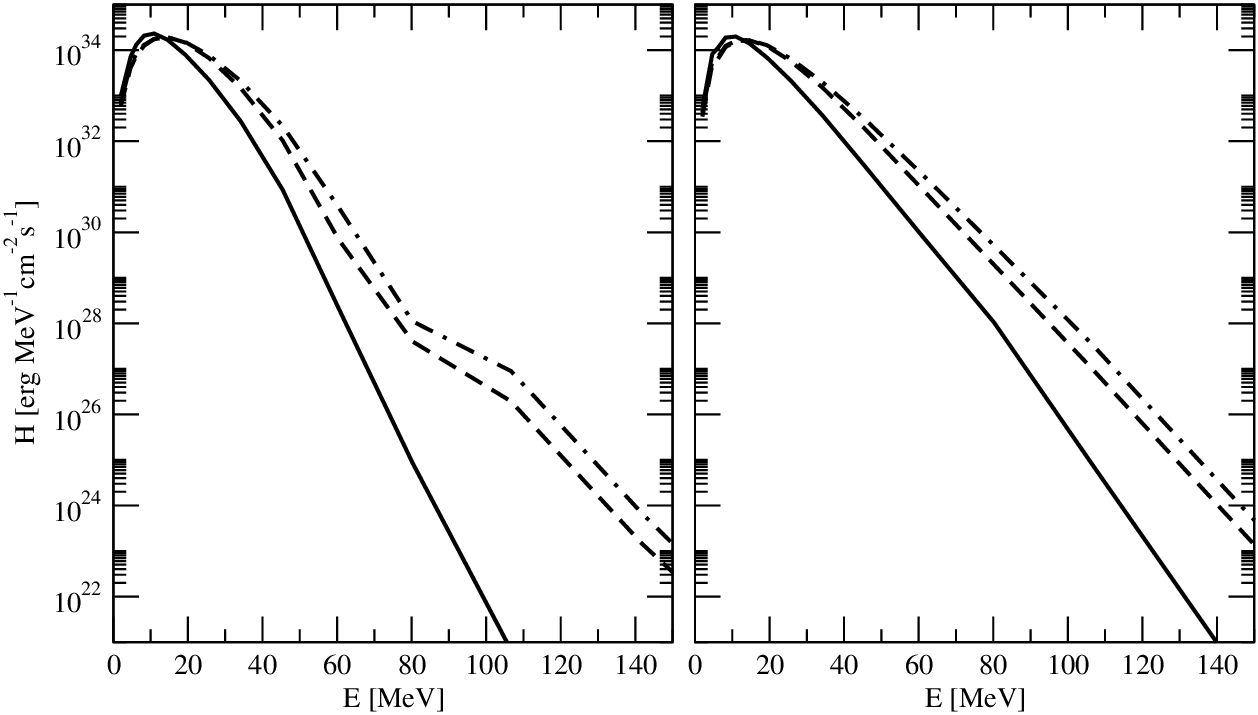}
\caption{Spectral energy flux of electron neutrinos (solid), electron
  antineutrinos (dashed) and $\mu$/$\tau$ (dash-dotted) neutrinos from the
  simulation of a cooling neutron star with \textsc{VERTEX-PROMETHEUS}. All
  quantities are measured at $r_\mathrm{cf}=400 \ \mathrm{km}$ in the comoving
  frame. The left panel shows spectra from $1 \ \mathrm{s}$ after
  bounce, obtained with the Doppler treatment of {Rampp} \& {Janka} (2002).
  The spectra in the right panel (from $t_\mathrm{pb}=1.5
  \ \mathrm{s}$) were obtained with our new conservative Doppler scheme
  using ``harmonic'' ($\alpha=1/2$) interpolation. Aside from some
  secular changes of the electron neutrino spectrum, the main
  difference is the elimination of the dip in the $\bar{\nu}_e$- and
  $\nu_\mu$/$\nu_\tau$ spectra at the neutrino energy $E\approx 80
  \ \mathrm{MeV}$.
\label{fig:spektren_cons_dopp}
}
\end{figure}

While the monochromatic moment equations
(\ref{eq:momeq_cfc_energy_j}, \ref{eq:momeq_cfc_energy_h}) and
(\ref{eq:momeq_cfc_number_j}, \ref{eq:momeq_cfc_number_h}) for neutrino
energy and number transport are analytically equivalent, a
flux-conservative discretization of either pair of equations does not
necessarily guarantee neutrino number (or, in the presence of
neutrino-matter interactions, lepton number) and energy conservation
simultaneously. The reason for this problem, which is inherent to
spectral neutrino transport in the comoving frame, can be understood
by comparing the analytic and discretized forms of the Doppler and
gravitational redshift terms (denoted by a subscript ``D'') in the
moment equations. In the neutrino energy equation these terms can be
written as
\begin{equation}
\left(\frac{\partial J}{\partial t}\right)_\mathrm{D}=\frac{\partial \varepsilon w J}{\partial \varepsilon} - w J,
\end{equation}
where the advection ``velocity'' in energy space $w$ is a function of
the velocities, metric functions, and the Eddington factors $f_H=H/J$,
$f_K=K/J$, and $f_L=L/J$. To obtain the corresponding term in the
equation for the monochromatic neutrino number density
$\mathcal{J}=J/\varepsilon$, we divide by $\varepsilon$ and apply the
product rule,
\begin{equation}
\nonumber
\left(\frac{\partial \mathcal{J}}{\partial t}\right)_\mathrm{D}=
\frac{\partial w J}{\partial \varepsilon} + \frac{w J}{\varepsilon} -\frac{wJ}{\varepsilon}
=
\frac{\partial w J}{\partial \varepsilon}
=
\frac{\partial \varepsilon w \mathcal{J}}{\partial \varepsilon}.
\end{equation}
The Doppler term in the neutrino number equation is thus a pure flux
derivative term and cancels out if the equation is integrated over
$\varepsilon$ to obtain the evolution equation for the total neutrino
number density.  On the other hand, if the energy equation is
discretized with backward differences in time and a flux-conservative
form for the energy advection term,
\begin{equation}
 \left(\frac{J_i^{n+1} \Delta \varepsilon_i-J_i^{n} \Delta \varepsilon_i}{\Delta t}\right)_\mathrm{D} =
\varepsilon_{i+1/2} w_{i+1/2} J_{i+1/2}^{n+1}- \varepsilon_{i-1/2} w_{i-1/2} J_{i-1/2}^{n+1}
-
w_i J_i^{n+1} \Delta \varepsilon_i,
\end{equation}
the discretized Doppler term for $\mathcal{J}$ reads
\begin{eqnarray}
\nonumber
 \left(\frac{\mathcal{J}_i^{n+1} \Delta \varepsilon_i-\mathcal{J}_i^n \Delta \varepsilon_i}{\Delta t}\right)_\mathrm{D}
&=&
\nonumber
\frac{\varepsilon_{i+1/2}}{\varepsilon_i} w_{i+1/2} J^{n+1}_{i+1/2}-
\frac{\varepsilon_{i-1/2}}{\varepsilon_i} w_{i-1/2} J^{n+1}_{i-1/2}
-
w_i J_i^{n+1} \frac{\Delta \varepsilon_i}{\varepsilon_i}
\\
&=&
\nonumber
w_{i+1/2} \varepsilon_{i+1/2} \mathcal{J}_{i+1/2}^{n+1}-
w_{i-1/2} \varepsilon_{i-1/2} \mathcal{J}_{i-1/2}^{n+1}
-
w_i J_i^{n+1} \frac{\Delta \varepsilon_i}{\varepsilon_i}
\\
\nonumber
&&
\label{eq:doppler_fdiff}
+
 \frac{\varepsilon_{i+1/2}-\varepsilon_i}{\varepsilon_i} w_{i+1/2} J_{i+1/2}^{n+1}
+\frac{\varepsilon_i-\varepsilon_{i-1/2}}{\varepsilon_i} w_{i-1/2} J_{i-1/2}^{n+1}
.
\\
\end{eqnarray}
Here, $\varepsilon_i$ and $\varepsilon_{i+1/2}$ denote the values of
the energy at cell centers and interfaces, respectively, and $\Delta
\varepsilon_i=\varepsilon_{i+1/2}-\varepsilon_{i-1/2}$ is the width of
the $i$-th zone in energy space. The source term in the neutrino
number equation thus does \emph{not} vanish in the discretized form of
the equations, unless the last three terms in
Eq.~(\ref{eq:doppler_fdiff}) cancel. Different methods have been
proposed to recover the property of neutrino number conservation
({Bruenn} 1985; {Liebend{\"o}rfer} {et~al.} 2004).  In the original version of VERTEX
({Rampp} \& {Janka} 2002) the problem is overcome by solving both the
monochromatic moment equations for the neutrino energy density and
flux ($J,H$), and the number density and flux
($\mathcal{J},\mathcal{H}$) for electron neutrinos and antineutrinos
\emph{simultaneously}, but there are significant drawbacks to this
approach: First, the mean energy $J/\mathcal{J}$ within a given energy
zone $[\varepsilon_{i-1/2},\varepsilon_{i+1/2}]$ is not constrained to
the cell-center value $\varepsilon_{i}$ and may even move beyond the
cell boundaries if $\mathcal{J}$ is treated as an independent
quantity. Moreover, the solution of the equations by the
Newton-Raphson method involves the inversion of square matrix blocks
of size $4\times N_\varepsilon+2$ instead of $2\times N_\varepsilon+2$
for neutrino energy transport alone, which makes the Newton-Raphson
iteration almost eight times more expensive.

Fortunately, the Doppler terms in the moment equations for neutrino
energy transport can be discretized in such a way as to guarantee
neutrino energy and number conservation simultaneously. For example,
in a rather simple approach the interface values $J_{i+1/2}^{n+1}$ can
be computed iteratively from $J_i$ and $J_{i-1/2}$ using the condition
\begin{equation}
\frac{\varepsilon_{i+1/2}-\varepsilon_i}{\varepsilon_i} w_{i+1/2} J_{i+1/2}^{n+1}=
w_i J_i \frac{\Delta \varepsilon_i}{\varepsilon_i}
-\frac{\varepsilon_i-\varepsilon_{i-1/2}}{\varepsilon_i} w_{i-1/2} J_{i-1/2}^{n+1}
,
\end{equation}
so that the last three terms in Eq.~(\ref{eq:doppler_fdiff}) would
indeed cancel. However, since this scheme allows large violations of
the monotonicity condition ($J_{i}<J_{i+1/2}<J_{i+1}$ or
$J_{i}>J_{i+1/2}>J_{i+1}$), the resulting algorithm would be rather
unstable. To construct a stable scheme, we need to impose a weaker
condition on the interface fluxes $F=\varepsilon w J$, i.e.\ we only
require the source term in the frequency-integrated neutrino number
equation to vanish:
\begin{equation}
\label{eq:source_j_tot}
 \left(\frac{\mathcal{J}_\mathrm{tot}^{n+1} -\mathcal{J}_\mathrm{tot}^n }{\Delta t}\right)_\mathrm{D} =\sum_i \left(
\frac{\varepsilon_{i+1/2}}{\varepsilon_i} w_{i+1/2} J_{i+1/2}^{n+1}-
\frac{\varepsilon_{i-1/2}}{\varepsilon_i} w_{i-1/2} J_{i-1/2}^{n+1}
-
w_i J_i^{n+1} \frac{\Delta \varepsilon_i}{\varepsilon_i}\right)=0.
\end{equation}
In order to find interface fluxes $F_{i+1/2}=\varepsilon_{i+1/2}
w_{i+1/2} J_{i+1/2}^{n+1}$ that fulfil this condition, we first
rearrange the sum in Eq.~(\ref{eq:source_j_tot}) in the following
manner:
\begin{eqnarray}
 \left(\frac{\mathcal{J}_\mathrm{tot}^{n+1} -\mathcal{J}_\mathrm{tot}^n }{\Delta t}\right)_\mathrm{D} &=&
\ldots
-\frac{F_{i-3/2}}{\varepsilon_{i-1}}
-w_i J_{i-1}^{n+1} \frac{\Delta \varepsilon_{i-1}}{\varepsilon_{i-1}}
+\frac{F_{i-1/2}}{\varepsilon_{i-1}} 
\\
\nonumber
&&
\phantom{\ldots}
-\frac{F_{i-1/2}}{\varepsilon_i}
-w_i J_i^{n+1} \frac{\Delta \varepsilon_{i}}{\varepsilon_{i}}
+\frac{F_{i+1/2}}{\varepsilon_i}
\\
\nonumber
&&
\phantom{\ldots}
-\frac{F_{i+1/2}}{\varepsilon_{i+1}}
-w_i J_{i+1}^{n+1} \frac{\Delta \varepsilon_{i+1}}{\varepsilon_{i-1}}
+\frac{F_{i+3/2}}{\varepsilon_{i+1}}
-\ldots
\\
\nonumber
&=&
\ldots
+\left(\frac{1}{\varepsilon_{i-2}}-\frac{1}{\varepsilon_{i-1}}\right)F_{i-3/2}
-w_i J_{i-1}^{n+1} \frac{\Delta \varepsilon_{i-1}}{\varepsilon_{i-1}}\\
&&
\phantom{\ldots}
+\left(\frac{1}{\varepsilon_{i-1}}-\frac{1}{\varepsilon_{i}}\right)F_{i-1/2}
-w_i J_{i}^{n+1} \frac{\Delta \varepsilon_{i}}{\varepsilon_{i}}\\
\nonumber
&&
\phantom{\ldots}
+\left(\frac{1}{\varepsilon_{i}}-\frac{1}{\varepsilon_{i+1}}\right)F_{i+1/2}
-w_i J_{i}^{n+1} \frac{\Delta \varepsilon_{i+1}}{\varepsilon_{i+1}}
+\ldots.
\end{eqnarray}
Next we split the fluxes into ``left'' and ``right'' components which
are assigned to the adjacent energy zones,
\begin{equation}
 F_{i+1/2}=F^\mathrm{L}_{i}+F^\mathrm{R}_{i+1},
\end{equation}
and again rewrite the frequency-integrated Doppler term in the
discretized neutrino number equation:
\begin{eqnarray}
\nonumber
 \left(\frac{\mathcal{J}_\mathrm{tot}^{n+1} -\mathcal{J}_\mathrm{tot}^n }{\Delta t}\right)_\mathrm{D} &=&
\ldots
+\left[
\left(
\frac{1}{\varepsilon_{i-2}}-\frac{1}{\varepsilon_{i-1}}\right) F^\mathrm{R}_{i-1}
-w_i J_{i-1}^{n+1} \frac{\Delta \varepsilon_{i-1}}{\varepsilon_{i-1}}
+\left(\frac{1}{\varepsilon_{i-1}}-\frac{1}{\varepsilon_{i}}\right) F^\mathrm{L}_{i-1}
\right]
\\
&&
\label{eq:source_j_tot_3}
\phantom{\ldots}
+\left[
\left(
\frac{1}{\varepsilon_{i-1}}-\frac{1}{\varepsilon_{i}}\right) F^\mathrm{R}_{i}
-w_i J_{i}^{n+1} \frac{\Delta \varepsilon_{i}}{\varepsilon_{i}}
+\left(\frac{1}{\varepsilon_{i}}-\frac{1}{\varepsilon_{i+1}}\right) F^\mathrm{L}_{i}
\right]\\
&&
\nonumber
\phantom{\ldots}
+\left[
\left(
\frac{1}{\varepsilon_{i}}-\frac{1}{\varepsilon_{i+1}}\right) F^\mathrm{R}_{i+1}
-w_i J_{i+1}^{n+1} \frac{\Delta \varepsilon_{i+1}}{\varepsilon_{i+1}}
+\left(\frac{1}{\varepsilon_{i+1}}-\frac{1}{\varepsilon_{i+2}}\right) F^\mathrm{L}_{i+1}
\right]+\ldots
.
\end{eqnarray}
If $F^\mathrm{R}$ and $F^\mathrm{L}$ are chosen such that each of the square brackets in
Eq.~(\ref{eq:source_j_tot_3}) vanishes, no source term in the
frequency-integrated neutrino number equation appears, and hence the
total lepton number is also conserved\footnote{Strictly speaking, this
  statement holds only as long as the non-linear moment equations are
  solved exactly. In practice, the time-critical Newton-Raphson iteration is
  terminated once a specified accuracy in the solution is reached, and
  a sufficiently low tolerance level must be chosen to guarantee
  long-time conservation of lepton number.}.

The condition we obtain for the ``left'' and ``right'' fluxes,
\begin{equation}
\left(
\frac{1}{\varepsilon_{i-1}}-\frac{1}{\varepsilon_{i}}\right) F^\mathrm{R}_{i}
+\left(\frac{1}{\varepsilon_{i}}-\frac{1}{\varepsilon_{i+1}}\right) F^\mathrm{L}_{i}
=w_i J_{i}^{n+1} \frac{\Delta \varepsilon_{i}}{\varepsilon_i},
\end{equation}
only fixes the (weighted) sum of $F_R$ and $F_L$, so we are still free
to specify the ratio $F_R/F_L$ to obtain steeper flux profiles in the
high-energy tail of the spectrum, where $J_i$ decreases rapidly with
the zone index $i$. To this end, we parameterize $F_R$ and $F_L$ by a
weighting factor $\xi_i$,
\begin{eqnarray}
\label{eq:doppler_f_l}
F^\mathrm{L}_i&=& \frac{w_i \Delta \varepsilon_{i}}{1-\varepsilon_{i}^{}
  \varepsilon_{i+1}^{-1}} J_i^{n+1} \xi_i, \\
\label{eq:doppler_f_r}
F^\mathrm{R}_i&=& \frac{w_i \Delta
  \varepsilon_{i}}{\varepsilon_{i}^{}\varepsilon_{i-1}^{-1}-1}
J_i^{n+1} \left(1-\xi_i\right).
\end{eqnarray}
Here, $\xi$ is defined as a function of the zeroth angular moment
$j=J h^3 c^2/\varepsilon^3$ of the neutrino distribution function at zone
interfaces (obtained as weighted geometric mean of $j$ in the adjacent
zones), and a steepness parameter $\sigma$,
\begin{equation}
 \xi_i=\frac{j_{i+1/2}^\sigma}{j_{i-1/2}^\sigma + j_{i+1/2}^\sigma}.
\end{equation}
Only the lowest energy zone constitutes an exception, since the
condition that $F(\varepsilon=0)$ should vanish requires
$\xi_1=1$. For the steepness parameter, we typically set $\sigma=1/2$
or $\sigma=1$; both choices result in a second-order accurate
scheme\footnote{Second-order accuracy can be proved by means of a
  Taylor series expansion, but this involves lengthy calculations that
  are omitted here.}.

The rationale for this seemingly complicated procedure for determining
the weighting factor can be illustrated as follows: First, we
specialize to the standard case of a logarithmic grid as used in VERTEX
\footnote{More precisely, the zone interfaces and centers are chosen as
follows: $\varepsilon_{i+1/2}=\Delta\varepsilon_0 \lambda^i$ (for
$i\in \{1,\ldots,N\}$), 
$\varepsilon_{-1/2}=0$, and
$\varepsilon_{i}=
\frac{1}{2}\left(\varepsilon_{i-1/2}+\varepsilon_{i+1/2}\right)$. Here,
$N$ is the number of energy zones, $\Delta\varepsilon_0$ is the width
of the first zone, and $\lambda>1$ is an adjustable parameter.}, so that
Eqs.~(\ref{eq:doppler_f_l}) and (\ref{eq:doppler_f_r}) simplify to,
\begin{eqnarray}
F^\mathrm{L}_i&=& w_i \varepsilon_{i+1/2} J_i^{n+1}  \xi_i ,\\
F^\mathrm{R}_i&=& w_i \varepsilon_{i-1/2} J_i^{n+1}  \left(1-\xi_i\right).
\end{eqnarray}
In regions where $J_i$ varies mildly with $i$, $\xi$ is close to
$1/2$, and hence the total interface flux is approximately given by
the arithmetic mean of the flux in the adjacent cells,
\begin{equation}
F_{i+1/2}=F^\mathrm{L}_i+F^\mathrm{R}_{i+1}\approx \frac{w_i \varepsilon_i J_i^{n+1}+w_{i+1} \varepsilon_{i+1} J_{i+1}^{n+1}}{2} . 
\end{equation}
On the other hand, in the high-energy tail of the spectrum $J_i$ drops
rapidly with increasing $i$, i.e.\  $J_{i+1} \ll J_{i}$. Assuming that
the spectrum can be approximated by a Fermi-Dirac distribution (with a
small chemical potential in the sense that $\mu \ll \varepsilon_i$),
$J_i$ is given in terms of the inverse temperature $\beta$ and a
normalization factor $k$ as
\begin{equation}
J_i  \approx k e^{-\beta \varepsilon_i} \varepsilon_i^3,
\end{equation}
and the weighting factor is approximately
\begin{equation}
 \xi\approx e^{-\beta \Delta \varepsilon_i} \ll 1.
\end{equation}
In this case the flux components read
\begin{eqnarray}
F^\mathrm{L}_i&=& w_i \varepsilon_{i+1/2} k e^{-\beta \left(\varepsilon_i +\sigma \Delta \varepsilon_i \right)} \varepsilon_i^3,
 \\
F^\mathrm{R}_i&=& w_i \varepsilon_{i-1/2} k e^{-\beta \varepsilon_i} \varepsilon_i^3.
\end{eqnarray}
Bearing in mind that $\varepsilon^3$ varies far more slowly than $e^{-\beta \varepsilon}$ in the steep tail of
the spectrum, we obtain
\begin{eqnarray}
F^\mathrm{L}_i \approx w_i \varepsilon_{i+1/2} k J_{i+1/2},
\qquad
\left| F^\mathrm{R}_i\right| \approx \left| w_i \varepsilon_{i-1/2} k J_{i} \right| 
\ll
\left| F^\mathrm{L}_{i-1} \right|,
\qquad
F_{i+1/2}\approx F^\mathrm{L}_i 
\end{eqnarray}
for $\sigma=1/2$, while for $\sigma=1$ we have,
\begin{eqnarray}
F^\mathrm{L}_i \approx w_i \varepsilon_{i+1/2} k  J_{i+1},
\qquad
F^\mathrm{R}_i \approx w_i \varepsilon_{i-1/2} k  J_{i} 
\qquad
F_{i+1/2}\approx \left(w_i+w_{i+1}\right) \varepsilon_{i+1/2}  k J_{i+1}.
\end{eqnarray}

Thus, in the case of $\sigma=1/2$, the effective interface value of
$J$ appearing in the fluxes is given by the weighted \emph{geometric}
mean of $J$ in the adjacent zones; for $\sigma=1$ it is twice the
minimum value in either zone, which corresponds more closely (but not
exactly) to the weighted \emph{harmonic} mean.  Since $\sigma=1$
generally reduces the absolute values of the fluxes, the resulting
scheme is somewhat more robust than for $\sigma=1/2$, i.e.\ the
Newton-Raphson iteration fails to convergence less frequently.

``Harmonic'' interpolation with $\sigma=1$ has another welcome side
effect in addition to the improved robustness of the scheme: The
method of geometric interpolation (also in the original scheme of
{Rampp} \& {Janka}, 2002) suffers from a mild ``kink'' instability, that may
produce an unphysical dip in the neutrino spectra. In a long-time
simulation of the neutron star cooling phase with
\textsc{VERTEX-PROMETHEUS}, in which our new scheme was switched on
about $1 \ \mathrm{s}$ after bounce, we found that ``harmonic''
interpolation eliminates such a dip, as shown in
Fig.~\ref{fig:spektren_cons_dopp}. Because of the favourable stability
properties and the smoother neutrino spectra, we preferably work with
$\sigma=1$, but $\sigma=1/2$ is also a viable choice.

\acknowledgements We are very grateful to A.~Marek and E.~M\"uller for
their input to various aspects of the reported project.  This work was
supported by the Deutsche Forschungsgemeinschaft through the
Transregional Collaborative Research Centers SFB/TR 27 ``Neutrinos and
Beyond'' and SFB/TR 7 ``GravitationalWave Astronomy'' and the Cluster
of Excellence EXC 153 ``Origin and Structure of the Universe''
(http://www.universe-cluster.de). The computations were performed on
on the IBM p690 and p575 of the Computer Center Garching (RZG) and the
NEC SX-8 at the HLRS in Stuttgart.  H.D.  acknowledges a Marie Curie
Intra-European Fellowship within the 6th European Community Framework
Programme (IEF 040464).

\bibliography{paper}

\end{document}